\pdfoutput=1
\documentclass[a4paper,11pt]{article}
\usepackage{jheppub}   %
\usepackage[T1]{fontenc}
\usepackage{slashed}
\usepackage{booktabs}
\usepackage{tabularx}
\usepackage{xcolor}
\usepackage{pifont}
\usepackage{upquote}   %
\usepackage{tikz}
\usepackage[section]{placeins}   %
\usetikzlibrary{decorations.pathmorphing,decorations.markings,arrows.meta}

\newcommand{\ii}{\mathrm{i}}
\newcommand{\dd}{\mathrm{d}}
\newcommand{\gfive}{\gamma_5}
\newcommand{\Lag}{\mathcal{L}}
\newcommand{\Tr}{\operatorname{Tr}}

\newcommand{\MSbar}{\overline{\mathrm{MS}}}
\newcommand{\order}[1]{\mathcal{O}(#1)}

\newcommand{\Rtwo}{R_2}

\newcommand{\SMEFTNLO}{SMEFT@NLO}

\newcommand{\vT}{v_T}
\newcommand{\vh}{\hat v}
\newcommand{\Lam}{\Lambda}
\newcommand{\vf}{\mathcal V}
\newcommand{\Cb}{\bar C}

\renewcommand{\Re}{\mathrm{Re}\,}
\renewcommand{\Im}{\mathrm{Im}\,}
\usepackage[normalem]{ulem}
\numberwithin{figure}{section}
\numberwithin{table}{section}

\definecolor{opverm}{RGB}{213,94,0}
\definecolor{readblue}{RGB}{0,114,178}
\tikzset{
  sfer/.style={line width=0.75pt, postaction={decorate},
    decoration={markings, mark=at position 0.55 with {\arrow{Stealth[length=3.6pt,width=3pt]}}}},
  slep/.style={sfer, line width=0.5pt},
  sglu/.style={line width=0.65pt, decorate, decoration={coil, aspect=0.55, amplitude=1.6pt, segment length=3.1pt}},
  spho/.style={line width=0.65pt, decorate, decoration={snake, amplitude=1.3pt, segment length=4pt}},
  shig/.style={line width=0.75pt, dashed},
  sopv/.style={circle, fill=opverm, inner sep=1.8pt},
  scur/.style={line width=2.4pt, draw=opverm},
  sread/.style={draw=readblue, line width=0.7pt, {Bar[width=5pt]}-{Stealth[length=4pt,width=3.4pt]}},
  sreadpt/.style={circle, draw=readblue, line width=0.8pt, minimum size=11pt, inner sep=0pt},
  sr2/.style={rectangle, fill=black, inner sep=2.1pt},
  sct/.style={circle, draw, fill=white, line width=0.6pt, inner sep=0pt, minimum size=7pt,
    path picture={\draw[line width=0.6pt] (path picture bounding box.north west) -- (path picture bounding box.south east)
      (path picture bounding box.south west) -- (path picture bounding box.north east);}},
  sblob/.style={circle, draw, fill=gray!30, line width=0.6pt, minimum size=11pt, inner sep=0pt},
  every picture/.append style={every node/.append style={font=\small}},
}

\preprint{MPP-2026-227}

\title{\boldmath Rational terms and UV counterterms of the SMEFT at next-to-leading order in QCD}

\author[a]{Giuseppe Ventura}
\affiliation[a]{Max-Planck-Institut f\"ur Physik (Werner-Heisenberg-Institut),\\
Boltzmannstra\ss e 8, 85748 Garching bei M\"unchen, Germany}
\emailAdd{gventura@mpp.mpg.de}

\abstract{We compute the rational terms $\Rtwo$ and the ultraviolet counterterms of the dimension-six Standard Model Effective Field Theory (SMEFT) at next-to-leading order (NLO) in QCD without any flavour assumption, for complex Wilson coefficients, massive quarks and a general CKM matrix. We also compute the on-shell renormalisation factors of the quark fields in the presence of flavour mixing. We work with an anticommuting $\gfive$ and read closed fermion loops with a single $\gfive$ at the operator insertion. The results, also for common flavour assumptions, are provided in electronic form and in a Mathematica package attached to the paper. We use them to build example models for \textsc{MadGraph5\_aMC@NLO} and give NLO predictions for representative processes.
}

\begin{document}
\maketitle
\raggedbottom   %

\section{Introduction}\label{sec:intro}
The Standard Model Effective Field Theory (SMEFT) is a widely used framework in which deviations from the Standard Model (SM) are
parametrised and constrained %
at colliders under the assumption that new particles are too heavy to be produced directly. %
At dimension six the operators are written in the Warsaw basis~\cite{Grzadkowski:2010es}. The one-loop
renormalisation-group equations of the SMEFT are known~\cite{Jenkins:2013zja,Jenkins:2013wua,Alonso:2013hga}, and results at two
loops~\cite{Bern:2020ikv,Born:2024mgz,DiNoi:2024ajj,DiNoi:2025tka,Thomsen:2025kka,Duhr:2025zqw,Duhr:2025yor,Duhr:2026btw,DiNoi:2025arz,DiNoi:2023ygk,Haisch:2024wnw,Haisch:2025vqj,Haisch:2025lvd} have
culminated in the two-loop running of the dimension-six SMEFT~\cite{Born:2026xkr}. Global interpretations of the data~\cite{Ethier:2021bye,Ellis:2020unq,deBlas:2025gyz,Armadillo:2026mvp} rely on precise theoretical predictions, which need both the running~\cite{Aoude:2022aro,DiNoi:2023onw,Maltoni:2024dpn,terHoeve:2025gey,Armadillo:2026mvp,Biekotter:2026dlb,Mantani:2026fao} and higher-order corrections. %
Among these, the next-to-leading-order (NLO) corrections in QCD reduce the scale dependence, modify the distributions through which operators are disentangled, and can substantially change the sensitivity to individual operators~\cite{Maltoni:2016yxb,Degrande:2020evl,Haisch:2025jqr,Bevilacqua:2026udk}.

Automated NLO event generators such as \textsc{MadGraph5\_aMC@NLO}~\cite{Alwall:2014hca} obtain the one-loop amplitudes from one-loop matrix-element generators, such as \textsc{MadLoop}~\cite{Hirschi:2011pa} %
and
\textsc{OpenLoops}~\cite{Cascioli:2011va,Buccioni:2019sur}, which evaluate the one-loop integrand numerically in four
dimensions, for instance with the method of Ossola, Papadopoulos and Pittau (OPP)~\cite{Ossola:2006us}. %
The part of the numerator in the remaining $D-4$ dimensions is therefore not included automatically.
Multiplied by the ultraviolet (UV) poles of the loop integrals, this part gives a finite and local contribution, the rational term $\Rtwo$~\cite{Ossola:2008xq}.
The one-loop generator takes it, together with the UV counterterms, as additional effective Feynman rules from the Universal FeynRules Output (UFO) model~\cite{Degrande:2011ua,Darme:2023jdn} of the theory, which contains all its Feynman rules. For the SM these rules are known~\cite{Draggiotis:2009yb,Garzelli:2009is,Pittau:2011qp},
and for renormalisable Lagrangians \textsc{NLOCT}~\cite{Degrande:2014vpa} derives them from a \textsc{FeynRules}~\cite{Alloul:2013bka}
implementation. %
At two loops the rational terms of UV origin are also local and process independent, so that they can be supplied in the same way~\cite{Pozzorini:2020hkx,Duhr:2023wsr}.

For the SMEFT at NLO in QCD the public sets of such vertices are those of the \SMEFTNLO{} model~\cite{Degrande:2020evl} and, for flavour-changing top-quark couplings, of the model of ref.~\cite{Degrande:2014tta}. Earlier models at NLO in QCD cover
electroweak Higgs production~\cite{Degrande:2016dqg}, top-quark operators~\cite{BuarqueFranzosi:2015jrv,Zhang:2016omx,BessidskaiaBylund:2016jvp,Degrande:2018fog} and an effective Lagrangian of the Higgs boson~\cite{Artoisenet:2013puc,Demartin:2014fia}, and the
rational terms of the effective coupling of the Higgs boson to gluons are given in ref.~\cite{Page:2013xla}.
The flavour
structure of \SMEFTNLO{} singles out the third generation, with massless light quarks, a $U(2)_q\times U(2)_u\times U(3)_d$ symmetry and a unit Cabibbo--Kobayashi--Maskawa (CKM)
matrix~\cite{Aguilar-Saavedra:2018ksv}, and CP-violating operators enter only in recent process-specific
extensions~\cite{Thomas:2024dwd,Rossia:2024rfo}. This suits top-quark physics, but it excludes
flavour-sensitive observables, as well as predictions for global fits without a flavour
symmetry or with other flavour assumptions~\cite{Allwicher:2023shc,Bartocci:2023nvp}. Tree-level models exist without flavour assumptions~\cite{Brivio:2017btx,Brivio:2020onw,Dedes:2023zws} and with user-defined flavour
symmetries~\cite{Hourtz:2026ijm}. One-loop matching has been automated~\cite{Fuentes-Martin:2022jrf,Carmona:2021xtq}, but the $\Rtwo$ and UV vertices that %
a one-loop generator needs are not
available for general flavour.

The $\Rtwo$ depends on the treatment of $\gfive$, for which we adopt naive dimensional
regularisation (NDR) with an anticommuting $\gfive$ in $D$ dimensions. A trace with a single $\gfive$ is then defined only once the vertex at which it is read, the reading point, is
fixed~\cite{Kreimer:1989ke,Korner:1991sx,Fuentes-Martin:2022vvu,Born:2026xkr}. The choice of the reading point, while changing the $\Rtwo$, leaves the UV poles untouched. The Dirac structures of a four-fermion vertex can also be reduced in different
ways, which define different evanescent schemes affecting the $\Rtwo$~\cite{Buras:1989xd,Dugan:1990df,Herrlich:1994kh,Dekens:2019ept,Fuentes-Martin:2022vvu}. In this work every closed trace with a single $\gfive$ is read at the operator insertion. This prescription preserves the colour and electroweak Ward identities and Bose symmetry, and needs no symmetry-restoring counterterm.

In this paper we compute the complete set of one-loop $\Rtwo$ vertices and UV poles of the dimension-six SMEFT at NLO in QCD. We work in
the broken phase and at first order in the Wilson coefficients, with the full baryon-number- and lepton-number-conserving Warsaw basis including
the CP-odd operators. The coefficients are general complex tensors, and all quark masses and the CKM matrix are kept symbolic. Each vertex is
computed with open generation indices, and its flavour dependence is a sum of matrix chains and traces of the coefficient tensors, the CKM matrix and the mass matrices. From the poles we construct the complete set of UV counterterm vertices, including the on-shell renormalisation factors of the fields and
masses, which with general flavour also account for the mixing of flavours and of masses~\cite{Denner:1990yz,Kniehl:1996bd,Espriu:2002xv,Degrande:2014tta,Zhang:2014rja}.
We also give the vertices for common flavour assumptions. These are the U35, minimal flavour violation (MFV), top and topU3l assumptions of
\textsc{SMEFTsim}~\cite{Brivio:2020onw} and the flavour assumption of \SMEFTNLO{}, which we extend to the CP-odd operators~\cite{Thomas:2024dwd,Rossia:2024rfo}. For each of them we
provide the map to the Warsaw basis. %

With these vertices, we build general-flavour example NLO models in the UFO
format~\cite{Degrande:2011ua,Darme:2023jdn} for \textsc{MadGraph5\_aMC@NLO} in the five-flavour scheme with the top quark as the only massive quark. We give NLO QCD predictions for representative processes, from top-quark flavour violation with charged leptons to a CP-odd asymmetry in $t\bar th$ production. The electronic files and the Mathematica package attached to this paper contain every ingredient needed to build such a model for any flavour assumption. The release of NLO models, with general flavour and with common flavour assumptions, is left to a subsequent publication.

Section~\ref{sec:setup} defines the Lagrangian, the mass basis and the set of effective vertices we compute. Section~\ref{sec:r2method} describes how these vertices are computed and section~\ref{sec:gamma5} states the treatment of $\gfive$ and of the four-fermion structures.
Section~\ref{sec:uv} turns the poles into the counterterms of an NLO model and section~\ref{sec:results} presents the results, the example
models and the validation. Section~\ref{sec:apps} presents the NLO predictions, and section~\ref{sec:conclusions} concludes.
Appendix~\ref{app:conventions} gives the details of the SMEFT in the broken phase, appendix~\ref{app:ct} of the computation of a vertex,
appendix~\ref{app:scheme} of the treatment of $\gfive$ and of the four-fermion structures, and appendix~\ref{app:os} of the renormalisation
factors. Appendix~\ref{app:smeftatnlo} compares our vertices with those of the \SMEFTNLO{} model, and appendix~\ref{app:anc} describes the electronic files and the Mathematica package that comes with them, with examples of its use.

\section{The SMEFT with general flavour}\label{sec:setup}
\subsection{Lagrangian and conventions}\label{sec:setup:lag}
We write the SMEFT Lagrangian, truncated at dimension six and restricted to baryon- and lepton-number-conserving operators, as
\begin{equation}\label{eq:lag}
  \Lag=\Lag_{\rm SM}+\Lag_6 ,\qquad
  \Lag_6=\frac1{\Lam^2}\Big[\sum_{i\in\mathcal H}C_i\,O_i+\sum_{i\in\mathcal N}\big(C_i\,O_i+C_i^*\,O_i^\dagger\big)\Big],
\end{equation}
where the $O_i$ are the operators of the Warsaw basis~\cite{Grzadkowski:2010es}, and $\mathcal H$ and $\mathcal N$ are the sets of Hermitian and
non-Hermitian operators, respectively. The sums also run over all generation indices of each operator. We work at first order in the Wilson coefficients, so that all the effective vertices computed in this work are of order $1/\Lam^2$. %
We do not absorb any gauge coupling, Yukawa coupling or factor $16\pi^2$ into the operators or the Wilson coefficients. The covariant derivative reads
\begin{equation}\label{eq:covd}
  D_\mu=\partial_\mu+\ii g_s\,T^AG^A_\mu+\ii g_2\,\frac{\tau^I}2W^I_\mu+\ii g_1\,\mathsf Y B_\mu ,
\end{equation}
with the sign of the Warsaw basis. We fix the Levi-Civita tensor and $\gfive$ by
\begin{equation}\label{eq:eps}
  \epsilon^{0123}=+1,\qquad \gfive=\ii\gamma^0\gamma^1\gamma^2\gamma^3,\qquad
  \Tr[\gamma^\mu\gamma^\nu\gamma^\rho\gamma^\sigma\gfive]=-4\ii\,\epsilon^{\mu\nu\rho\sigma},
\end{equation}
with $P_{L,R}=(1\mp\gfive)/2$, and define the dual field strengths as
\begin{equation}\label{eq:dual_fs}
  \widetilde X_{\mu\nu}=-\frac12\epsilon_{\mu\nu\rho\sigma}X^{\rho\sigma}
\end{equation}
with the Levi-Civita tensor of eq.~\eqref{eq:eps}.\footnote{This is the definition $\widetilde X_{\mu\nu}=\frac12\epsilon_{\mu\nu\rho\sigma}X^{\rho\sigma}$ of ref.~\cite{Grzadkowski:2010es}, with $\epsilon_{0123}=+1$, since $\epsilon_{0123}=-\epsilon^{0123}$.} Our tree-level Feynman rules are those of the general-flavour model of \textsc{SMEFTsim}~3~\cite{Brivio:2017btx,Brivio:2020onw}
implemented in \textsc{FeynRules}~\cite{Alloul:2013bka}, whose normalisation is that of eq.~\eqref{eq:lag}. All our Wilson coefficients refer to the Warsaw basis.\footnote{As we adopt the Warsaw basis, the only difference from the operators of the \textsc{SMEFTsim} model files is the sign of
$O_{Hud}$, whose \textsc{SMEFTsim} coefficient is $-C_{Hud}$. }

\subsection{Wilson coefficients and CP}\label{sec:setup:wc}
The Wilson coefficients are general complex tensors, constrained only by the hermiticity of $\Lag_6$ and by the symmetries of the operators.
The coefficients of the bosonic operators are real, those of the Hermitian two-fermion currents satisfy $C_{pr}=C^*_{rp}$ and those of the
Hermitian four-fermion operators
\begin{equation}\label{eq:herm4f}
  C_{prst}=C^*_{rpts},
\end{equation}
while the coefficients of the non-Hermitian operators are arbitrary complex tensors. For the operators built from two identical currents,
$O_{qq}^{(1,3)}$, $O_{uu}$, $O_{dd}$, $O_{ll}$ and $O_{ee}$, only the symmetric part of the coefficient contributes, and we
impose~\cite{Alonso:2013hga,Brivio:2020onw}
\begin{equation}\label{eq:sym4f}
  C_{prst}=C_{stpr}.
\end{equation}
Every complex coefficient appears in $\Lag_6$ together with its complex conjugate.
If $C$ multiplies $O$, then $C^*$ multiplies $O^\dagger$, and
\begin{equation}\label{eq:hermCP}
  C\,O+C^*O^\dagger=\mathrm{Re}\,C\,\big(O+O^\dagger\big)+\ii\,\mathrm{Im}\,C\,\big(O-O^\dagger\big).
\end{equation}
This occurs for the non-Hermitian operators, and for the Hermitian ones whenever $O^\dagger$ is a different flavour entry, as for
$(O^{(1)}_{Hq})_{12}^\dagger=(O^{(1)}_{Hq})_{21}$. CP exchanges $O$ and $O^\dagger$, and the real part of $C$ multiplies a CP-even operator and
the imaginary part a CP-odd one.
The real coefficients of the operators with a dual
field strength multiply CP-odd operators. Counted in this way the 2499 real parameters of the basis split into 1350 CP-even and 1149 CP-odd
ones~\cite{Alonso:2013hga}. Of these, 2245 enter the effective vertices of this work through their Feynman rules, 1205 CP-even and 1040
CP-odd.
The CP-odd parts of the vertices contain the Levi-Civita tensor and $\gfive$ traces, whose treatment in $D$ dimensions is discussed in section~\ref{sec:gamma5}.

\subsection{Regularisation}\label{sec:setup:reg}
We work in the Feynman gauge and in $D=4-2\epsilon$ dimensions. The loop momentum, the metric and the Dirac
matrices inside the loop are $D$-dimensional, while the external momenta, polarisations, spinors and the Lorentz indices of the external legs
are four-dimensional. This is the 't~Hooft--Veltman (HV) regularisation~\cite{tHooft:1972tcz} of
refs.~\cite{Draggiotis:2009yb,Gnendiger:2017pys}, with $\lambda_{\rm HV}=1$ in the notation of ref.~\cite{Draggiotis:2009yb}. The SM rational terms of one-loop generators are given in
the same regularisation~\cite{Draggiotis:2009yb,Garzelli:2009is,Degrande:2014vpa}. The HV regularisation fixes only the dimension of the external
states. It should not be confused with the treatment of $\gfive$ by 't~Hooft, Veltman, Breitenlohner and Maison, the BMHV scheme~\cite{tHooft:1972tcz,Breitenlohner:1977hr}, which we do not use. Inside the loops $\gfive$ is anticommuting, as in NDR, and section~\ref{sec:gamma5} gives our prescriptions for $\gfive$ and for the Levi-Civita tensor. We write every
one-loop one-particle-irreducible (1PI) Green's function as
\begin{equation}\label{eq:units}
  \Gamma=\frac{\ii}{16\pi^2}\,\mathcal V
\end{equation}
and quote $\mathcal V$ throughout. Its poles are the coefficients of $1/\epsilon$, and the factor $(4\pi)^\epsilon e^{-\epsilon\gamma_E}$ is
absorbed into the renormalisation scales as in the $\MSbar$ scheme. All numbers are for $N_c=3$ colours.

\subsection{The broken phase}\label{sec:setup:ewsb}
We work in the broken phase, with physical masses, measured input parameters and canonically normalised fields.\footnote{The implementation of \textsc{SMEFTsim} rescales $h$ but not $G^0$. Its field $G^0$ is
$1+\delta k_{G^0}$ times ours, so that its vertices follow from ours by setting $\delta k_{G^0}$ to zero.} The Higgs doublet is
\begin{equation}\label{eq:doublet}
  \varphi=\begin{pmatrix}-\ii G^+\\[2pt] \frac1{\sqrt2}\big(\vT+(1+\delta k_H)\,h+\ii\,(1+\delta k_{G^0})\,G^0\big)\end{pmatrix},
\end{equation}
where $\vT$ is the vacuum expectation value. In the SM it is fixed by the Fermi constant, $\vT=\vh$ with $\vh=(\sqrt2G_F)^{-1/2}$. At dimension six
$O^{(3)}_{Hl}$ and $O_{ll}$ also contribute to muon decay, so that $\vT=\vh(1+\delta G_F/2)$. The operators $O_{H\Box}$ and $O_{HD}$ change the normalisation of the kinetic term of $h$, and the
factor $1+\delta k_H$ restores it. The operator $O_{HD}$ changes that of $G^0$ as well, which $1+\delta k_{G^0}$ restores. In a term that already carries a Wilson coefficient, $\vT$ and $\vh$ differ only at
$\order{\Lam^{-4}}$, and we write $v$ for both. The shifts are
\begin{equation}\label{eq:dGFdkH}
\begin{gathered}
  \delta G_F=\frac{v^2}{\Lam^2}\Big((C^{(3)}_{Hl})_{11}+(C^{(3)}_{Hl})_{22}-(C_{ll})_{1221}\Big),\\[4pt]
  \delta k_H=\frac{v^2}{\Lam^2}\Big(C_{H\Box}-\frac{C_{HD}}4\Big),\qquad
  \delta k_{G^0}=-\frac{v^2}{\Lam^2}\,\frac{C_{HD}}4,
\end{gathered}
\end{equation}
and appendix~\ref{app:inputs} gives the relations of the input scheme.

In the QCD sector the operator $O_{HG}$ rescales the gluon kinetic term when $\varphi^\dagger\varphi$ is replaced by $v^2/2$. The term is
made canonical by
\begin{equation}\label{eq:gaugeredef}
  G^A_\mu\to\Big(1+C_{HG}\frac{v^2}{\Lam^2}\Big)G^A_\mu,\qquad
  g_s\to\Big(1-C_{HG}\frac{v^2}{\Lam^2}\Big)g_s,
\end{equation}
under which $g_sG^A_\mu$ does not change at order $1/\Lam^2$. The coefficient $C_{HG}$ therefore does not enter the covariant derivative, the
quark--gluon couplings or the pure-gluon couplings, and it appears only in the Feynman rules of $O_{HG}$ itself. Throughout, $g_s$ is the
coupling of the canonically normalised gluon field, the one that is renormalised in section~\ref{sec:uv} and run by the event generator. The
operators $O_{HW}$ and $O_{HB}$ are treated in the same way.

The quark fields of the Lagrangian, denoted with a prime, $q'$, $u'$ and $d'$, are in a generic flavour basis, in which their mass
matrices after electroweak symmetry breaking are not diagonal. The mass basis is obtained by unitary rotations,
\begin{equation}\label{eq:flavrot}
  u'_L=U_{u_L}u_L,\qquad d'_L=U_{d_L}d_L,\qquad u'_R=U_{u_R}u_R,\qquad d'_R=U_{d_R}d_R,
\end{equation}
which make both mass matrices diagonal at the scale $\mu_{\rm eft}$ at which the Wilson coefficients are given. The gauge and kinetic terms of the Lagrangian have the flavour
symmetry $U(3)_q\times U(3)_u\times U(3)_d$, so that three of the rotations of eq.~\eqref{eq:flavrot} can be absorbed into a redefinition of
the Yukawa matrices and the Wilson coefficients. Since the doublet contains both $u'_L$ and $d'_L$ and is rotated as a whole, the fourth
cannot, and only one of its components can be a mass eigenstate. Rotating it with $U_{u_L}$ gives
the up basis, used in \textsc{SMEFTsim}~\cite{Brivio:2020onw} and called \texttt{Warsaw up} in the Wilson coefficient exchange format (WCxf)~\cite{Aebischer:2017ugx},
\begin{equation}\label{eq:warsawup}
  q_p=\begin{pmatrix}u_{L,p}\\ V_{pr}\,d_{L,r}\end{pmatrix},\qquad l_p=\begin{pmatrix}\nu_{L,p}\\ e_{L,p}\end{pmatrix},\qquad
  u_p=u_{R,p},\quad d_p=d_{R,p},\quad e_p=e_{R,p},
\end{equation}
with the CKM matrix~\cite{Dedes:2017zog,Aebischer:2017ugx}
\begin{equation}\label{eq:ckmdef}
  V=U^\dagger_{u_L}U_{d_L} .
\end{equation}
Rotating it with $U_{d_L}$ gives instead the down basis, $q=(V^\dagger u_L,\,d_L)$, called \texttt{Warsaw} in WCxf. The Wilson coefficients are
defined in the chosen basis. We use the up basis throughout, and all our results can be obtained in the down basis, or in any other, with the
Mathematica package attached to this paper, described in appendix~\ref{app:anc}. The Yukawa matrices are fixed so that the mass terms contain
the physical masses, as shown in appendix~\ref{app:opsmass}.%

The CKM matrix is fixed by the rotations that diagonalise the mass matrices, which include the contributions of $O_{uH}$
and $O_{dH}$. It is unitary, with three angles and one phase once the quark fields are rephased, and it is renormalised in
section~\ref{sec:uv:ckm}. In the SMEFT it is not the coupling of the $W$ boson
to the quarks, which at order $1/\Lambda^2$ also receives contributions from $O^{(3)}_{Hq}$ and $O_{Hud}$~\cite{Dedes:2017zog,Brivio:2020onw}, and its
numerical value has to be extracted from data within the SMEFT~\cite{Descotes-Genon:2018foz,deBlas:2025xhe}.

The rational terms and poles keep all six quark masses non-zero and symbolic%
. The
example NLO models of section~\ref{sec:ufo} and the applications keep only the top quark massive. The leptons are
massless and do not mix.

The flavour indices of the Wilson coefficients refer to the fields of eq.~\eqref{eq:warsawup}. Since the lower component of $q_p$ is
$V_{pr}d_{L,r}$, every left-handed down-type quark that enters an operator through a doublet $q$ comes with a factor of the CKM matrix, $V$ for
$d_L$ and $V^\dagger$ for $\bar d_L$. For instance, the chromomagnetic dipole operator $O_{uG}$ contains
\begin{equation}\label{eq:uGexample}
  \Lag_6\supset\frac1{\Lam^2}\Big[\bar u_L\sigma^{\mu\nu}T^A\,C_{uG}\,u_R\,\frac{v+h-\ii G^0}{\sqrt2}
  -\ii\,\bar d_L\sigma^{\mu\nu}T^A\,(V^\dagger C_{uG})\,u_R\,G^-\Big]G^A_{\mu\nu}+\text{h.c.},
\end{equation}
where the first term, from the upper component of the doublet, couples the up-type quarks to the gluon, and the second, from its lower
component, couples them to a down-type quark, the gluon and a charged Goldstone boson. We keep these factors explicit in all results and do
not absorb them into the coefficients. The results are expressed in terms of $V$ and of the quark masses, and the rational terms and the poles are polynomials in the masses.
The case without flavour mixing, $V=1$, and the limit of massless light quarks therefore follow by direct substitution. In contrast, the on-shell renormalisation factors of section~\ref{sec:uv}, which contain logarithms of the masses, are given for a chosen set of massless quarks.

We fix the electroweak parameters in the $\{M_W,M_Z,G_F\}$ scheme~\cite{Berthier:2015oma,Brivio:2017vri,Brivio:2020onw}. In this scheme the operators $O^{(3)}_{Hl}$, $O_{ll}$, $O_{H\Box}$, $O_{HD}$ and $O_{HWB}$ change the relation between the inputs and the parameters of the Lagrangian.  In our vertices they enter only through the shifts $\delta G_F$, $\delta k_H$ and $\delta k_{G^0}$, the shifts $\delta g_2$ and $\delta g_1$ of the gauge couplings, given in
appendix~\ref{app:inputs}, and $C_{HWB}$. The operator $O_H$ changes only the Higgs potential and does not contribute.

\subsection{The set of effective vertices}\label{sec:scope}
We compute one-loop 1PI Green's functions with one insertion of a dimension-six operator, in dimensional
regularisation, in the broken phase. From each Green's function we extract two quantities. The first is the rational term $\Rtwo$ defined in
section~\ref{sec:r2:def}, the part of the one-loop amplitude missed by a computation with four-dimensional numerators, and the second is the
coefficient of the UV pole, which fixes the UV counterterm. We call the two together an effective vertex. We compute every effective vertex that satisfies two conditions. %
First, it receives a contribution from a 1PI diagram whose loop is made only of quark, gluon and ghost
propagators. Second, at least one of the external legs is a quark or a gluon. These are all the effective vertices needed for the $\order{\alpha_s}$ correction to a process with one insertion of a dimension-six operator. Quark loops without a factor $g_s$ are included only when at least one leg is coloured.
\newcommand{\scopetri}[2]{%
  \coordinate (G1) at (0,0.65); \coordinate (G2) at (0,-0.65); \coordinate (V) at (1.1,0);
  \draw[#1] (-0.85,1.0) -- (G1); \draw[#1] (-0.85,-1.0) -- (G2);
  \draw[sfer] (G1) -- (V); \draw[sfer] (V) -- (G2); \draw[sfer] (G2) -- (G1);
  #2
  \node[sopv] at (V) {};}
\newcommand{\scopebub}[2]{%
  \coordinate (V) at (0,0); \coordinate (H) at (1.3,0);
  \draw[#1] (-0.8,-0.8) -- (V); \draw[#1] (V) -- (-0.8,0.8);
  \draw[sfer] (V) arc (180:0:0.65); \draw[sfer] (H) arc (0:-180:0.65);
  \draw[#2] (H) -- (2.1,0);
  \node[sopv] at (V) {};}
\newcommand{\scopepic}[1]{\begin{tikzpicture}[scale=0.5,baseline={(0,-0.12)}]\useasboundingbox (-1.0,-1.35) rectangle (2.2,1.35);#1\end{tikzpicture}}
\newcommand{\scopeyes}{\textcolor{green!45!black}{\ding{51}}}
\newcommand{\scopeno}{\textcolor{red!70!black}{\ding{55}}}
\begin{table}[!htbp]
\centering\small
\renewcommand{\arraystretch}{1.15}
\begin{tabularx}{\textwidth}{@{}clcccc>{\raggedright\arraybackslash}X@{}}
\toprule
 & \multicolumn{1}{c}{diagram} & vertex & QCD loop & coloured leg & in the set & remark\\
\midrule
(a) & \scopepic{
  \coordinate (O) at (0,0); \coordinate (A) at (1.1,0.65); \coordinate (B) at (1.1,-0.65);
  \draw[sglu] (-0.9,0) -- (O);
  \draw[sfer] (1.9,-1.05) -- (B); \draw[sfer] (B) -- (O); \draw[sfer] (O) -- (A); \draw[sfer] (A) -- (1.9,1.05);
  \draw[sglu] (A) -- (B); \node[sopv] at (O) {};}
  & $q\bar qg$, $\order{C_{uG}}$ & \scopeyes & \scopeyes & \scopeyes & QCD correction to the dipole vertex\\
(b) & \scopepic{\scopetri{sglu}{\draw[shig] (V) -- (2.0,0);}} & $ggh$, $\order{C_{uH}}$ & \scopeyes & \scopeyes & \scopeyes & loop-induced $gg\to h$\\
(c) & \scopepic{\scopetri{sglu}{\draw[slep] (1.9,-0.8) -- (V); \draw[slep] (V) -- (1.9,0.8);}}
  & $\ell\bar\ell gg$, $\order{C^{(1)}_{lq}}$ & \scopeyes & \scopeyes & \scopeyes & loop-induced $gg\to\ell\bar\ell$\\
(d) & \scopepic{\scopebub{sfer}{shig}} & $t\bar th$, $\order{C^{(1)}_{qu}}$ & \scopeyes & \scopeyes & \scopeyes & four-quark loop without gluons\\
(e) & \scopepic{\scopetri{spho}{\draw[shig] (V) -- (2.0,0);}} & $h\gamma\gamma$, $\order{C_{uH}}$ & \scopeyes & \scopeno & \scopeno & colourless legs\\
(f) & \scopepic{\scopetri{shig}{\draw[shig] (V) -- (2.0,0);}} & $hhh$, $\order{C_{uH}}$ & \scopeyes & \scopeno & \scopeno & colourless legs\\
(g) & \scopepic{\scopebub{slep}{spho}} & $\ell\bar\ell\gamma$, $\order{C^{(1)}_{lq}}$ & \scopeyes & \scopeno & \scopeno & colourless legs\\
\bottomrule
\end{tabularx}
\caption{Representative effective vertices and the two conditions that define the set. ``QCD loop'' is the first condition, a
loop of quark, gluon and ghost propagators through the operator insertion, marked by the dot. ``Coloured leg'' is the second condition, at least one external quark or gluon. The
vertices of rows (e)--(g), whose external legs are all colourless, are not computed.}\label{tab:scope}
\end{table}

Table~\ref{tab:scope} applies the two conditions to representative vertices. Besides the QCD correction to a vertex of the operator, row~(a),
the set contains the closed quark loops of loop-induced processes such as $gg\to h$ and $gg\to\ell\bar\ell$, rows~(b) and~(c), and the closed
quark loops without gluons of the four-quark operators, row~(d). The second condition removes the vertices whose external legs are all
colourless, rows~(e)--(g). In the SM such vertices are counted among the electroweak rational terms~\cite{Garzelli:2009is}, and we do not
compute them. The set is defined by the two diagrammatic conditions, and not by a power of the strong coupling, since for example the closed quark loops of row~(d), which we compute, carry no power of $g_s$%
.

Power counting limits the number of external legs. The rational term and the UV pole of a 1PI Green's function are local, polynomials in the external momenta, the masses and $v$, as explained in section~\ref{sec:r2:def}. A 1PI Green's function with $f$ external fermions and $b$ external bosons has mass dimension
$4-\frac32f-b$, and its $\order{C}$ part carries the factor $C/\Lam^2$, so the polynomial that multiplies this factor has mass dimension
$6-\frac32f-b$. A local term at $\order{C}$ can therefore exist only if
\begin{equation}\label{eq:pc:omega}
  \omega=6-\tfrac32f-b\;\ge\;0 ,
\end{equation}
where $\omega$ is the superficial degree of divergence of a diagram with one insertion, and the
local part is a polynomial of degree $\omega$~\cite{Weinberg:1959nj}. Table~\ref{tab:scope:pc} lists the classes that satisfy this bound. A class at the bound, such
as the four-fermion or the six-boson vertices, has a constant local term, and each unit of $\omega$ above zero allows one more power of momenta,
masses or $v$.\label{sec:r2:pc} Vertices that vanish by colour, Furry's theorem, parity or CP
are counted in the set with a zero result. Appendix~\ref{app:pc} gives the details of the counting.

An effective vertex is labelled by its external fields, with open generation indices. An operator contributes to a vertex directly, through its own Feynman rules, or indirectly, through the
shifts of the electroweak inputs of section~\ref{sec:setup:ewsb}, as in the QCD correction to $q\bar qZ$ through $\delta g_2$ and
$\delta g_1$. In total the set contains 396 effective vertices, a vertex and its charge conjugate counted separately, with 4743 direct and 25
indirect contributions. The four-quark contributions are given in the reduction scheme
defined in section~\ref{sec:schemes}.

\begin{table}[!htbp]
\centering\small
\begin{tabular}{@{}cccrr>{\raggedright\arraybackslash}p{0.46\textwidth}@{}}
\toprule
$f$ & $b$ & $6-\frac32f-b$ & vertices & contributions & examples\\
\midrule
0 & 2 & 4 & 1 & 6 & $gg$ at $\order{C_G}$ and $\order{C_{uG}}$\\
0 & 3 & 3 & 5 & 36 & $ggg$ at $\order{C_G}$, $ggh$ at $\order{C_{HG}}$ and $\order{C_{uH}}$\\
0 & 4 & 2 & 19 & 153 & $gggg$ at $\order{C_G}$, $gghh$ at $\order{C_{HG}}$\\
0 & 5 & 1 & 43 & 314 & $ggggg$ at $\order{C_G}$, $ggghh$ at $\order{C_{HG}}$\\
0 & 6 & 0 & 47 & 186 & $gggggg$ at $\order{C_G}$, $gggghh$ at $\order{C_{HG}}$\\
2 & 0 & 3 & 2 & 28 & $q\bar q$ at $\order{C_{uG}}$, $q\bar q$ of four-quark operators\\
2 & 1 & 2 & 14 & 256 & $q\bar qg$ at $\order{C_{uG}}$, $q\bar qh$ at $\order{C_{HG}}$\\
2 & 2 & 1 & 60 & 1008 & $q\bar qgg$, $q\bar qgh$ at $\order{C_{uG}}$\\
2 & 3 & 0 & 196 & 2694 & $q\bar qggg$, $q\bar qZgg$ at $\order{C_G}$\\
4 & 0 & 0 & 9 & 62 & four-quark and semileptonic $q\bar q\ell\bar\ell$\\
\midrule
\multicolumn{3}{@{}l}{total} & 396 & 4743 & \\
\multicolumn{3}{@{}l}{all other $(f,b)$} & 0 & 0 & $6-\frac32f-b<0$, for instance $q\bar qgggg$ or $q\bar qq\bar qg$\\
\bottomrule
\end{tabular}
\caption{Classes of effective vertices with $f$ external fermions and $b$ external bosons allowed by eq.~\eqref{eq:pc:omega},
with the number of vertices of the class, the number of direct operator
contributions to them, and examples. The 25 indirect contributions through the input shifts are not included.}\label{tab:scope:pc}
\end{table}

\section{Computation of the one-loop effective vertices}\label{sec:r2method}
In this section we discuss the methods of our computations. Throughout, the diagrams are generated with \textsc{QGRAF}~\cite{Nogueira:1991ex} from the Feynman rules of the Lagrangian of section~\ref{sec:setup}, and the
Dirac and colour algebra, the contractions and the integration are carried out with
\textsc{FORM}~\cite{Vermaseren:2000nd,Ruijl:2017dtg}.

\subsection{Rational terms \texorpdfstring{$\Rtwo$}{R2}}\label{sec:r2:def}
A one-loop amplitude takes the general form
\begin{equation}\label{eq:r2:integral}
  \mathcal A=\int\!\frac{\dd^Dk}{(2\pi)^D}\,\frac{N(k)}{D_0D_1\cdots D_{n-1}},\qquad D_i=(k+p_i)^2-m_i^2 ,
\end{equation}
summed over the diagrams, where $N(k)$ is the numerator of a diagram with $n$ propagators. It can be reduced to scalar box, triangle,
bubble and tadpole integrals and a rational remainder $R$~\cite{tHooft:1978jhc,Passarino:1978jh}.

In dimensional regularisation the loop momentum, the Dirac matrices and the metric live in $D=4-2\epsilon$ dimensions. We split each of them into
a four-dimensional part, written with a bar, and a part in the remaining $D-4=-2\epsilon$ dimensions, written with a tilde,\footnote{The bars and
tildes play the role of the bars and hats of the literature on the BMHV scheme~\cite{Breitenlohner:1977hr,Belusca-Maito:2020ala}, while in refs.~\cite{Ossola:2008xq,Draggiotis:2009yb} a bar denotes the
$D$-dimensional loop momentum.} so that $k=\bar k+\tilde k$ and
\begin{equation}\label{eq:r2:split}
  N(k)=\bar N(\bar k)+\tilde N(\bar k,\tilde k,\epsilon) .
\end{equation}
A one-loop matrix-element generator such as \textsc{MadLoop} or \textsc{OpenLoops} evaluates only the four-dimensional numerator $\bar N$.
From it, it obtains the coefficients of the scalar integrals, with the OPP method~\cite{Ossola:2006us,Ossola:2007ax} or with tensor
reduction~\cite{Passarino:1978jh,Denner:2016kdg}. The denominators stay $D$-dimensional, and their dependence on $\tilde k^2$ gives a first rational
term $R_1$, which the generator also computes~\cite{Ossola:2008xq}. The part $\tilde N$ of the numerator is not computed by the generator, and its contribution to the amplitude is the rational term $\Rtwo$~\cite{Ossola:2008xq,Binoth:2006hk,Draggiotis:2009yb},
\begin{equation}\label{eq:r2def}
  \Rtwo\equiv\int\!\frac{\dd^Dk}{(2\pi)^D}\,\frac{\tilde N}{D_0\cdots D_{n-1}}\Big|_{\epsilon\to0},
\end{equation}
so that $R=R_1+\Rtwo$.

Every term of $\tilde N$ is proportional to $\epsilon$ or to $\tilde k^2$, and gives a non-zero result only when it multiplies a UV pole of
the integral, since the rational terms of infrared origin cancel in one-loop amplitudes~\cite{Bredenstein:2008zb}. For this reason $\Rtwo$ is
finite and local. It is a polynomial in the masses and in the external momenta of each 1PI Green's function, of a degree limited by power
counting, as discussed in section~\ref{sec:scope}. A renormalisable theory therefore has $\Rtwo$ only in vertices with up to four external legs~\cite{Draggiotis:2009yb,Pittau:2011qp},
while in the SMEFT vertices with up to six legs contribute, as eq.~\eqref{eq:pc:omega} shows. These local terms are supplied to the generator as
effective tree-level vertices, the $\Rtwo$ vertices, together with the UV counterterm vertices that cancel the poles and that we construct in
section~\ref{sec:uv}~\cite{Draggiotis:2009yb,Garzelli:2009is,Pittau:2011qp,Degrande:2014vpa}.

The $\Rtwo$ vertices are not gauge invariant by themselves~\cite{Ossola:2008xq,Pittau:2011qp}. In this work we compute them in the Feynman gauge of QCD. In our case, they do
not depend on the electroweak gauge, since no electroweak boson propagates in the loops of the vertices defined in section~\ref{sec:scope}. The
$\Rtwo$ vertices do not depend on the renormalisation scheme, but they depend on the regularisation, through the dimension of the external
states, the treatment of $\gfive$ and of the Levi-Civita tensor, and the definition of the four-fermion Dirac structures in $D$ dimensions. These
choices are discussed in sections~\ref{sec:setup:reg} and~\ref{sec:gamma5}. When two choices differ by the Feynman rule of a gauge-invariant local
operator, the difference is a finite shift of the coefficient of that operator. The regularisation is in this sense part of the definition of the
$\MSbar$ Wilson coefficients~\cite{Buras:1989xd,Dugan:1990df,Herrlich:1994kh,Fuentes-Martin:2022vvu}.

\subsection{Tensor decomposition}\label{sec:tensorbasis}
Like a tree-level Feynman rule, an effective vertex carries a Lorentz index for every external vector boson, a spinor index for every external fermion and a colour index for
every coloured leg. We write an effective vertex, and the tree-level Feynman rule, which we call the tree, in the form
\begin{equation}\label{eq:tb:vertex}
  \mathcal V=\sum_{\alpha,\beta}c_{\alpha\beta}\;\mathcal C_\alpha\otimes\mathcal T_\beta ,
\end{equation}
where the $\mathcal C_\alpha$ form a basis of colour structures and the $\mathcal T_\beta$ a basis of Lorentz--Dirac structures. The
latter are built from the metric, the independent external momenta, one Levi-Civita tensor and, on each open fermion line, one of the sixteen elements of the basis
\begin{equation}\label{eq:tb:dirac}
  \Gamma_A\in\bigl\{1,\ \gfive,\ \gamma^\mu,\ \gamma^\mu\gfive,\ \sigma^{\mu\nu}\bigr\},\qquad A=1,\dots,16.
\end{equation}
The coefficients $c_{\alpha\beta}$ are polynomials in the masses and in the scalar products of the external momenta, and they are obtained by projection.
To every structure we associate a projector $P_{\alpha\beta}$, which contracts all the indices of
a tensor with those of $\mathcal C_\alpha\otimes\mathcal T_\beta$ and closes every fermion line into a trace, and eq.~\eqref{eq:tb:vertex} gives
\begin{equation}\label{eq:tb:gram}
  P_{\alpha\beta}(\mathcal V)=\sum_{\alpha',\beta'}G^{\mathcal C}_{\alpha\alpha'}\,G^{\mathcal T}_{\beta\beta'}\,c_{\alpha'\beta'} ,\qquad
  G^{\mathcal C}_{\alpha\alpha'}=P_\alpha(\mathcal C_{\alpha'}),\qquad G^{\mathcal T}_{\beta\beta'}=P_\beta(\mathcal T_{\beta'}),
\end{equation}
with a numerical colour Gram matrix and a Lorentz--Dirac Gram matrix whose entries are polynomials in the scalar products. Since the external
momenta, polarisations and spinors are four-dimensional, we keep a maximal set of structures independent in four
dimensions.
The projectors carry four-dimensional metric tensors on the external indices,
while the Dirac algebra and the loop momentum inside the diagram stay $D$-dimensional. The same projectors act on $N$ and on $\bar N$, and
the rational term and the pole of each projected coefficient are computed as described in section~\ref{sec:r2:ct}.

The basis of structures is complete in four dimensions but not in $D$~\cite{Buras:1989xd,Dugan:1990df,Herrlich:1994kh,Fuentes-Martin:2022vvu}. However, at one loop and with four-dimensional external states, the identities that hold only
in four dimensions between structures built from external indices and momenta, such as the Schouten identity of section~\ref{sec:g5:eps},
hold exactly. The Levi-Civita tensor of an operator on an open quark line, whose indices are summed inside the loop, needs a separate
prescription, given in the same section. Similarly, in the four-fermion
vertices the Dirac indices that connect the two quark lines are summed inside the loop and are $D$-dimensional. The reduction of their
products to the four-dimensional basis leaves evanescent terms, whose treatment defines the scheme of section~\ref{sec:schemes}, as in
ref.~\cite{Fuentes-Martin:2022vvu}. The steps of the computation, from the list of structures to the solution of the system, are described in more detail
in appendix~\ref{app:tb:solve}.

\providecommand{\figclosurelegs}{%
  \node[sblob, minimum size=26pt] (G) at (0,2.1) {$\mathcal V$};
  \draw[sfer] (-1.9,0.1) -- (G.235);
  \draw[sfer] (G.262) -- (-0.6,0.1);
  \draw[line width=0.6pt, fill=gray!8] (-2.5,-1.0) rectangle (2.5,0.1);
  \node[font=\small] at (-1.9,-0.2) {$s_1c_1$};
  \node[font=\small] at (-0.6,-0.2) {$s_2c_2$};
}
\begin{figure}[!htbp]
\centering
\begin{tikzpicture}
\begin{scope}
  \figclosurelegs
  \draw[sglu] (G.278) -- (0.6,0.1);
  \draw[sglu] (G.305) -- (1.9,0.1);
  \node[font=\small] at (0.6,-0.2) {$\mu_1a_1$};
  \node[font=\small] at (1.9,-0.2) {$\mu_2a_2$};
  \node at (0,-0.68) {$\bigl(P_{\alpha\beta}\bigr)^{c_1c_2,\,a_1a_2}_{s_1s_2,\,\mu_1\mu_2}$};
  \node at (0,-1.5) {(a) projection};
\end{scope}
\begin{scope}[xshift=7.1cm]
  \figclosurelegs
  \draw[sglu] (G.278) -- (0.6,0.26);
  \draw[sglu] (G.305) -- (1.9,0.26);
  \node[circle, draw, fill=white, line width=0.6pt, inner sep=1.6pt] at (0.6,0.26) {};
  \node[circle, draw, fill=white, line width=0.6pt, inner sep=1.6pt] at (1.9,0.26) {};
  \node[font=\small, anchor=west] at (0.7,0.34) {$e_1^{\mu_1}$};
  \node[font=\small, anchor=west] at (2.0,0.34) {$e_2^{\mu_2}$};
  \node[font=\small] at (0.6,-0.2) {$a_1$};
  \node[font=\small] at (1.9,-0.2) {$a_2$};
  \node at (0,-0.68) {$\bigl(\Gamma_A\bigr)_{s_1s_2}\bigl(\mathcal C_\alpha\bigr)^{c_1c_2a_1a_2}$};
  \node at (0,-1.5) {(b) open indices};
\end{scope}
\end{tikzpicture}
\caption{Two ways of reading off the coefficients of the vertex $q\bar qgg$. (a) The projector $P_{\alpha\beta}$ contracts all indices with
those of a structure $\mathcal C_\alpha\otimes\mathcal T_\beta$. (b) The Lorentz index of each gluon is contracted with an auxiliary vector $e_i$
(open circle), the spinor and colour indices with $\Gamma_A$ and $\mathcal C_\alpha$.}\label{fig:tb:closure}
\end{figure}%
The projection becomes expensive for vertices with many external legs, since every structure needs its own projector,
as in figure~\hyperref[fig:tb:closure]{\ref*{fig:tb:closure}(a)}, and the number of Lorentz structures grows quickly. The four-gluon vertex of $O_G$ alone needs 846
projectors. The Lorentz indices can instead be handled without projectors, by contracting the index of every external vector
boson with an auxiliary four-dimensional vector $e_i$ as in figure~\hyperref[fig:tb:closure]{\ref*{fig:tb:closure}(b)}. Each external index appears
once in every diagram, so that the result is linear in each $e_i$. Inside a diagram, $e_i$ can meet the loop momentum, as in $e_i\cdot k$, and is
integrated like an external momentum. After the integration every diagram depends on the auxiliary vectors only through $e_i\cdot e_j$,
$e_i\cdot p_k$ and Levi-Civita tensors, and the replacements $e_i\cdot e_j\to g^{\mu_i\mu_j}$ and $e_i\cdot p_k\to p_k^{\mu_i}$ give the vertex
with its Lorentz indices open. An open fermion line is closed into a trace with a matrix $\Gamma_A$ of eq.~\eqref{eq:tb:dirac}, with its Lorentz indices also
contracted with auxiliary vectors. For most vertices the two routes take a similar time. For the four-gluon vertex of $O_G$ the auxiliary vectors are nine times faster than
the projection, and they made the five- and six-gluon vertices feasible in a reasonable computing time. For the vertices that we computed with both routes, the two routes gave the same tree, pole and rational term.

\subsection{\texorpdfstring{Calculation%
}{Calculation} of the rational terms and UV poles}\label{sec:r2:ct}
After the contraction with a projector or with the auxiliary vectors of section~\ref{sec:tensorbasis}, the %
numerator of each diagram is %
 a scalar function of the loop momentum. We split it as in eq.~\eqref{eq:r2:split} and integrate both parts with the standard one-loop integrals, including the integrals with powers of $\tilde k^2$ given in refs.~\cite{Ossola:2008xq,Draggiotis:2009yb}. We evaluate them for any power of the loop momentum in the numerator with Feynman parameters, as described in appendix~\ref{app:ct:fp}. %
After the
integration, every term of $\tilde N$ carries a factor $D-4$, either from the $D$-dimensional algebra, as in $\gamma^\mu\gamma_\mu=D$, or from the reduction of the $(D-4)$-dimensional loop momentum $\tilde k$, as in $\tilde k^{\,2}\to(D-4)\,k^2/D$. The integral of
$\tilde N$ is therefore $D-4$ times an integral with only UV poles, since in our framework the external momenta are kept off shell and there are no infrared poles.
Writing the pole of this integral as $Q/\epsilon$ and using $D-4=-2\epsilon$, we obtain
\begin{equation}\label{eq:r2:A}
  \int\!\frac{\dd^Dk}{(2\pi)^D}\,\frac{\tilde N}{D_0\cdots D_{n-1}}=(D-4)\Big[\frac{Q}{\epsilon}+\order{\epsilon^0}\Big]
  \qquad\Longrightarrow\qquad \Rtwo=-2\,Q .
\end{equation}
The full integral is then
\begin{equation}\label{eq:r2:P}
  \int\!\frac{\dd^Dk}{(2\pi)^D}\,\frac{N}{D_0\cdots D_{n-1}}=\frac{\mathcal P}{\epsilon}+\bar F+\Rtwo+\order{\epsilon},
\end{equation}
where the UV pole $\mathcal P/\epsilon$ and the finite part $\bar F$ come from the integral of $\bar N$, which the one-loop generator evaluates itself, and whose UV pole is cancelled by the counterterm vertices of section~\ref{sec:uv}. Both $\mathcal P$ and $Q$ need only the UV poles of the scalar integrals. They are polynomials in the external momenta and the masses up to the degree allowed by power counting, so that
terms of higher degree can be dropped before the integration. An independent computation, in which every propagator is expanded with a single auxiliary infrared
mass~\cite{Misiak:1994zw,Chetyrkin:1997fm}, gives the same $\Rtwo$ and poles. Since this method is much slower, we used it only as an additional check on a representative subset of vertices. Details of the integration and of the power counting are given in appendix~\ref{app:ct}.

\subsection{Flavour}\label{sec:flavour}
We keep the generation indices of all quarks open, so that one computation gives an effective vertex for every choice of the external
generations.
The gluons and the ghosts do not see flavour. The quark propagators and the couplings of the quarks to the photon, the $Z$ boson,
the Higgs boson and $G^0$ are diagonal in the mass basis of eq.~\eqref{eq:warsawup}, so that a quark line changes generation only through the operator insertion or through the coupling to a $W^\pm$ or a
$G^\pm$ boson.

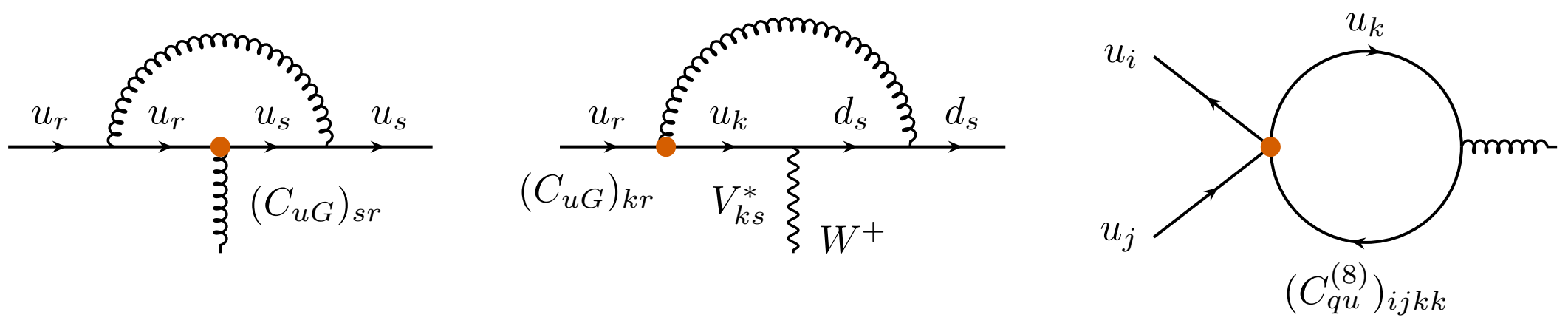
\begin{figure}[!htbp]
\centering
\begin{tikzpicture}[scale=0.95, every node/.style={font=\small}]
\begin{scope}
  \coordinate (A) at (-1.0,0); \coordinate (O) at (0,0); \coordinate (B) at (1.0,0);
  \draw[sfer] (-2.0,0) -- (A); \draw[sfer] (A) -- (O); \draw[sfer] (O) -- (B); \draw[sfer] (B) -- (2.0,0);
  \draw[sglu] (A) arc (180:0:1.0);
  \draw[sglu] (O) -- (0,-1.0);
  \node[sopv] at (O) {};
  \node[above] at (-1.6,0.02) {$u_r$}; \node[above] at (-0.5,0.02) {$u_r$};
  \node[above] at (0.5,0.02) {$u_s$}; \node[above] at (1.6,0.02) {$u_s$};
  \node[anchor=west] at (0.12,-0.55) {$(C_{uG})_{sr}$};
  \node at (0,-2.0) {(a)};
\end{scope}
\begin{scope}[xshift=5.3cm]
  \coordinate (O) at (-1.1,0); \coordinate (W) at (0.1,0); \coordinate (B) at (1.2,0);
  \draw[sfer] (-2.1,0) -- (O); \draw[sfer] (O) -- (W); \draw[sfer] (W) -- (B); \draw[sfer] (B) -- (2.1,0);
  \draw[sglu] (O) arc (180:0:1.15);
  \draw[spho] (W) -- (0.1,-1.0);
  \node[sopv] at (O) {};
  \node[above] at (-1.65,0.02) {$u_r$}; \node[above] at (-0.5,0.02) {$u_k$};
  \node[above] at (0.65,0.02) {$d_s$}; \node[above] at (1.7,0.02) {$d_s$};
  \node[anchor=north east] at (-1.05,-0.1) {$(C_{uG})_{kr}$};
  \node[anchor=east] at (0.0,-0.55) {$V^*_{ks}$};
  \node[anchor=west] at (0.2,-0.85) {$W^+$};
  \node at (0,-2.0) {(b)};
\end{scope}
\begin{scope}[xshift=10.5cm]
  \coordinate (O) at (-0.6,0); \coordinate (R) at (1.2,0);
  \draw[sfer] (-1.7,-0.85) -- (O); \draw[sfer] (O) -- (-1.7,0.85);
  \draw[sfer] (O) arc (180:0:0.9); \draw[sfer] (R) arc (0:-180:0.9);
  \draw[sglu] (R) -- (2.1,0);
  \node[sopv] at (O) {};
  \node[anchor=east] at (-1.7,-0.85) {$u_j$}; \node[anchor=east] at (-1.7,0.85) {$u_i$};
  \node[above] at (0.3,0.9) {$u_k$};
  \node[below] at (0.3,-0.95) {$(C^{(8)}_{qu})_{ijkk}$};
  \node at (0.3,-2.0) {(c)};
\end{scope}
\end{tikzpicture}
\caption{Flavour flow in three one-loop diagrams, with the operator insertion drawn as a dot and the generation indices of the
quark lines shown next to them. (a) A diagram of the quark--gluon vertex of a gluon dipole. (b) The same line with an external $W$ boson.
The generation $k$ of the internal up-type quark is summed. (c) A penguin diagram of a four-quark operator, whose closed loop gives the sum over $k$.}
\label{fig:flavour}
\end{figure}
Figure~\ref{fig:flavour} shows how the flavour factors appear. When the operator is the only flavour-changing point, as in diagram (a),
the quark line enters with generation $r$ and leaves with generation $s$,
and the diagram is proportional to $(C_{uG})_{sr}$. When a $W$ boson is attached to the same line, as in diagram (b), the quark between
the operator and the $W$ vertex has a generation $k$ that is summed over, which gives
\begin{equation}\label{eq:fl:chain}
  \sum_k V^*_{ks}\,a(m_k)\,C_{kr}=\bigl(V^\dagger a(M_u)\,C\bigr)_{sr} ,
\end{equation}
where $a$ is a polynomial in the mass of the internal quark and $M_u$ the diagonal up-type mass matrix. In diagram (c) a four-quark operator
closes a quark loop, whose generation is summed,
\begin{equation}\label{eq:fl:trace}
  \sum_k b(m_k)\,C_{ijkk} ,
\end{equation}
with $b$ again a polynomial. A closed loop of a two-quark operator gives in the same way a trace of the mass matrices and the coefficient matrix.
Although the CKM matrix is unitary, we do not use this property to simplify these products, which therefore stay as explicit sums over the generations. In this way every expression keeps its full dependence on $V$, and the package function \texttt{CKMUnitarity} applies the unitarity when it is wanted.

\section{The treatment of \texorpdfstring{$\gamma_5$}{gamma5} and of four-fermion operators}\label{sec:gamma5}
The rational term of a one-loop effective vertex depends on choices that leave its UV pole invariant. These are the continuation of $\gfive$
to $D$ dimensions, the dimension of the Levi-Civita tensor of an operator on an open quark line, and the definition of the four-fermion Dirac
structures that vanish only in four dimensions.

\subsection{\texorpdfstring{$\gamma_5$}{gamma5} in dimensional regularisation and the reading point}\label{sec:g5:def}\label{sec:g5:problem}\label{sec:g5:conseq}
An anticommuting $\gfive$, the cyclicity of the trace and a non-vanishing trace of $\gfive$ with four gamma matrices cannot simultaneously hold in $D$
dimensions~\cite{Breitenlohner:1977hr,Kreimer:1989ke,Korner:1991sx}. We keep
an anticommuting $\gfive$ and give up cyclicity, following refs.~\cite{Kreimer:1989ke,Korner:1991sx}. A trace with a single $\gfive$ is then defined only once the vertex at which it starts, the reading point, is
fixed~\cite{Kreimer:1989ke,Korner:1991sx}. An odd number of $\gfive$ reduces to a single one by anticommutation, and the reading point matters only for traces with at least six other gamma matrices, as discussed in section~\ref{sec:g5:affected}.

\providecommand{\gvloop}{%
  \draw[sfer] (180:1) arc (180:100:1);
  \draw[sfer] (100:1) arc (100:20:1);
  \draw[densely dotted, line width=0.75pt] (20:1) arc (20:-110:1);
  \draw[sfer] (250:1) arc (250:180:1);
  \foreach \a in {180,100,20,250} {\fill (\a:1) circle (1.6pt); \draw[line width=0.6pt] (\a:1) -- (\a:1.55);}
  \node[sreadpt] at (180:1) {};
  \node[anchor=east] at (180:1.62) {$\Omega_r$};
  \node[anchor=south east] at (100:1.55) {$\Omega_{r+1}$};
  \node[anchor=south west] at (20:1.55) {$\Omega_{r+2}$};
  \node[anchor=north east] at (250:1.55) {$\Omega_{r-1}$};
  \node at (140:1.32) {$\Pi_r$}; \node at (60:1.34) {$\Pi_{r+1}$}; \node at (215:1.36) {$\Pi_{r-1}$};}
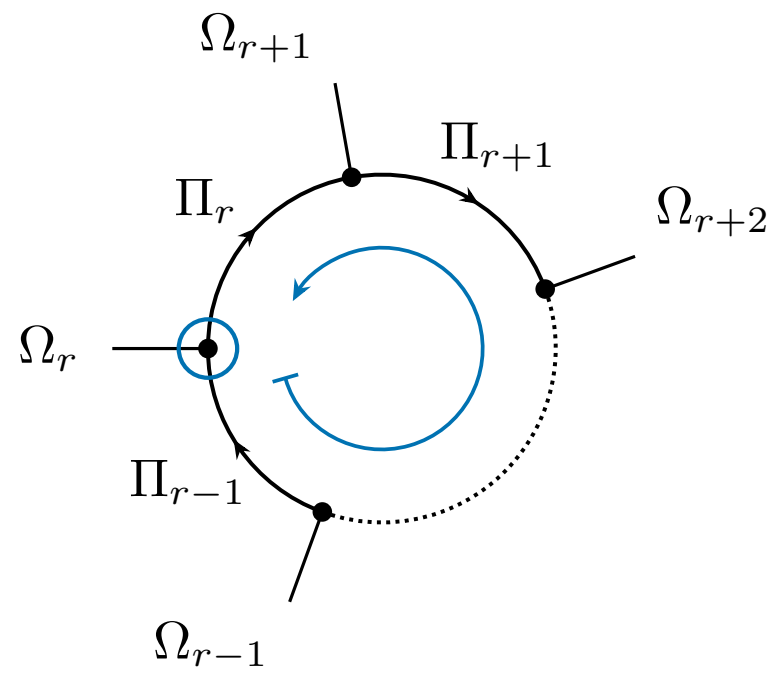
\begin{figure}[!htbp]
\centering
\begin{tikzpicture}[scale=1.15]
  \gvloop
  \draw[sread] (0,0) ++(196:0.58) arc (196:512:0.58);
\end{tikzpicture}
\caption{A closed quark loop read at vertex $r$ (blue circle). The vertices are numbered along the fermion flow, $\Omega_j$ is the Dirac
matrix of vertex $j$ and $\Pi_j$ the numerator of the propagator from $\Omega_j$ to $\Omega_{j+1}$. The blue arc follows the matrices of the
string $T_r$ of eq.~\eqref{eq:g5:string}, written against the fermion flow as the Feynman rules give them, from the first (bar) to the last
(arrow), which is the matrix of the reading vertex.}
\label{fig:g5:string}
\end{figure}

We number the vertices of a closed quark loop along the fermion flow and call $\Omega_j$ the Dirac matrix in the Feynman rule of vertex $j$. The numerator of the propagator from $\Omega_j$ to $\Omega_{j+1}$ is
$\Pi_j=\slashed q_j+m_j$. The trace read at vertex $r$ starts at $\Omega_r$ and proceeds along the fermion flow~\cite{Korner:1991sx,Born:2026xkr}. Since
the matrices of the Feynman rules are written against the fermion flow, this is the string that ends with $\Omega_r$, as shown in
figure~\ref{fig:g5:string},
\begin{equation}\label{eq:g5:string}
  T_r=\Tr\bigl[\Pi_{r-1}\,\Omega_{r-1}\cdots\Pi_{r+1}\,\Omega_{r+1}\,\Pi_r\,\Omega_r\bigr]\Big|_{\rm no\ cyclicity},
\end{equation}
and cyclicity is never applied. Ref.~\cite{Fuentes-Martin:2022vvu} states the same prescription in this form, and for example ref.~\cite{Korner:1991sx} reads the closed loops of the SM at the axial vertex.

Every $\gfive$ is moved by anticommutation to the right end of the string, to the right of the matrices of $\Omega_r$, and pairs of $\gfive$ are removed. The remaining product of
$D$-dimensional gamma matrices is written as a sum of antisymmetrised products $\Gamma^{\mu_1\cdots\mu_k}$ times metric
tensors~\cite{Korner:1991sx,Kuusela:2019iok}, as in $\gamma^\mu\gamma^\nu=\Gamma^{\mu\nu}+g^{\mu\nu}$, and the trace with $\gfive$ is defined for
each antisymmetrised product by
\begin{equation}\label{eq:g5:basis}
  \Tr[\Gamma^{\mu_1\mu_2\mu_3\mu_4}\gfive]=-4\ii\,\epsilon^{\mu_1\mu_2\mu_3\mu_4},\qquad \Tr[\Gamma^{\mu_1\cdots\mu_k}\gfive]=0\ \ (k\ne4).
\end{equation}
In four dimensions an antisymmetrised product of more than four gamma
matrices vanishes, and eq.~\eqref{eq:g5:basis} reproduces the familiar traces. In $D$ dimensions such products do not vanish, and setting
their trace with $\gfive$ to zero is part of the definition. For a product of $2m$ gamma matrices with $\gfive$ the definition gives~\cite{Kreimer:1989ke,Korner:1991sx}
\begin{equation}\label{eq:g5:even}
\begin{split}
  \Tr[\gamma^{\mu_1}\cdots\gamma^{\mu_{2m}}\gfive]=-4\ii\sum_{\sigma}\mathrm{sgn}(\sigma)\,
  &g^{\mu_{\sigma_1}\mu_{\sigma_2}}\cdots g^{\mu_{\sigma_{2m-5}}\mu_{\sigma_{2m-4}}}\\
  &\times\epsilon^{\mu_{\sigma_{2m-3}}\mu_{\sigma_{2m-2}}\mu_{\sigma_{2m-1}}\mu_{\sigma_{2m}}},
\end{split}
\end{equation}
where the sum runs over the ways of placing four of the indices in the Levi-Civita tensor and pairing the others into metric tensors, and $\mathrm{sgn}(\sigma)$ is the sign of the corresponding permutation $\sigma$ of $1,\ldots,2m$. For six gamma matrices, for example, the sum has 15 terms. Figure~\ref{fig:g5:read} shows the definition on the triangle of the $ggZ$ vertex at
$\order{C_{Hu}}$.

Two readings of the same trace differ by terms of order $D-4$. The pole of a one-loop 1PI Green's function comes from the
numerator at $D=4$ and is the same for every reading. The difference is $D-4$ times UV-divergent tensor integrals and leaves a
finite local remainder, which is part of $\Rtwo$. The reading point also decides where the anomaly sits. Read at the axial vertex, a loop
with one axial and two vector vertices conserves the vector currents and puts the anomaly in the axial current~\cite{Adler:1969gk,Bell:1969ts}.
For QCD loops the gluon currents are then conserved, while other readings need a local counterterm~\cite{Bardeen:1969md}, as discussed in
appendix~\ref{app:g5su3}.
\newcommand{\gvread}[2]{\draw[sread] (0.51,0) ++(#1+28:0.36) arc (#1+28:#1+332:0.36); \node[sreadpt] at (#2) {};}
\newcommand{\gvtri}{%
  \coordinate (G1) at (0,0.9); \coordinate (G2) at (0,-0.9); \coordinate (V) at (1.5,0);
  \draw[sfer] (G1) -- (V); \draw[sfer] (V) -- (G2); \draw[sfer] (G2) -- (G1);}
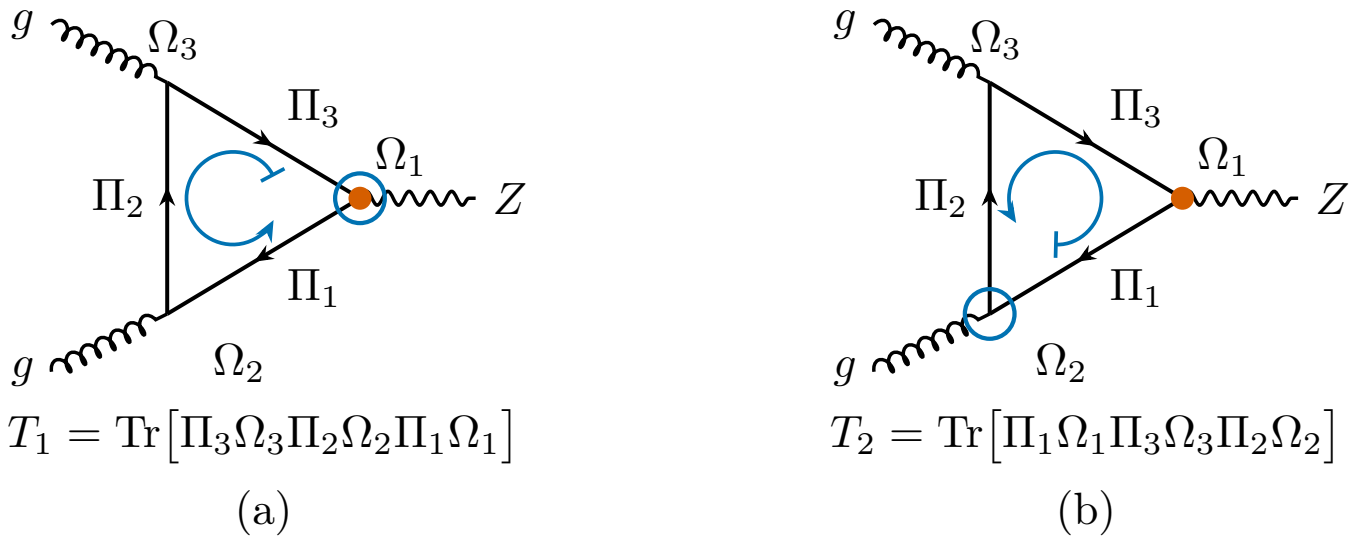
\begin{figure}[!htbp]
\centering
\begin{tikzpicture}[scale=0.85, every node/.append style={transform shape}]
\begin{scope}
  \gvtri
  \draw[sglu] (-0.9,1.35) -- (G1); \draw[sglu] (-0.9,-1.35) -- (G2); \draw[spho] (V) -- (2.4,0);
  \node[sopv] at (V) {};
  \node[anchor=south west] at (1.48,0.04) {$\Omega_1$}; \node[anchor=north west] at (0.22,-1.0) {$\Omega_2$};
  \node[anchor=south] at (0.05,0.95) {$\Omega_3$};
  \node[anchor=north west] at (0.8,-0.42) {$\Pi_1$}; \node[anchor=east] at (-0.03,0) {$\Pi_2$}; \node[anchor=south west] at (0.8,0.42) {$\Pi_3$};
  \node[anchor=west] at (2.4,0) {$Z$}; \node[anchor=east] at (-0.9,1.35) {$g$}; \node[anchor=east] at (-0.9,-1.35) {$g$};
  \gvread{0}{V}
  \node at (0.75,-1.85) {$T_1=\Tr\bigl[\Pi_3\Omega_3\Pi_2\Omega_2\Pi_1\Omega_1\bigr]$};
  \node at (0.75,-2.45) {(a)};
\end{scope}
\begin{scope}[xshift=6.4cm]
  \gvtri
  \draw[sglu] (-0.9,1.35) -- (G1); \draw[sglu] (-0.9,-1.35) -- (G2); \draw[spho] (V) -- (2.4,0);
  \node[sopv] at (V) {};
  \node[anchor=south west] at (1.48,0.04) {$\Omega_1$}; \node[anchor=north west] at (0.22,-1.0) {$\Omega_2$};
  \node[anchor=south] at (0.05,0.95) {$\Omega_3$};
  \node[anchor=north west] at (0.8,-0.42) {$\Pi_1$}; \node[anchor=east] at (-0.03,0) {$\Pi_2$}; \node[anchor=south west] at (0.8,0.42) {$\Pi_3$};
  \node[anchor=west] at (2.4,0) {$Z$}; \node[anchor=east] at (-0.9,1.35) {$g$}; \node[anchor=east] at (-0.9,-1.35) {$g$};
  \gvread{240.5}{G2}
  \node at (0.75,-1.85) {$T_2=\Tr\bigl[\Pi_1\Omega_1\Pi_3\Omega_3\Pi_2\Omega_2\bigr]$};
  \node at (0.75,-2.45) {(b)};
\end{scope}
\end{tikzpicture}
\caption{The reading point, for the triangle of the $ggZ$ vertex at $\order{C_{Hu}}$. The vertices of the loop are numbered along the fermion flow
from the operator vertex $\Omega_1$ (dot), and $\Pi_r$ is the
propagator that follows $\Omega_r$. The reading vertex is circled in blue, and the blue arc shows the order of the Dirac string of eq.~\eqref{eq:g5:string}. (a) The reading at the operator, which is our prescription. (b) The reading at a gluon vertex. The two strings are cyclic
permutations of each other, equal in four dimensions but not in $D$. }
\label{fig:g5:read}
\end{figure}

We read every closed trace with a single $\gfive$ at the operator insertion. At one loop and at $\mathcal{O}(\Lambda^{-2})$ this reading is always available, since in a one-loop 1PI diagram every vertex,
the operator insertion included, lies on the loop. In the indirect contributions the shifted SM vertex plays the role of the operator
insertion. It has the Dirac and flavour structure of the SM vertex, and its axial coupling is rescaled by
a factor common to all quarks. These loops behave like their SM counterparts, and their Levi-Civita parts cancel within each
generation, as the gauge anomalies do~\cite{Bouchiat:1972iq,Gross:1972pv}.

\subsection{Reading-point dependence and gauge invariance}\label{sec:g5:affected}\label{sec:g5:ward}
An effective vertex depends on the reading point only if one of its diagrams contains a closed quark loop with a single $\gfive$ and at least six gamma matrices~\cite{Kreimer:1989ke,Korner:1991sx}, and then only through the Levi-Civita part of its $\Rtwo$. A scan of every result finds three classes,
listed vertex by vertex in table~\ref{tab:app:g5classes} of appendix~\ref{app:g5list}. The first are the vertices of the Higgs-current operators $O_{Hq}^{(1,3)}$, $O_{Hu}$
and $O_{Hd}$ with two or three gluons and a $Z$ or Goldstone boson, in which the operator vertex is the axial vertex of the loop. The second are
the vertices of the gluon dipoles $O_{uG}$ and $O_{dG}$, for the real and imaginary parts of their coefficients, with two or three gluons and scalars or a $Z$ boson. Here the loop contains two sources of $\gfive$, the
operator and the SM vertex.
The third are the vertices of four-fermion operators with two or three gauge bosons, in which one of the two vector currents of the operator closes into a quark loop. In contrast, the vertices with an operator insertion of the electroweak dipoles, of $O_{uH}$ and $O_{dH}$, or of the scalar and tensor
four-fermion operators do not depend on the reading point.

As an example consider $g^A_{\mu_1}(p_1)\,g^B_{\mu_2}(p_2)\,G^0$ with an insertion of the gluon dipoles. Each diagram is a closed loop of
one flavour with either the dipole vertex $q\bar qgG^0$ alone or the dipole vertex $q\bar qg$ together with the coupling of
$G^0$, shown in figure~\ref{fig:g5:ggG0}. Read at the operator, the Levi-Civita part of the rational term is
\begin{equation}\label{eq:g5:ggG0}
  \mathcal V^{\Rtwo}_{ggG^0}\Big|_{\rm LC}=\frac{16\,g_s v}{3\Lambda^2}\,\delta^{AB}\,\epsilon^{\mu_1\mu_2p_1p_2}
  \Bigl[\Re\Tr\bigl(C_{uG}\,\hat Y_u\bigr)-\Re\Tr\bigl(V^\dagger C_{dG}\,\hat Y_d\bigr)\Bigr],
\end{equation}
where $\hat Y_{u,d}=\sqrt2M_{u,d}/v$ are the diagonal Yukawa matrices and the trace runs over generations. Read at the $G^0$ vertex it is
$-\frac12$ times eq.~\eqref{eq:g5:ggG0}.
\newcommand{\gvlegs}{%
  \draw[sglu] (-0.9,1.35) -- (G1); \draw[sglu] (-0.9,-1.35) -- (G2); \draw[shig] (V) -- (2.3,0);
  \node[sopv] at (G1) {};
  \node[anchor=south west] at (0.22,1.0) {$\Omega_1$}; \node[anchor=south west] at (1.48,0.04) {$\Omega_2$};
  \node[anchor=north west] at (0.22,-1.0) {$\Omega_3$};
  \node[anchor=south west] at (0.8,0.42) {$\Pi_1$}; \node[anchor=north west] at (0.8,-0.42) {$\Pi_2$}; \node[anchor=east] at (-0.03,0) {$\Pi_3$};
  \node[anchor=west] at (2.3,0) {$G^0$}; \node[anchor=east] at (-0.9,1.35) {$g_1$}; \node[anchor=east] at (-0.9,-1.35) {$g_2$};}
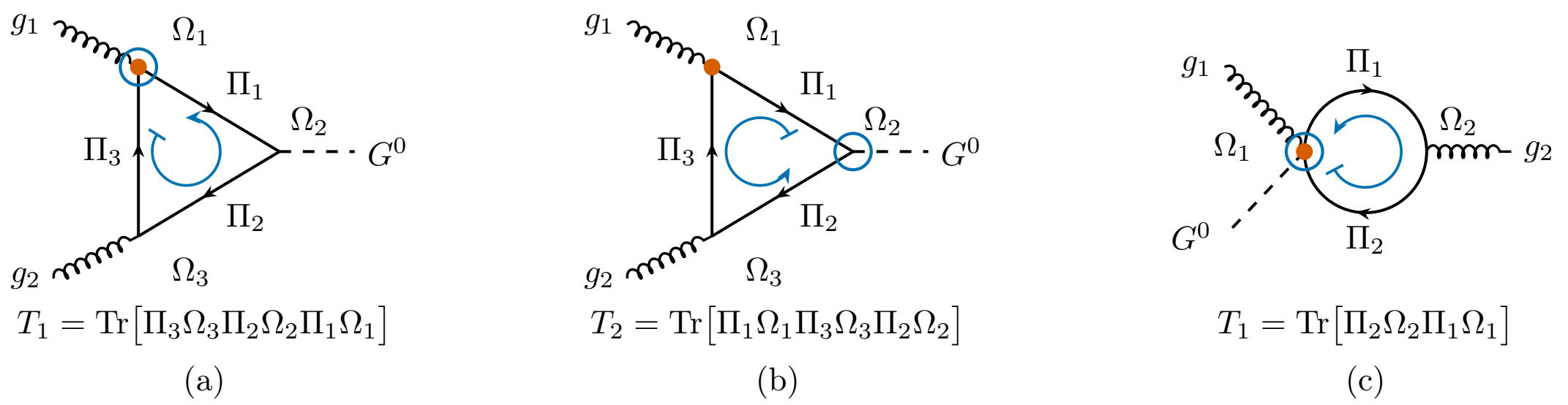
\begin{figure}[!htbp]
\centering
\begin{tikzpicture}[scale=0.85, every node/.append style={transform shape}]
\begin{scope}
  \gvtri \gvlegs \gvread{119.5}{G1}
  \node at (0.7,-1.85) {$T_1=\Tr\bigl[\Pi_3\Omega_3\Pi_2\Omega_2\Pi_1\Omega_1\bigr]$};
  \node at (0.7,-2.45) {(a)};
\end{scope}
\begin{scope}[xshift=6.1cm]
  \gvtri \gvlegs \gvread{0}{V}
  \node at (0.7,-1.85) {$T_2=\Tr\bigl[\Pi_1\Omega_1\Pi_3\Omega_3\Pi_2\Omega_2\bigr]$};
  \node at (0.7,-2.45) {(b)};
\end{scope}
\begin{scope}[xshift=12.4cm]
  \coordinate (V) at (0,0); \coordinate (H) at (1.3,0);
  \draw[sglu] (-0.85,0.9) -- (V); \draw[shig] (V) -- (-0.85,-0.9); \draw[sglu] (H) -- (2.2,0);
  \draw[sfer] (V) arc (180:0:0.65); \draw[sfer] (H) arc (0:-180:0.65);
  \node[sopv] at (V) {};
  \draw[sread] (0.65,0) ++(-152:0.38) arc (-152:152:0.38); \node[sreadpt] at (V) {};
  \node[anchor=east] at (-0.85,0.9) {$g_1$}; \node[anchor=east] at (-0.85,-0.9) {$G^0$}; \node[anchor=west] at (2.2,0) {$g_2$};
  \node[anchor=east] at (-0.42,0.05) {$\Omega_1$}; \node[anchor=south west] at (1.3,0.04) {$\Omega_2$};
  \node[anchor=south] at (0.65,0.66) {$\Pi_1$}; \node[anchor=north] at (0.65,-0.66) {$\Pi_2$};
  \node at (0.65,-1.85) {$T_1=\Tr\bigl[\Pi_2\Omega_2\Pi_1\Omega_1\bigr]$};
  \node at (0.65,-2.45) {(c)};
\end{scope}
\end{tikzpicture}
\caption{The diagrams of $ggG^0$ with an insertion of the gluon dipoles.
In the triangle the dipole vertex $\Omega_1$ and the pseudoscalar Yukawa coupling $\Omega_2\propto\gfive$ of $G^0$ are the two sources of
$\gfive$. (a) Our reading, at the operator. (b) The reading at the $G^0$ vertex, which gives $-\frac12$ times the
Levi-Civita part of eq.~\eqref{eq:g5:ggG0}. (c) The bubble with the contact vertex $q\bar qgG^0$ of the dipole, which does not have enough gamma
matrices for the reading to matter.}
\label{fig:g5:ggG0}
\end{figure}

Different reading points give rational terms that differ by a finite local term. If this term is the Feynman rule of a gauge-invariant
operator, the choice of reading point is a convention for the coefficient of that operator~\cite{Fuentes-Martin:2022vvu,Born:2026xkr,DiNoi:2025uan}.
Otherwise a reading that violates the Ward or Slavnov--Taylor identities needs finite counterterms that restore the
symmetry~\cite{Bardeen:1969md,Belusca-Maito:2020ala,Cornella:2022hkc,OlgosoRuiz:2024dzq}, which we want to avoid. The apparent gauge anomalies of the operators with a Higgs current are discussed in refs.~\cite{Feruglio:2020kfq,Bonnefoy:2020tyv,Cata:2020crs}. With the reading at the operator no
such counterterm is needed. The Ward and Slavnov--Taylor identities hold in the three classes of vertices above, because the gauge bosons
attached to the operator vertex couple through the gauge-invariant structure of the operator itself. This is not the case for every reading. We verified the identities explicitly for the vertices of two and three gluons with Goldstone bosons and a closed top-quark loop, at linear order in $(C_{uG})_{33}$ and $(C_{Hu})_{33}$, and the gauge independence of
$gg\to Zh$. Appendix~\ref{app:g5su3} gives the
argument and shows which identities other readings satisfy or violate.

\subsection{Levi-Civita contractions}\label{sec:g5:eps}
\begin{figure}[!htbp]
\centering
\begin{tikzpicture}[scale=0.85, every node/.append style={transform shape}]
\begin{scope}[xshift=6.0cm]
  \coordinate (Q1) at (-1.6,-0.7); \coordinate (A) at (-0.75,-0.7); \coordinate (B) at (0.75,-0.7); \coordinate (Q2) at (1.6,-0.7);
  \coordinate (O) at (0,0.55);
  \draw[sfer] (Q1) -- (A); \draw[sfer] (A) -- (B); \draw[sfer] (B) -- (Q2);
  \draw[sglu] (A) -- (O); \draw[sglu] (B) -- (O); \draw[sglu] (O) -- (0,1.45);
  \node[sopv] at (O) {};
  \node[anchor=east] at (Q1) {$q$}; \node[anchor=west] at (Q2) {$q$}; \node[anchor=south] at (0,1.45) {$g$};
  \node[anchor=west] at (0.12,0.6) {$\epsilon$ of $O_{\widetilde G}$};
  \node at (0,-1.55) {(b)};
\end{scope}
\begin{scope}
  \coordinate (O) at (0,0.55); \coordinate (A) at (-0.75,-0.7); \coordinate (B) at (0.75,-0.7);
  \draw[sglu] (O) -- (A); \draw[sglu] (A) -- (B); \draw[sglu] (B) -- (O);
  \draw[sglu] (O) -- (0,1.45); \draw[sglu] (A) -- (-1.5,-1.2); \draw[sglu] (B) -- (1.5,-1.2);
  \node[sopv] at (O) {};
  \node[anchor=south] at (0,1.45) {$g$}; \node[anchor=east] at (-1.5,-1.2) {$g$}; \node[anchor=west] at (1.5,-1.2) {$g$};
  \node[anchor=west] at (0.12,0.6) {$\epsilon$ of $O_{\widetilde G}$};
  \node at (0,-1.55) {(a)};
\end{scope}
\end{tikzpicture}
\caption{(a) Purely gluonic vertex with an insertion of $O_{\widetilde G}$, whose Levi-Civita structures obey the Schouten identity.
(b) Open quark line at $\order{C_{\widetilde G}}$, where a projector closing the quark line can give a second Levi-Civita tensor, contracted with
the four-dimensional one of the operator.}\label{fig:g5:epspair}
\end{figure}
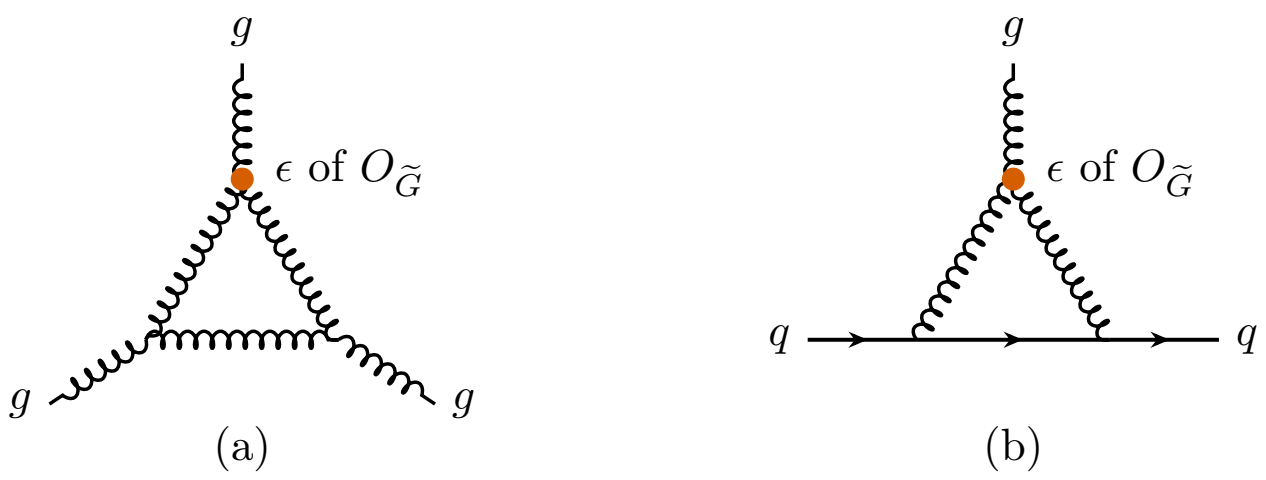

Levi-Civita tensors enter the vertices through the operators with a dual field strength, $O_{\widetilde G}$ and
$O_{H\widetilde G}$, and through the traces with one $\gfive$. Since the projectors of section~\ref{sec:tensorbasis} carry four-dimensional
external indices, the relations that hold only in four dimensions between structures built from external indices hold exactly. An
example is the Schouten identity
\begin{equation}\label{eq:schouten}
  g^{\mu\nu}\epsilon^{\alpha\beta\gamma\delta}=g^{\mu\alpha}\epsilon^{\nu\beta\gamma\delta}+g^{\mu\beta}\epsilon^{\alpha\nu\gamma\delta}
  +g^{\mu\gamma}\epsilon^{\alpha\beta\nu\delta}+g^{\mu\delta}\epsilon^{\alpha\beta\gamma\nu},
\end{equation}
which relates the Levi-Civita structures of the purely gluonic vertices of $O_{\widetilde G}$, as in figure~\hyperref[fig:g5:epspair]{\ref*{fig:g5:epspair}(a)}. A
$D$-dimensional projection would violate it by bosonic evanescent terms of the kind discussed in
ref.~\cite{Fuentes-Martin:2022vvu}.

A scheme choice instead remains where the Levi-Civita tensor of the operator is contracted with indices summed inside the loop.
This happens on an open quark line, which begins and ends on external quarks, as in figure~\hyperref[fig:g5:epspair]{\ref*{fig:g5:epspair}(b)}. The tensor of the
operator is contracted with the gamma matrices of the line, and when the projector closes the line with a $\gfive$ the trace gives a second
Levi-Civita tensor. With the convention of eq.~\eqref{eq:eps}, their product is a determinant of metric tensors,
\begin{equation}\label{eq:eps:det}
  \epsilon^{\mu_1\mu_2\mu_3\mu_4}\,\epsilon_{\nu_1\nu_2\nu_3\nu_4}=-\det\bigl[\delta^{\mu_i}_{\nu_k}\bigr]_{i,k=1,\ldots,4},
\end{equation}
in which the indices summed between the two tensors run over four values or over $D$, so that two summed indices give $2$ or $(D-3)(D-2)$.
The difference is proportional to $D-4$, multiplies the pole of the loop integral and leaves a finite rational term. In four dimensions the
contraction reduces to the Chisholm identity
\begin{equation}\label{eq:chisholm}
  \epsilon^{\mu\nu\rho\sigma}\gamma_\sigma=-\ii\,\Gamma^{\mu\nu\rho}\gfive ,\qquad
  \Gamma^{\mu\nu\rho}=\gamma^\mu\gamma^\nu\gamma^\rho-g^{\mu\nu}\gamma^\rho+g^{\mu\rho}\gamma^\nu-g^{\nu\rho}\gamma^\mu ,
\end{equation}
which rewrites the tensor of the operator on the line as an antisymmetrised product of gamma matrices and $\gfive$.

We treat this Levi-Civita tensor as in ref.~\cite{Born:2026xkr}. The tensor is removed from the quark line with the identity of eq.~\eqref{eq:chisholm}, keeping the factors that the identity gives in four dimensions, and the remainder of order $\epsilon$ is an evanescent structure, which we drop in the same way as the evanescent four-fermion structures of section~\ref{sec:schemes}. For our vertices this is equivalent to treating the Levi-Civita tensor as strictly four-dimensional. %
 This fixes the
rational terms of all vertices with this tensor on an open quark line, and leaves for example the rational term of the $q\bar qh$ vertex of
$O_{H\widetilde G}$ at zero. The $D$-dimensional choice differs, for the vertices of $O_{H\widetilde G}$, by $-\frac{11}3$ times the pole structure, which
can be absorbed into a finite shift of the imaginary parts of $C_{uH}$ and $C_{dH}$.

\subsection{Evanescent four-fermion structures}\label{sec:schemes}\label{sec:sch:eva}\label{sec:g5:choice}
At one loop a gluon exchanged between the two currents of a four-quark operator turns the current on each quark line into a product of three
gamma matrices. The UV pole of the loop is removed by counterterms with the Dirac structures of the tree, so that these products need to be
reduced to those structures. In four dimensions this is unique, for example
$\gamma^\mu\gamma^\nu\gamma^\rho P_L\otimes\gamma_\mu\gamma_\nu\gamma_\rho P_L=16\,\gamma^\mu P_L\otimes\gamma_\mu P_L$. In $D$ dimensions it is not, and one writes instead~\cite{Buras:1989xd,Dekens:2019ept,Fuentes-Martin:2022vvu}
\begin{equation}\label{eq:sch:ex}
  \gamma^\mu\gamma^\nu\gamma^\rho P_L\otimes\gamma_\mu\gamma_\nu\gamma_\rho P_L=4\,(4-b_{\rm ev}\,\epsilon)\,\gamma^\mu P_L\otimes\gamma_\mu P_L+E,
\end{equation}
with an evanescent structure $E$ that vanishes in four dimensions. The four-dimensional coefficient $16$ is fixed. The choice of the value of $b_{\rm ev}$ defines an evanescent scheme, and it decides how much of the term of order $\epsilon$ goes into the coefficient and how much into $E$~\cite{Buras:1989xd,Dugan:1990df,Herrlich:1994kh}. There are
analogous parameters $a_{\rm ev}$, $c_{\rm ev}$, $d_{\rm ev}$, $e_{\rm ev}$ and $f_{\rm ev}$ for the other Dirac structures in the notation of
ref.~\cite{Dekens:2019ept}. The choice of these values matters
because the coefficient multiplies the $1/\epsilon$ pole of the loop, and $-4b_{\rm ev}\epsilon\times1/\epsilon$ is a finite local term in $\Rtwo$. We fix the terms of order $\epsilon$ by $D$-dimensional traces in which a basis element closes the two lines into one trace, following the trace
method of refs.~\cite{Buras:1989xd,Herrlich:1994kh,Tracas:1982gp} in the form of ref.~\cite{Dekens:2019ept}. This sets $a_{\rm ev},\dots,f_{\rm ev}$ %
 to one. We drop $E$ after subtracting its pole, which is equivalent to an evanescent counterterm~\cite{Dugan:1990df,Herrlich:1994kh,Dekens:2019ept,Naterop:2023dek}. This is also the scheme of ref.~\cite{Fuentes-Martin:2022vvu}. Appendix~\ref{app:evtests} gives the
basis and the trace equations, which reproduce the values of refs.~\cite{Buras:1989xd,Buras:2000if,Fuentes-Martin:2022vvu}.

\section{Renormalisation}\label{sec:uv}
To compute a cross section at NLO a one-loop generator also needs counterterm vertices, which remove the UV poles of the one-loop amplitude, computed as described in section~\ref{sec:r2method}, and
fix its finite part in a definite renormalisation scheme. The pole of a 1PI Green's function is an off-shell quantity, and three
kinds of counterterms cancel it. The first renormalises the SM parameters of the tree-level vertex, the masses and $g_s$, whose counterterms also
receive contributions from the operators. The second renormalises the fields of the external legs. The third renormalises the Wilson
coefficients of an off-shell basis, the Green's basis, including its redundant operators, which can be removed by equations of motion~\cite{Grzadkowski:2010es,Gherardi:2020det}. With general flavour the quark two-point functions also mix generations,
as we discuss in section~\ref{sec:uv:mix}, and the CP-odd top-quark dipole generates a CP-odd top-quark mass, which we discuss in section~\ref{sec:uv:cpodd}.

\subsection{Counterterm vertices}\label{sec:uv:os}
For a vertex whose tree-level Feynman rule has the
coupling $c$, the counterterm vertex is obtained as
\begin{equation}\label{eq:uv:rule}
  \delta c=\sum_{a}\frac{\partial c}{\partial p_a}\,\delta p_a
  +\frac12\Big(\sum_{\rm out}\delta\bar Z\,c+\sum_{\rm in}c\,\delta Z\Big)
  -\Gamma^{\rm UV}_{C}\Big(\frac1\epsilon-L_{\rm eft}\Big),
\end{equation}
truncated at $\order{\Lambda^{-2}}$. The three terms follow the order above. The first renormalises the parameters $p_a$ on which the tree depends, such as the masses, the Yukawa couplings and $g_s$, with
counterterms $\delta p_a$. The second renormalises the fields of the external legs, with field renormalisation factors $\delta Z$ for the incoming and
$\delta\bar Z$ for the outgoing ones, which for quarks are matrices in flavour space acting on the generation indices of $c$. The third renormalises the
Wilson coefficients. Appendix~\ref{sec:method:example} applies this rule to the dipole quark--gluon vertex, as an example.

The coefficient $\Gamma^{\rm UV}_C$ of the third term is fixed by the condition that $\delta c$ cancels
the leftover UV pole of the 1PI Green's function $\Gamma=\ii\mathcal V/(16\pi^2)$ once the first two terms are accounted for. It multiplies
$1/\epsilon-L_{\rm eft}$, with $L_{\rm eft}=\ln(\mu_{\rm eft}^2/\mu_R^2)$, because the Wilson coefficients are $\MSbar$ quantities at their own
renormalisation scale $\mu_{\rm eft}$. This scale is chosen independently of the scale $\mu_R$ at which $g_s$ is renormalised, and
$L_{\rm eft}$ accounts for the difference between the two.

Since the Green's functions are off shell, $\Gamma^{\rm UV}_C$ also contains
poles with the form of redundant operators. We apply no field redefinition at this point. These poles stay in the counterterm vertices as they are,
subtracted in $\MSbar$ at $\mu_{\rm eft}$ like the others. In on-shell amplitudes they act as the Warsaw operators into which
the field redefinitions of ref.~\cite{Gherardi:2020det} map them. Their counterterms are kept on internal
lines as well, as discussed in section~\ref{sec:uv:mix}, and the shifts of the SM masses and fields generated by the equations of motion
are absorbed by the on-shell mass and field counterterms. For on-shell amplitudes our Wilson coefficients
are therefore the $\MSbar$ Warsaw coefficients at $\mu_{\rm eft}$, which run with the known one-loop anomalous
dimensions~\cite{Jenkins:2013zja,Jenkins:2013wua,Alonso:2013hga}.

The Wilson coefficients are defined in the SMEFT with six quark flavours, while the counterterms $\delta p_a$ and the renormalisation
factors $\delta Z$ depend on the number of light flavours chosen for the NLO computation. The light quarks are massless and contribute to the
running of $g_s$, while the heavier quarks keep their masses and are decoupled from it. The one-loop generator also needs on-shell field
renormalisation factors, so that the external fields have unit residue and its amplitudes require no corrections on the external legs. For most
collider applications the choice is the five-flavour scheme, in which the top-quark mass and the external fields are renormalised on shell, and
$g_s$ in $\MSbar$ with five light flavours and the top quark decoupled at zero momentum. The top quark is also the only massive quark in the
example models used for the applications of section~\ref{sec:apps}.

\subsection{Quark self-energies with flavour mixing}\label{sec:uv:mix}
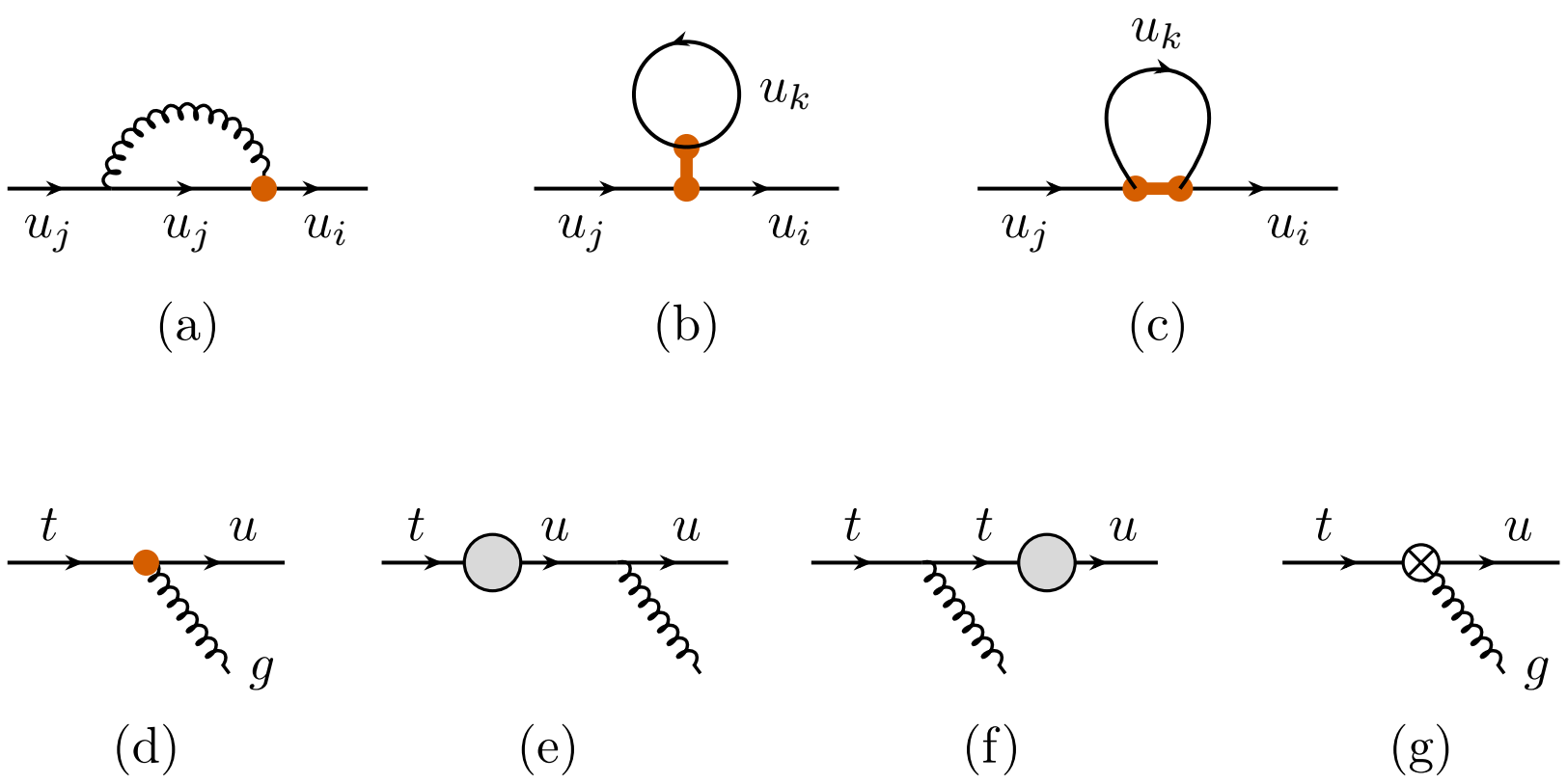
\begin{figure}[!htbp]
\centering
\begin{tikzpicture}[scale=0.95]
\begin{scope}
  \coordinate (A) at (-0.55,0); \coordinate (B) at (0.55,0);
  \draw[sfer] (-1.3,0) -- (A); \draw[sfer] (A) -- (B); \draw[sfer] (B) -- (1.3,0);
  \draw[sglu] (A) arc (180:0:0.55); \node[sopv] at (B) {};
  \node[anchor=north] at (-1.0,-0.05) {$u_j$}; \node[anchor=north] at (0,-0.05) {$u_j$}; \node[anchor=north] at (1.0,-0.05) {$u_i$};
  \node at (0,-1.0) {(a)};
\end{scope}
\begin{scope}[xshift=3.6cm]
  \coordinate (A) at (0,0); \coordinate (B) at (0,0.3);
  \draw[sfer] (-1.1,0) -- (A); \draw[sfer] (A) -- (1.1,0);
  \draw[scur] (A) -- (B); \node[sopv] at (A) {}; \node[sopv] at (B) {};
  \draw[sfer] (B) arc (-90:270:0.38);
  \node[anchor=north] at (-0.75,-0.05) {$u_j$}; \node[anchor=north] at (0.75,-0.05) {$u_i$}; \node[anchor=west] at (0.38,0.68) {$u_k$};
  \node at (0,-1.0) {(b)};
\end{scope}
\begin{scope}[xshift=7.0cm]
  \coordinate (A) at (-0.16,0); \coordinate (B) at (0.16,0);
  \draw[sfer] (-1.3,0) -- (A); \draw[sfer] (B) -- (1.3,0);
  \draw[scur] (A) -- (B); \node[sopv] at (A) {}; \node[sopv] at (B) {};
  \draw[sfer] (A) .. controls (-1.0,1.15) and (1.0,1.15) .. (B);
  \node[anchor=north] at (-0.95,-0.05) {$u_j$}; \node[anchor=north] at (0.95,-0.05) {$u_i$}; \node[anchor=south] at (0,0.87) {$u_k$};
  \node at (0,-1.0) {(c)};
\end{scope}
\begin{scope}[yshift=-2.7cm, xshift=-0.3cm]
  \coordinate (O) at (0,0);
  \draw[sfer] (-1.0,0) -- (O); \draw[sfer] (O) -- (1.0,0); \draw[sglu] (O) -- (0.6,-0.8);
  \node[sopv] at (O) {};
  \node[anchor=south] at (-0.7,0.02) {$t$}; \node[anchor=south] at (0.7,0.02) {$u$}; \node[anchor=west] at (0.6,-0.8) {$g$};
  \node at (0,-1.35) {(d)};
\end{scope}
\begin{scope}[yshift=-2.7cm, xshift=2.6cm]
  \coordinate (S) at (-0.4,0); \coordinate (O) at (0.5,0);
  \draw[sfer] (-1.2,0) -- (S); \draw[sfer] (S) -- (O); \draw[sfer] (O) -- (1.3,0); \draw[sglu] (O) -- (1.1,-0.8);
  \node[sblob] at (S) {};
  \node[anchor=south] at (-0.95,0.02) {$t$}; \node[anchor=south] at (0.05,0.02) {$u$}; \node[anchor=south] at (1.0,0.02) {$u$};
  \node at (0,-1.35) {(e)};
\end{scope}
\begin{scope}[yshift=-2.7cm, xshift=5.8cm]
  \coordinate (O) at (-0.5,0); \coordinate (S) at (0.4,0);
  \draw[sfer] (-1.3,0) -- (O); \draw[sfer] (O) -- (S); \draw[sfer] (S) -- (1.2,0); \draw[sglu] (O) -- (0.1,-0.8);
  \node[sblob] at (S) {};
  \node[anchor=south] at (-1.0,0.02) {$t$}; \node[anchor=south] at (-0.05,0.02) {$t$}; \node[anchor=south] at (0.95,0.02) {$u$};
  \node at (0,-1.35) {(f)};
\end{scope}
\begin{scope}[yshift=-2.7cm, xshift=8.9cm]
  \coordinate (O) at (0,0);
  \draw[sfer] (-1.0,0) -- (O); \draw[sfer] (O) -- (1.0,0); \draw[sglu] (O) -- (0.6,-0.8);
  \node[sct] at (O) {};
  \node[anchor=south] at (-0.7,0.02) {$t$}; \node[anchor=south] at (0.7,0.02) {$u$}; \node[anchor=west] at (0.6,-0.8) {$g$};
  \node at (0,-1.35) {(g)};
\end{scope}
\end{tikzpicture}
\caption{Flavour mixing in the quark self-energies. Top row, the self-energy $\Sigma_{ij}$ of eq.~\eqref{eq:mix:decomp} at $\order{C}$,
drawn for up-type quarks, through (a) a gluon dipole, (b) a closed quark loop of a four-quark operator, drawn as its two currents (dots) joined by a bar,
and (c) an open quark loop.  Bottom row, the decay $t\to ug$
through $(C_{uG})_{13}$. (d) The tree diagram. (e) and (f) The diagrams with a $t\to u$ self-energy (blob) on an external leg. (g) The counterterm vertex, which includes the off-diagonal field renormalisation factors of eq.~\eqref{eq:mix:dZ}, and supplies the
effect of (e) and (f).}
\label{fig:uv:mix}
\end{figure}
With general flavour the one-loop QCD self-energy of a quark is not diagonal at $\order{\Lambda^{-2}}$. Only two kinds of operator mix
the flavours at first order in the Wilson coefficients, as shown in the top row of figure~\ref{fig:uv:mix}. An off-diagonal dipole operator
mixes them through a gluon loop, as shown in diagram (a), and a four-quark operator through tadpole diagrams (b) and (c), which give a mass
mixing term proportional to the cube of the mass in the loop. The operators $O_{uH}$ and $O_{dH}$ do not mix the flavours, since the
Yukawa matrices absorb their contributions to the quark masses, as discussed in appendix~\ref{app:opsmass}.

The decay $t\to ug$ with $(C_{uG})_{13}$ shows why this matters. Its tree-level amplitude is the dipole vertex, diagram (d). At one loop the same dipole can
also turn the top quark into an up quark on an external leg, before or after the SM vertex emits the gluon, diagrams (e) and (f). A one-loop
generator does not compute self-energy insertions on external legs, so that their effect has to be supplied by a counterterm, diagram (g),
through an off-diagonal renormalisation of the quark fields. The same was done for flavour-changing neutral top-quark processes at NLO in
ref.~\cite{Degrande:2014tta}.

We write the self-energy of an incoming quark $j$ and an outgoing quark $i$, with the full propagator $\ii(\slashed p-m-\Sigma)^{-1}$, as
\begin{equation}\label{eq:mix:decomp}
  \Sigma_{ij}(p)=\slashed pP_L\Sigma^L_{ij}(p^2)+\slashed pP_R\Sigma^R_{ij}(p^2)+P_L\Sigma^{SL}_{ij}(p^2)+P_R\Sigma^{SR}_{ij}(p^2).
\end{equation}
The scalar parts $\Sigma^{SL}$ and $\Sigma^{SR}$ flip the chirality of the quark. In the SM only a mass can do this, so that they are
proportional to the quark masses, $\Sigma^{SL}_{ij}=m_i\Sigma^S_{ij}$ and $\Sigma^{SR}_{ij}=m_j\Sigma^S_{ij}$ with a common
$\Sigma^S_{ij}$~\cite{Denner:1990yz,Denner:1991kt}. The dipoles and the four-quark operators flip the chirality themselves. Besides the terms
proportional to the quark masses, their scalar parts contain terms that do not vanish for massless external quarks. The two scalar parts are independent functions and cannot be written with a common $\Sigma^S_{ij}$. The on-shell conditions for this general form are solved in refs.~\cite{Kniehl:1996bd,Espriu:2002xv},
and in the SMEFT in ref.~\cite{Endo:2020kie}. We renormalise the left-handed fields as
\begin{equation}\label{eq:mix:Z}
  \psi^0_{L,i}=\psi_{L,i}+\frac12\,\delta Z^L_{ij}\,\psi_{L,j} ,\qquad
  \bar\psi^0_{L,i}=\bar\psi_{L,i}+\frac12\,\bar\psi_{L,j}\,\delta\bar Z^L_{ji} ,
\end{equation}
and the right-handed ones in the same way. With a diagonal and real renormalised mass matrix, the condition that an on-shell quark $j$ does
not turn into a quark $i\neq j$ of different mass gives
\begin{equation}\label{eq:mix:dZ}
  \delta Z^L_{ij}=-\frac{2\big[m_j^2\Sigma^L_{ij}+m_im_j\Sigma^R_{ij}+m_i\Sigma^{SL}_{ij}+m_j\Sigma^{SR}_{ij}\big](m_j^2)}{m_i^2-m_j^2},
\end{equation}
with $\delta Z^R_{ij}$ by $L\leftrightarrow R$ and $SL\leftrightarrow SR$. For quarks of equal mass, such as the massless quarks of the
five-flavour scheme, eq.~\eqref{eq:mix:dZ} cannot be used, and the on-shell conditions are solved as described in
appendix~\ref{app:os:lsz}. The renormalisation factor $\delta\bar Z_{ij}$ of an outgoing quark follows from the
same condition at $p^2=m_i^2$. It differs from $\delta Z^\dagger$ when the self-energy has an imaginary part, as in the example below, and both
enter the counterterm vertices.

As an example, we consider the top quark and the up quark mixed by the dipole. In the following, the indices of the
renormalisation factors and of the CKM matrix are generation numbers, as for the Wilson coefficients, so that $\delta Z^L_{13}$ has an
outgoing up quark and an incoming top quark. The left-handed field renormalisation factors are
\begin{equation}\label{eq:mix:tq}
\begin{split}
  \delta Z^L_{31}&=\frac{\sqrt2g_sm_tv}{2\pi^2\Lam^2}\,(C_{uG})^*_{13}\Big(\frac1\epsilon-L_t+\frac13\Big),\\
  \delta Z^L_{13}&=-\frac{\sqrt2g_sm_tv}{4\pi^2\Lam^2}\,(C_{uG})_{13}\Big(\frac3\epsilon-L_t-2L_{\rm eft}-\frac23-\ii\pi\Big),
\end{split}
\end{equation}
with $L_t=\ln(m_t^2/\mu_R^2)$. The right-handed renormalisation factors follow by $(C_{uG})_{13}\to(C_{uG})^*_{31}$. The two factors are taken at different
momenta. The factor $\delta Z^L_{13}$ turns an incoming top quark into an up quark and is taken at $p^2=m_t^2$, above the threshold of the up quark and
the gluon, which gives the complex term $-\ii\pi$. The factor $\delta Z^L_{31}$ turns an incoming up quark into a top quark and is taken at $p^2=0$,
below every threshold.

The term $-2L_{\rm eft}$ of $\delta Z^L_{13}$, and $2/\epsilon$ of its pole, come from a second counterterm of the off-diagonal two-point
function. The
pole of the scalar part of the dipole self-energy contains a term proportional to $p^2$,
\begin{equation}\label{eq:mix:p2}
  \Sigma^{SR}_{ij}\supset\frac{k_{ij}\,p^2}{\epsilon} ,\qquad k_{ij}=\frac{\sqrt2g_sv}{4\pi^2\Lam^2}\,(C_{uG})_{ij} ,
\end{equation}
and $\Sigma^{SL}_{ij}$ the same with $(C_{uG})^*_{ji}$. The field renormalisation factors multiply $\slashed p$ and the mass, and cannot remove the term proportional to $p^2$. This term is
the Feynman rule of a redundant operator of the Green's basis, of the type $\bar qD^2u\,\tilde\varphi$, whose coefficient we renormalise in
$\MSbar$ at $\mu_{\rm eft}$ as discussed in section~\ref{sec:uv:os}. Its counterterm adds $-k_{ij}\,p^2(1/\epsilon-L_{\rm eft})$ to
$\Sigma^{SR}_{ij}$, and the field renormalisation factors are fixed by the on-shell condition with this counterterm included. In eq.~\eqref{eq:mix:dZ} with
$i=1$ and $j=3$ it enters through $m_t\Sigma^{SR}_{13}(m_t^2)$, and gives
\begin{equation}\label{eq:mix:p2dZ}
  \delta Z^L_{13}\supset-2k_{13}\,m_t\Big(\frac1\epsilon-L_{\rm eft}\Big)
  =-\frac{\sqrt2g_sm_tv}{4\pi^2\Lam^2}\,(C_{uG})_{13}\Big(\frac2\epsilon-2L_{\rm eft}\Big) .
\end{equation}
At $p^2=0$, where $\delta Z^L_{31}$ is fixed, the redundant counterterm vanishes. The fields are thus renormalised on shell and the Wilson
coefficients in $\MSbar$.

Ref.~\cite{Zhang:2014rja} has no separate counterterm for the redundant operator and lets the field renormalisation factor absorb
the whole self-energy, which gives the bracket $1/\epsilon-L_t-\frac23-\ii\pi$, without the terms of eq.~\eqref{eq:mix:p2dZ}. This is
sufficient for the decays $t\to ug$ and $t\to cg$, where the self-energy appears only on external legs at $p^2=m_t^2$. In a framework in which every counterterm vertex
removes the full pole of its Green's function, as in ours or in \SMEFTNLO{}, the two-point counterterm also enters internal quark lines,
where $p^2\neq m_t^2$. There the term proportional to $p^2$ is removed only by the counterterm of the redundant operator, since the field renormalisation factors
do not depend on the momentum.

\subsection{Light-quark masses from four-quark operators}\label{sec:uv:lightmass}
A four-quark operator that couples left- and right-handed quarks, such as $O^{(1,8)}_{qu}$ or $O^{(1,8)}_{quqd}$, gives a mass term to a light quark through a massive quark loop as in
diagram (c) of figure~\ref{fig:uv:mix} with $u_i=u_j$ and $u_k=t$. This mass is proportional to $m_t^3$, and it is the broken-phase equivalent of the mixing of these operators into $O_{uH}$ and
$O_{dH}$~\cite{Jenkins:2013wua}. In the five-flavour scheme the light quarks must stay
massless, so we cancel this generated mass term with an on-shell counterterm. We write the counterterm as $\delta m^*P_R+\delta m\,P_L$ with a complex $\delta m$. For $(C^{(1,8)}_{qu})_{3ii3}$ and $i=1,2$ it is
\begin{equation}\label{eq:mix:mq}
  \delta m_{ii}=-\frac{\big(C^{(1)}_{qu}+C_FC^{(8)}_{qu}\big)_{3ii3}\,m_t^3}{16\pi^2\Lam^2}\Big(\frac4\epsilon+2-4L_t\Big),
\end{equation}
with $C_F=4/3$. Since the tadpole does not depend on the momentum, this counterterm cancels the tadpole %
 at every $p^2$ and sets the pole mass of the
light quark to zero at $\order{\Lambda^{-2}}$. For example, with an $\MSbar$ counterterm alone the light quark would acquire a pole mass of about 65~MeV for $(C^{(1)}_{qu})_{3ii3}=1$, $\Lam=1$~TeV and $\mu_R=m_t$, which would be inconsistent with the flavour scheme. The same counterterm removes the mass mixing of the up and charm quarks. The diagonal renormalisation factors and the closed forms of the self-energies are given in
appendices~\ref{app:os:lsz} and~\ref{app:os:selfenergies}.

\subsection{Renormalisation of the CKM matrix}\label{sec:uv:ckm}
\begin{figure}[!htbp]
\centering
\begin{tikzpicture}[scale=0.95]
\begin{scope}
  \coordinate (S) at (-0.4,0); \coordinate (O) at (0.5,0);
  \draw[sfer] (-1.2,0) -- (S); \draw[sfer] (S) -- (O); \draw[sfer] (O) -- (1.3,0); \draw[spho] (O) -- (1.1,-0.8);
  \node[sblob] at (S) {};
  \node[anchor=south] at (-0.95,0.02) {$t$}; \node[anchor=south] at (0.05,0.02) {$u$}; \node[anchor=south] at (1.0,0.02) {$b$};
  \node[anchor=west] at (1.1,-0.8) {$W^+$}; \node[anchor=north east] at (0.48,-0.06) {\scriptsize$V^*_{13}$};
  \node at (0,-1.35) {(a)};
\end{scope}
\begin{scope}[xshift=4.2cm]
  \coordinate (O) at (0,0);
  \draw[sfer] (-1.0,0) -- (O); \draw[sfer] (O) -- (1.0,0); \draw[spho] (O) -- (0.6,-0.8);
  \node[sct] at (O) {};
  \node[anchor=south] at (-0.7,0.02) {$t$}; \node[anchor=south] at (0.7,0.02) {$b$}; \node[anchor=west] at (0.6,-0.8) {$W^+$};
  \node at (0,-1.35) {(b)};
\end{scope}
\end{tikzpicture}
\caption{The decay $t\to bW^+$ through $(C_{uG})_{13}$. (a) The dipole self-energy (blob) turns the top quark into an up quark, which decays
through the SM vertex with $V^*_{13}$. (b) The counterterm vertex with the off-diagonal field renormalisation factors, the SM $u\to bW^+$ vertex times
$\frac12\delta Z^L_{13}$.}
\label{fig:uv:ckm}
\end{figure}
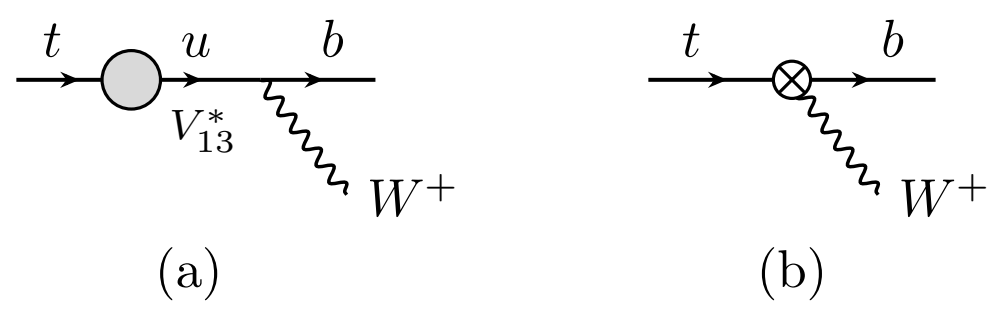

The dipoles and the four-quark operators generate off-diagonal mass terms at one loop, and the basis in which the quark mass
matrices are diagonal is restored through the off-diagonal field renormalisation factors of section~\ref{sec:uv:mix}. This rotation drops out of the vertices of the gluon, the photon and the $Z$ boson, which
couple to the three generations in the same way, but not out of the vertex of the $W$ boson, where the up-type and the down-type quarks
are rotated differently. There it is absorbed by the counterterm of the CKM matrix, defined as in section~\ref{sec:setup:ewsb}. The UV pole of the $W$ vertex is removed by this counterterm together with those
of the Wilson coefficients in eq.~\eqref{eq:uv:rule}, in particular of $C^{(3)}_{Hq}$ and $C_{Hud}$. %
 An example is the decay $t\to bW^+$ at first order in $(C_{uG})_{13}$, shown in figure~\ref{fig:uv:ckm}. In diagram (a) the
dipole turns the top quark into an up quark, which decays with $V^*_{13}$. Diagram (b) is its counterterm, the SM vertex $u\to bW^+$
times $\frac12\delta Z^L_{13}$ of eq.~\eqref{eq:mix:tq}. The leftover pole is absorbed by the renormalisation of the CKM matrix entry $V_{33}$ together with the counterterm of the redundant operator of eq.~\eqref{eq:mix:p2}.

We renormalise the CKM matrix of the SMEFT in $\MSbar$ at $\mu_{\rm eft}$, so that the input is $V(\mu_{\rm eft})$, the Lagrangian parameter of section~\ref{sec:setup:ewsb}%
. An on-shell matrix
$V^{\rm OS}$ can also be defined in the SMEFT by applying the prescription of ref.~\cite{Denner:1990yz} to the off-diagonal
field renormalisation factors above. The example models of section~\ref{sec:ufo} use the $\MSbar$ matrix. The on-shell definition is gauge dependent in the electroweak
theory~\mbox{\cite{Gambino:1998ec,Kniehl:2006bs}}, but at this order it does not depend on the QCD gauge, since the part of the gluon propagator that depends on the gauge vanishes when contracted with the dipole vertex, and the four-quark tadpoles contain no gluon.

Since the dipoles and the four-quark operators change the mass matrices with the scale, the $\MSbar$ matrix $V(\mu_{\rm eft})$
runs at linear order in the SMEFT. Its running is the one that follows from the mixing of the dipoles and the four-quark operators, for instance $C_{uG}$ and $C^{(1,8)}_{qu}$, into
$C_{uH}$ and $C_{dH}$~\cite{Jenkins:2013wua,Alonso:2013hga} when the quark mass matrix is diagonalised at $\mu_{\rm eft}$. The effects of the
running of the Yukawa matrices on flavour were studied in ref.~\cite{Aebischer:2020lsx}. The on-shell matrix also depends on
$\mu_{\rm eft}$, but only through the dipoles, because they induce the redundant operator of eq.~\eqref{eq:mix:p2}, which is
renormalised in $\MSbar$ at $\mu_{\rm eft}$. Appendix~\ref{app:os:ckm} gives the two counterterms and the running of the $\MSbar$ matrix with all quarks massive. The electronic files described in appendix~\ref{app:anc} give the complete counterterm of $V$, with the dipoles of both sectors, the
four-quark operators and all quark masses, in either the $\MSbar$ or the on-shell scheme.

\subsection{The CP-odd top-quark dipole and the chiral rotation}\label{sec:uv:cpodd}
At one loop the imaginary part of a diagonal dipole, ${\rm Im}\,(C_{uG})_{ii}$ or ${\rm Im}\,(C_{dG})_{ii}$, generates a CP-odd mass term for
the quark $i$, $\Sigma_{ii}\supset\ii\gfive\Sigma_5(p^2)$, which in terms of the scalar parts of eq.~\eqref{eq:mix:decomp} is
\begin{equation}\label{eq:cp:sigma5}
  \Sigma_5(p^2)=\frac\ii2\Big[\Sigma^{SL}_{ii}(p^2)-\Sigma^{SR}_{ii}(p^2)\Big],
\end{equation}
and vanishes on shell for massless quarks. In the five-flavour scheme
it therefore appears only for the top quark, from ${\rm Im}\,C_{tG}$, with $C_{tG}\equiv(C_{uG})_{33}$. Its on-shell counterterm $\delta m_5$ enters the mass term of the top quark as
\begin{equation}\label{eq:cp:mass}
  \Lag\supset-\bar t\,\big(m_t+\delta m_t-\ii\gfive\,\delta m_5\big)\,t ,
\end{equation}
and is fixed by the on-shell condition
\begin{equation}\label{eq:cp:os}
  \Sigma_5(m_t^2)-\delta m_5-\delta K_5\,m_t^2=0 ,
\end{equation}
where $\delta K_5$ is the $\MSbar$ counterterm of the $p^2$ term of $\Sigma_5$, as derived in appendix~\ref{app:os:cpodd}. A one-loop
generator cannot use the counterterm $-\ii\gfive\delta m_5$ directly. The top-quark mass and Yukawa coupling are real parameters, which enter the propagators, the spinors and the scalar
couplings, and a complex mass counterterm would require CP-odd counterterms on every vertex that contains $m_t$ or $y_t$. As for the phases of the quark masses in the SM, a chiral rotation of the quark field removes this mass term. For example, the same rotation removes the CP-violating quark masses generated at one loop in the matching of the SMEFT to the low-energy effective theory~\cite{Dekens:2019ept} or in models with new CP-violating phases~\cite{Pospelov:2005pr}. We therefore remove the CP-odd mass by a phase rotation of the bare right-handed
top quark by a small angle of first order in $g_s\,{\rm Im}\,C_{tG}$,
\begin{equation}\label{eq:cp:rot}
  t_R\to e^{\ii\theta}\,t_R ,\qquad
  \theta=-\frac{\delta m_5}{m_t}=\frac{\sqrt2\,g_sm_tv\,{\rm Im}\,C_{tG}}{2\pi^2\Lam^2}
  \Big(\frac1\epsilon+\frac16-\frac{L_t}2-\frac{L_{\rm eft}}2\Big),
\end{equation}
which turns the mass term into $-m_t\bar tt-\ii\theta m_t\bar t\gfive t+\order{\theta^2}$, so that the renormalised mass stays real and on shell.
With an anticommuting $\gfive$ the kinetic term is invariant under the redefinition of eq.~\eqref{eq:cp:rot} also in $D$ dimensions, so that the vector and axial currents of the
top quark and its gauge couplings are unchanged, and the left-handed doublets are not touched. Only the scalar couplings of the top quark
change, by
\begin{equation}\label{eq:cp:shifts}
\begin{split}
  \delta\Lag={}&-\ii\theta\,m_t\,\bar t\gfive t-\ii\theta\,\frac{m_t}{v}\,h\,\bar t\gfive t-\theta\,\frac{m_t}{v}\,G^0\,\bar tt\\
  &-\theta\,\frac{\sqrt2\,m_t}{v}\sum_i\Big(V_{3i}\,G^+\,\bar tP_Ld_i+V^*_{3i}\,G^-\,\bar d_iP_Rt\Big),
\end{split}
\end{equation}
while the dimension-six vertices change only at $\order{\Lambda^{-4}}$. The shifts of eq.~\eqref{eq:cp:shifts} are added as effective counterterm vertices. Like every chiral rotation~\cite{Adler:1969gk,Bell:1969ts}, eq.~\eqref{eq:cp:rot}
shifts the coefficients of $G\widetilde G$ and $B\widetilde B$ through its anomalous
Jacobian~\cite{Fujikawa:1979ay,Bhattacharya:2015rsa}. For $G\widetilde G$ the shift is of order $g_s^3\,{\rm Im}\,C_{tG}$~\cite{Dekens:2019ept}, which is
beyond the order of this work. Both $G\widetilde G$ and $B\widetilde B$ are also total derivatives, whose Feynman rules vanish for every vertex of the
set.

In the five-flavour scheme other CP-odd coefficients do not need this treatment. The CP-odd Yukawa coefficient ${\rm Im}\,(C_{uH})_{33}$ never enters the
self-energies. The tadpole mass of eq.~\eqref{eq:mix:mq} is also complex when the four-quark
coefficient has an imaginary part, but it needs no rotation, since the renormalised mass of the light quark vanishes and the counterterm of eq.~\eqref{eq:mix:mq} removes the tadpole mass completely. 

In general, with additional massive quarks, the imaginary parts of diagonal dipoles and the complex four-quark tadpoles generate CP-odd masses of the quarks, which are removed by an analogous chiral
rotation. The Mathematica package applies it to every massive quark.

\section{Results, example models and validation}\label{sec:results}
This section presents the computed one-loop effective vertices with some examples, their form in common flavour assumptions including the
colour-octet Wilson coefficients, the example models we built from them and the tests we performed.

\subsection{Overview%
}\label{sec:res:content}\label{sec:res:reading}\label{sec:res:recipe}
We computed the rational terms and the UV poles of all effective vertices of the set defined in section~\ref{sec:scope}, the 1PI vertices with a loop of quarks, gluons and ghosts and at least one quark or gluon leg, at first order in the Wilson coefficients. For each vertex %
 we give the tree-level Feynman rule when it exists, the rational term $\Rtwo$ and the
coefficient of the $1/\epsilon$ pole before any renormalisation, with the four-quark structures reduced as in section~\ref{sec:schemes}. Of the 4743 direct contributions, 2502 have a non-zero rational term or pole. Table~\ref{tab:res:classes} groups them by class. The electroweak dipoles, for instance, have no
rational term or pole in any of their vertices. However, their counterterm vertices do not vanish, since the quark field renormalisation factors of
eq.~\eqref{eq:uv:rule} still act on their tree, and the pole of these factors gives their QCD running. The 25 indirect contributions through
the input shifts are given in terms of $\delta G_F$, $\delta g_2$, $\delta g_1$, $\delta k_H$, $\delta k_{G^0}$ and $C_{HWB}$, so that the set can be combined with any choice of electroweak inputs.

\begin{table}[!htbp]
\centering\small\setlength{\tabcolsep}{4pt}
\begin{tabular}{@{}>{\raggedright\arraybackslash}p{0.42\textwidth}rr@{}}
\toprule
class & all & $\neq0$\\
\midrule
$X^3$: $C_G$, $C_{\widetilde G}$ & 70 & 69\\
$X^2H^2$: $C_{HG}$, $C_{H\widetilde G}$ & 158 & 90\\
$\psi^2H^3$: $C_{uH}$, $C_{dH}$ & 110 & 72\\
$\psi^2XH$: $C_{uG}$, $C_{dG}$ & 594 & 396\\
$\psi^2XH$: electroweak dipoles & 318 & 0\\
$\psi^2H^2D$: $C^{(1,3)}_{Hq}$, $C_{Hu}$, $C_{Hd}$, $C_{Hud}$ & 451 & 365\\
four-quark, vector currents & 2392 & 1112\\
four-quark, $C^{(1,8)}_{quqd}$ & 514 & 318\\
semileptonic & 136 & 80\\
\midrule
total & 4743 & 2502\\
\bottomrule
\end{tabular}
\caption{Direct contributions per class, all of them (column ``all'') and those with a non-zero rational term or pole (column $\neq0$), counted before the unitarity of
the CKM matrix is used. The 25 indirect contributions through the input shifts are not included.}\label{tab:res:classes}\label{tab:val:coverage}
\end{table}

The tree vertex, the rational term and the pole are written in the form
of eq.~\eqref{eq:tb:vertex},
\begin{equation}\label{eq:anc:vertex}
  \vf^{X}=\sum_{\alpha,\beta}c^X_{\alpha\beta}\;\mathcal C_\alpha\otimes\mathcal T_\beta ,\qquad X\in\{\text{tree},\Rtwo,\text{UV}\},
\end{equation}
where each coefficient $c^X_{\alpha\beta}$ is a sum of terms, each a function of the masses and the momenta times a flavour factor that is a product of
Wilson coefficients, CKM elements and masses such as those of eqs.~\eqref{eq:fl:chain} and~\eqref{eq:fl:trace}. The effective vertex is
\begin{equation}\label{eq:anc:norm}
  \Gamma=\frac{\ii}{16\pi^2}\Bigl[\vf^{\Rtwo}+\frac{\vf^{\rm UV}}{\epsilon}\Bigr]+\ldots,
\end{equation}
so that the $\Rtwo$ Feynman rule in the conventions of refs.~\cite{Ossola:2008xq,Draggiotis:2009yb} is $\frac{\ii}{16\pi^2}\vf^{\Rtwo}$. The vertices are attached to this paper in electronic format and in a PDF document which lists them ordered by vertex and by Wilson
coefficient%
. The conventions for the legs, the indices and the momenta, the Levi-Civita symbol of the files and the names of the Wilson coefficients are given
in appendix~\ref{app:anc:format}.

The pole of a 1PI Green's function does not depend on the renormalisation scheme, while the counterterm depends on choices such as the number of light quarks and the renormalisation prescription for the masses, the fields, the
CKM matrix and $g_s$, which a model makes as discussed in section~\ref{sec:uv}. For a specific choice of which quarks are massless and of the scheme of the CKM matrix, $\MSbar$ or on shell, the complete counterterm vertex %
 is returned by the function \texttt{UVCounterterm} of the Mathematica package attached to this paper, described in
appendix~\ref{app:anc}. The on-shell renormalisation factors of section~\ref{sec:uv} and appendix~\ref{app:os} that enter this
counterterm are attached in a second PDF document and in electronic format. In the rest of this section, section~\ref{sec:res:examples} gives examples of effective vertices, section~\ref{sec:res:flavour} the vertices in common flavour assumptions, section~\ref{sec:ufo} the example models built from them and section~\ref{sec:validation} the validation.

\subsection{Example effective vertices}\label{sec:res:examples}
We now give four examples of effective vertices, in the normalisation $\vf$ of eq.~\eqref{eq:units}. They show the main features discussed in the previous sections. The first mixes two quark generations through a dipole, the second contains a closed quark loop with a single $\gfive$, the third is a penguin with CKM factors, and the fourth is a four-quark vertex that depends on the evanescent scheme. %

The rational term of the two-point vertex at $\order{C_{uG}}$ of an
incoming up-type quark of generation $i$ and an outgoing one of generation $j$, from the gluon loop of figure~\hyperref[fig:uv:mix]{\ref*{fig:uv:mix}(a)}, is
\begin{equation}\label{eq:sigma}
\begin{split}
  \mathcal V^{\Rtwo}_{ji}(p)=\frac{\sqrt2\,g_s v}{\Lambda^2}\Bigl\{&\Bigl[2p^2-\tfrac73\bigl(m_i^2+m_j^2\bigr)\Bigr]\,
  2\bigl((C_{uG})_{ji}P_R+(C^*_{uG})_{ij}P_L\bigr)\\
  &-\tfrac43\,\slashed p\,\Bigl[\bigl(m_j(C_{uG})_{ji}+m_i(C^*_{uG})_{ij}\bigr)P_R\\
  &\qquad\quad+\bigl(m_i(C_{uG})_{ji}+m_j(C^*_{uG})_{ij}\bigr)P_L\Bigr]\Bigr\}.
\end{split}
\end{equation}
Its pole is
\begin{equation}\label{eq:sigmaUV}
\begin{split}
  \mathcal V^{\rm UV}_{ji}(p)=\frac{\sqrt2\,g_s v}{\Lambda^2}\Bigl\{&\Bigl[-2p^2+2\bigl(m_i^2+m_j^2\bigr)\Bigr]\,
  2\bigl((C_{uG})_{ji}P_R+(C^*_{uG})_{ij}P_L\bigr)\\
  &+2\,\slashed p\,\Bigl[\bigl(m_j(C_{uG})_{ji}+m_i(C^*_{uG})_{ij}\bigr)P_R\\
  &\qquad\quad+\bigl(m_i(C_{uG})_{ji}+m_j(C^*_{uG})_{ij}\bigr)P_L\Bigr]\Bigr\}.
\end{split}
\end{equation}
The term of the pole proportional to $p^2$ is the Feynman rule of the redundant operator $\bar qD^2u\,\tilde\varphi$ of section~\ref{sec:uv:mix}. For $i=j$ the scalar part contains
$(C_{uG})_{ii}P_R+(C^*_{uG})_{ii}P_L={\rm Re}\,(C_{uG})_{ii}+\ii\gfive\,{\rm Im}\,(C_{uG})_{ii}$. The imaginary part of a diagonal dipole therefore enters the CP-odd mass term $\ii\gfive\Sigma_5$ of
eq.~\eqref{eq:cp:sigma5}%
.

The second example is the %
 loop-induced vertex $ggh$ at first order in the imaginary parts of the gluon dipoles, ${\rm Im}\,C_{uG}$ and ${\rm Im}\,C_{dG}$. On the structure $\delta^{AB}\epsilon^{\mu_1\mu_2p_1p_2}$ it reads
\begin{align}\label{eq:res:ggh}
  \vf^{\Rtwo}_{ggh}\big|_{\Im C_{uG},\,\Im C_{dG}}&=\frac{16\sqrt2}{3}\,\frac{g_s}{\Lambda^2}\Bigl[\sum_am_{u,a}\,\Im (C_{uG})_{aa}
    +\Im\sum_{a,b}V^*_{ab}(C_{dG})_{ab}\,m_{d,b}\Bigr],\nonumber\\
  \vf^{\rm UV}_{ggh}\big|_{\Im C_{uG},\,\Im C_{dG}}&=-8\sqrt2\,\frac{g_s}{\Lambda^2}\Bigl[\sum_am_{u,a}\,\Im (C_{uG})_{aa}
    +\Im\sum_{a,b}V^*_{ab}(C_{dG})_{ab}\,m_{d,b}\Bigr],
\end{align}
with the trace over the generations of the up-type loop and the CKM-rotated trace of the down-type loop. The Yukawa couplings of the diagrams
with a Higgs--quark vertex have been replaced by $y_\psi=\sqrt2\,m_\psi/v$ for each quark $\psi$, which holds at this order. With the loop read at the
operator the rational term is $-\frac23$ times the pole. Another reading point multiplies the rational term by $-\frac12$
at the Higgs vertex and by $\frac14$ at a gluon vertex, and the difference can be absorbed by a finite shift of $C_{H\widetilde G}$, given in appendix~\ref{app:g5su3}.

\begin{figure}[!htbp]
\centering
\begin{tikzpicture}[scale=0.95]
\begin{scope}
  \coordinate (A) at (-0.2,0); \coordinate (B) at (0.2,0);
  \draw[sfer] (-1.4,0) -- (A); \draw[sfer] (B) -- (1.4,0);
  \draw[sfer] (A) to[out=90,in=180] (0,0.55); \draw[sfer] (0,0.55) to[out=0,in=90] (B);
  \draw[sglu] (0,0.55) -- (0,1.35);
  \draw[scur] (A) -- (B); \node[sopv] at (A) {}; \node[sopv] at (B) {};
  \node[anchor=north] at (-1.0,-0.05) {$d_r$}; \node[anchor=north] at (1.0,-0.05) {$d_s$};
  \node[anchor=east] at (-0.38,0.55) {$d$}; \node[anchor=west] at (0,1.35) {$g$};
  \node at (0,-0.95) {(a)};
\end{scope}
\begin{scope}[xshift=3.7cm]
  \coordinate (A) at (0,0); \coordinate (B) at (0,0.22);
  \draw[sfer] (-1.1,0) -- (A); \draw[sfer] (A) -- (1.1,0);
  \draw[scur] (A) -- (B); \node[sopv] at (A) {}; \node[sopv] at (B) {};
  \draw[sfer] (B) arc (-90:270:0.36); \draw[sglu] (0,0.94) -- (0,1.52);
  \node[anchor=north] at (-0.75,-0.05) {$d_r$}; \node[anchor=north] at (0.75,-0.05) {$d_s$}; \node[anchor=west] at (0,1.52) {$g$};
  \node at (0,-0.95) {(b)};
\end{scope}
\begin{scope}[xshift=7.7cm]
  \coordinate (A) at (-0.17,0); \coordinate (B) at (0.17,0);
  \draw[sfer] (-1.3,-0.9) -- (A); \draw[sfer] (A) -- (-1.3,0.9); \draw[sfer] (1.3,-0.9) -- (B); \draw[sfer] (B) -- (1.3,0.9);
  \draw[scur] (A) -- (B); \node[sopv] at (A) {}; \node[sopv] at (B) {};
  \draw[sglu] (-0.95,-0.62) to[bend left=40] (-0.95,0.62);
  \node[anchor=east] at (-1.3,-0.9) {$u_i$}; \node[anchor=east] at (-1.3,0.9) {$\bar u_j$};
  \node[anchor=west] at (1.3,-0.9) {$d_r$}; \node[anchor=west] at (1.3,0.9) {$\bar d_s$};
  \node at (0,-1.65) {(c)};
\end{scope}
\begin{scope}[xshift=12.2cm]
  \coordinate (A) at (-0.17,0); \coordinate (B) at (0.17,0);
  \draw[sfer] (-1.3,-0.9) -- (A); \draw[sfer] (A) -- (-1.3,0.9); \draw[sfer] (1.3,-0.9) -- (B); \draw[sfer] (B) -- (1.3,0.9);
  \draw[scur] (A) -- (B); \node[sopv] at (A) {}; \node[sopv] at (B) {};
  \draw[sglu] (-0.95,-0.62) to[bend right=40] (0.95,-0.62);
  \node[anchor=east] at (-1.3,-0.9) {$u_i$}; \node[anchor=east] at (-1.3,0.9) {$\bar u_j$};
  \node[anchor=west] at (1.3,-0.9) {$d_r$}; \node[anchor=west] at (1.3,0.9) {$\bar d_s$};
  \node at (0,-1.65) {(d)};
\end{scope}
\end{tikzpicture}
\caption{Diagrams of the vertices of eqs.~\eqref{eq:res:penguin} and~\eqref{eq:res:fourq}, with a four-quark operator drawn as its two
currents (dots) joined by a bar. (a) The open and (b) the closed penguin of $d_r\bar d_sg$ at $\order{C^{(1)}_{qq}}$. (c) and (d) Two of the six
gluon exchanges of $u_i\bar u_jd_r\bar d_s$ at $\order{C^{(1,8)}_{ud}}$.}
\label{fig:res:ex}
\end{figure}
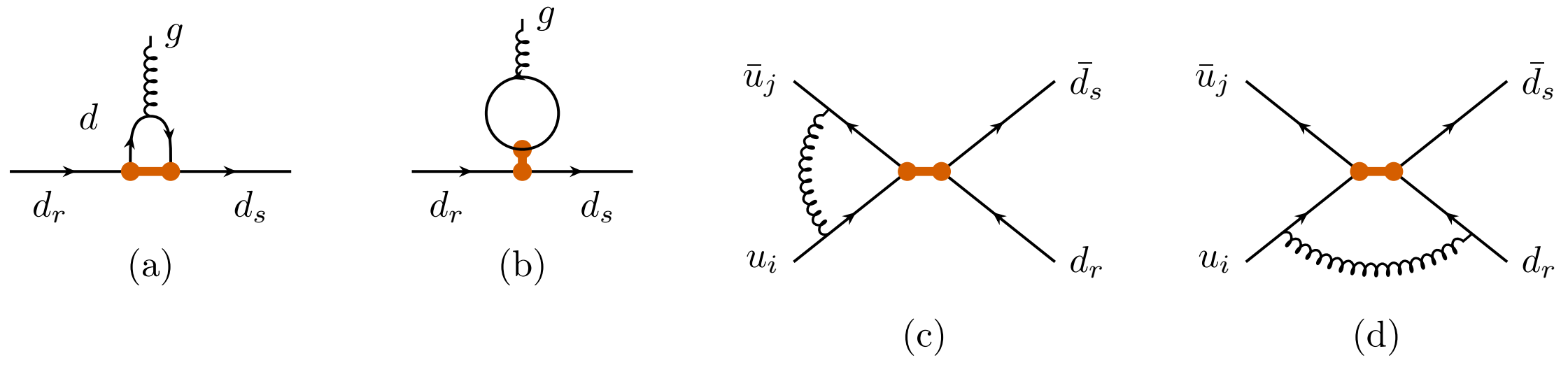
The third example is the %
 penguin $d_r\bar d_sg$ at $\order{C^{(1)}_{qq}}$. With $p$ the momentum of the gluon, it reads
\begin{equation}\label{eq:res:penguin}
\begin{split}
  \vf^{\Rtwo,\mu}_{d_r\bar d_sg}\big|_{C^{(1)}_{qq}}&=\frac{2g_s}{\Lambda^2}\,T^A\Bigl[\Bigl(-p^2\gamma^\mu+\tfrac23\,p^\mu\slashed p\Bigr)
  X_{sr}+2\gamma^\mu X^{(m)}_{sr}\Bigr]P_L ,\\
  \vf^{\rm UV,\mu}_{d_r\bar d_sg}\big|_{C^{(1)}_{qq}}&=\frac{4g_s}{3\Lambda^2}\,T^A\bigl(p^2\gamma^\mu-p^\mu\slashed p\bigr)X_{sr}P_L ,
\end{split}
\end{equation}
with the flavour factors
\begin{equation}\label{eq:res:penguinX}
\begin{aligned}
  X_{sr}&=\sum_{a,b,c,d}V^*_{as}\,(C^{(1)}_{qq})_{abcd}\,\bigl(VV^\dagger\bigr)_{bc}\,V_{dr},\\
  X^{(m)}_{sr}&=\sum_{a,b,c,d}V^*_{as}\,(C^{(1)}_{qq})_{abcd}\,\bigl(VM_d^2V^\dagger\bigr)_{bc}\,V_{dr},
\end{aligned}
\end{equation}
where the unitarity relation $(VV^\dagger)_{bc}=\delta_{bc}$ is not used, as stated in section~\ref{sec:flavour}. This is the open penguin of figure~\hyperref[fig:res:ex]{\ref*{fig:res:ex}(a)}, whose loop runs through a down-type
quark of any generation. The closed penguin, diagram (b), vanishes for a colour-singlet operator. The pole is transverse while the rational term is
not, and the flavour factor with the masses enters the rational term only.

Finally, the %
 four-quark vertex $u_i\bar u_jd_r\bar d_s$ at $\order{C^{(1,8)}_{ud}}$ needs two colour structures and one Lorentz--Dirac structure,
\begin{equation}\label{eq:res:fourqS}
  \mathcal C_1=\delta_{c_2c_1}\delta_{c_4c_3},\qquad \mathcal C_8=T^A_{c_2c_1}T^A_{c_4c_3},\qquad
  \mathcal T=[\gamma^\rho P_R]_{s_2s_1}[\gamma_\rho P_R]_{s_4s_3}.
\end{equation}
The tree, the rational term and the pole are
\begin{equation}\label{eq:res:fourq}
\begin{split}
  \vf^{\rm tree}_{u_i\bar u_jd_r\bar d_s}&=-\frac{\ii}{\Lambda^2}\Bigl[(C^{(1)}_{ud})_{jisr}\,\mathcal C_1+(C^{(8)}_{ud})_{jisr}\,\mathcal C_8\Bigr]\otimes\mathcal T,\\
  \vf^{\Rtwo}_{u_i\bar u_jd_r\bar d_s}\big|_{C^{(1,8)}_{ud}}&=\frac{g_s^2}{\Lambda^2}\Bigl[(C^{(1)}_{ud})_{jisr}\bigl(\tfrac{16}3\mathcal C_1+2\mathcal C_8\bigr)
    +(C^{(8)}_{ud})_{jisr}\bigl(\tfrac49\mathcal C_1+\tfrac{14}3\mathcal C_8\bigr)\Bigr]\otimes\mathcal T,\\
  \vf^{\rm UV}_{u_i\bar u_jd_r\bar d_s}\big|_{C^{(1,8)}_{ud}}&=\frac{g_s^2}{\Lambda^2}\Bigl[(C^{(1)}_{ud})_{jisr}\bigl(-\tfrac83\mathcal C_1+6\mathcal C_8\bigr)
    +(C^{(8)}_{ud})_{jisr}\bigl(\tfrac43\mathcal C_1-\tfrac{14}3\mathcal C_8\bigr)\Bigr]\otimes\mathcal T.
\end{split}
\end{equation}
Each current of the operators $O^{(1,8)}_{ud}$ connects two quarks of the same type, so no $t$-channel diagram joins an up-type to a
down-type quark, unlike the case of four identical quarks. Only the gluon exchanges between the two currents, figure~\hyperref[fig:res:ex]{\ref*{fig:res:ex}(d)},
mix the singlet into the octet. They also leave three gamma matrices on each quark line, which are reduced as in section~\ref{sec:schemes},
so their rational term depends on the evanescent scheme.

\subsection{Effective vertices in common flavour assumptions}\label{sec:res:octet}\label{sec:res:flavour}
Most NLO models restrict the 2499 real parameters of the SMEFT at dimension six with a flavour assumption,
which writes the Wilson coefficients in terms of a manageable number of parameters. We also give our vertices in the common assumptions of
table~\ref{tab:res:assumptions}, the four of \textsc{SMEFTsim}~\cite{Brivio:2020onw} and that of \SMEFTNLO{}~\cite{Degrande:2020evl}.
For all parameters except the colour-octet ones discussed below, this consists in substituting the coefficient tensors with the parameters of the assumption through the relations defined by the flavour restriction. The electronic files give this map for every assumption,
and the package function \texttt{ApplyFlavour} applies it, as described in appendix~\ref{app:anc}. For \SMEFTNLO{} the map is that
of ref.~\cite{SMEFTatNLO:definitions}, and appendix~\ref{app:maps} lists its differences from our conventions.
\begin{table}[!htbp]
\centering\small
\begin{tabularx}{\textwidth}{@{}l>{\raggedright\arraybackslash}Xcr@{}}
\toprule
assumption & flavour symmetry & CKM & parameters\\
\midrule
\textsc{SMEFTsim} U35 & $U(3)^5$, flavour universality, with the chirality-flipping coefficients proportional to the Yukawa matrices~\cite{Brivio:2020onw} & general & 85 (25)\\
\textsc{SMEFTsim} MFV & minimal flavour violation, CP-conserving~\cite{DAmbrosio:2002vsn,Brivio:2020onw} & general & 130 (0)\\
\textsc{SMEFTsim} top & $U(2)_q\times U(2)_u\times U(2)_d\times U(1)^3_{\ell+e}$~\cite{Brivio:2020onw,Aguilar-Saavedra:2018ksv} & $V=1$ & 275 (71)\\
\textsc{SMEFTsim} topU3l & $U(2)_q\times U(2)_u\times U(2)_d\times U(3)_\ell\times U(3)_e$~\cite{Brivio:2020onw} & $V=1$ & 182 (53)\\
\SMEFTNLO{} & $U(2)_q\times U(2)_u\times U(3)_d\times[U(1)_\ell\times U(1)_e]^3$~\cite{Degrande:2020evl}, extended here to the CP-odd operators & $V=1$ & 84 (10)\\
\bottomrule
\end{tabularx}
\caption{The flavour assumptions for which the electronic files give the map to the Warsaw basis, with their
flavour symmetry, the CKM matrix and the number of real parameters, with the number of CP-odd ones in brackets.}\label{tab:res:assumptions}
\end{table}

For the results described here, the flavour assumptions are imposed on the coefficients at $\mu_{\rm eft}$ in the up basis of
eq.~\eqref{eq:warsawup}. The assumptions that single out the third generation, top, topU3l and \SMEFTNLO{}, are used with $V=1$. In U35 and MFV the flavour
symmetry is broken only by the Yukawa matrices, whose insertions in the coefficients are included. In the up basis of eq.~\eqref{eq:warsawup}, the down-type Yukawa matrix contains
the CKM matrix, $Y_d^\dagger=V\hat Y_d$, and as in ref.~\cite{Brivio:2020onw} we keep $V$ general in these insertions.

In a theory with a flavour symmetry, the symmetry usually holds instead at the matching scale, on the coefficients in the flavour
basis~\cite{Faroughy:2020ina,deBlas:2025xhe}, and the coefficients at $\mu_{\rm eft}$ follow from it by the running and by the rotation to the mass basis. Imposing the assumption at a
$\mu_{\rm eft}$ different from the matching scale is therefore equivalent only if the running preserves it~\cite{Bartocci:2023nvp}. For instance, the QCD running is the same for all flavours and preserves the flavour symmetry of every assumption. With only the top
quark massive, for example, the complete one-loop running preserves top and topU3l, so that for them the scale at which the assumption is
imposed does not matter. In contrast, for U35 the terms of order $y_t^2$ of the running would generate coefficients outside the assumption,
such as Higgs currents of the third generation from closed top-quark loops of the four-quark operators, which are neglected when the
assumption is imposed at $\mu_{\rm eft}$.

Only the colour-octet parameters of top, topU3l and \SMEFTNLO{} need more than a substitution. A parameter such as $c_{QQ}^{(8)}$
multiplies an operator that is not in the Warsaw basis, $\Lag\supset c^{(8)}_{QQ}O^{(8)}_{QQ}/\Lambda^2$ with
\begin{equation}\label{eq:res:OQQ8}
  O^{(8)}_{QQ}=(\bar q_3\gamma_\mu T^Aq_3)(\bar q_3\gamma^\mu T^Aq_3).
\end{equation}
Ref.~\cite{Aguilar-Saavedra:2018ksv}
and \SMEFTNLO{} include a factor $\frac12$ in this term, so that their $c^{(8)}_{QQ}$ is twice ours. In four dimensions a Fierz identity writes
the octet operator in terms of Warsaw operators,
\begin{equation}\label{eq:res:fierzQQ8}
  O^{(8)}_{QQ}=\frac1{12}\,(O^{(1)}_{qq})_{3333}+\frac14\,(O^{(3)}_{qq})_{3333},
\end{equation}
and both \textsc{SMEFTsim} and \SMEFTNLO{} use the right-hand side. The two sides give the same trees and UV poles, but in $D$ dimensions the
identity holds only up to an evanescent operator, so that their rational terms differ by a local term~\cite{Aebischer:2022aze}. We write the rational
terms of each side as the tree-level vertices of a local Lagrangian, and define the difference $\Delta\Lag$ by
\begin{equation}\label{eq:res:octetL}
  \Lag^{\Rtwo}_{\rm octet}=\Lag^{\Rtwo}_{\rm Fierz}+\Delta\Lag ,
\end{equation}
where $\Lag^{\Rtwo}_{\rm Fierz}$ follows from our vertices of the Warsaw operators with the substitution of eq.~\eqref{eq:res:fierzQQ8}. For $c^{(8)}_{QQ}$
\begin{equation}\label{eq:anc:octshift}
  \Delta\Lag=\frac{1}{16\pi^2}\frac{c_{QQ}^{(8)}}{\Lambda^2}\Bigl[-\frac{4g_s}3\,\mathcal G_{q_3}+\frac{g_1}9\,\mathcal B_{q_3}+g_2\,\mathcal W_{q_3}
  +\frac{y_t^2}2\,(O^{(1)}_{Hq})_{33}-\frac{3y_t^2}2\,(O^{(3)}_{Hq})_{33}\Bigr],
\end{equation}
with the penguin operators of the Green's basis
\begin{equation}\label{eq:res:penguinops}
\begin{gathered}
  \mathcal G_\psi=(\bar\psi\gamma^\nu T^A\psi)(D^\mu G_{\mu\nu})^A,\qquad \mathcal B_\psi=(\bar\psi\gamma^\nu\psi)\,\partial^\mu B_{\mu\nu},\\
  \mathcal W_\psi=(\bar\psi\gamma^\nu\tau^I\psi)(D^\mu W_{\mu\nu})^I .
\end{gathered}
\end{equation}

The shift is computed from loops of quarks and gluons only, with $y_t$ as the only non-zero Yukawa coupling and $V=1$. Through the equations of
motion the penguin operators become four-quark operators and Higgs currents, so that $\Delta\Lag$ is a finite shift of these coefficients, a
difference in the definition of the operator basis in $D$ dimensions. It agrees with the evanescent shifts of ref.~\cite{Fuentes-Martin:2022vvu},
which in addition contain four-quark terms proportional to $y_t^2$ from loops with a Higgs or Goldstone boson, outside the loops considered here.

The electronic files give $\Delta\Lag$ for the colour-octet forms of $O^{(1)}_{qq}$, $O^{(3)}_{qq}$, $O_{uu}$ and $O_{dd}$, with the two currents in
the same generation or in two different ones, and so for every octet parameter of the three assumptions. By default \texttt{ApplyFlavour}
gives the rational terms of the Fierz combination, as used by \textsc{SMEFTsim} and \SMEFTNLO{}, and with the option \texttt{"OctetShift"} it
adds the vertices of $\Delta\Lag$. The package functions \texttt{OctetShift} and \texttt{OctetShiftVertex}, described in appendix~\ref{app:anc},
return $\Delta\Lag$ and its Feynman rules for any legs.

\subsection{Example models}\label{sec:ufo}
For the predictions of section~\ref{sec:apps} we turn the vertices into example NLO models in the
UFO format~\cite{Degrande:2011ua,Darme:2023jdn} for \textsc{MadGraph5\_aMC@NLO}~\cite{Alwall:2014hca}. These models contain the
\textsc{SMEFTsim} Lagrangian at tree level, our rational terms as $\Rtwo$ vertices and
the counterterms of eq.~\eqref{eq:uv:rule} as UV vertices. As discussed in sections~\ref{sec:setup:reg} and~\ref{sec:r2:def}, the vertices are computed in the
Feynman gauge of QCD with four-dimensional external states, the setting in which \textsc{MadLoop} uses rational
terms~\cite{Degrande:2014vpa}, while the electroweak gauge can be chosen freely.

The models use the five-flavour scheme of section~\ref{sec:uv:os}, with massless light quarks, the bottom quark included. The external
fields, including their flavour mixing, are renormalised on shell, as \textsc{MadLoop} requires, and so are the top-quark mass and
its Yukawa coupling. The CKM matrix is the $\MSbar$ matrix of section~\ref{sec:uv:ckm} and the
electroweak inputs are $\{M_W,M_Z,G_F\}$.

\subsection{Validation}\label{sec:validation}\label{sec:val:poles}\label{sec:val:r2}
We validated the vertices by comparing them with existing results in the literature, with internal consistency tests, and in complete one-loop amplitudes.%

The poles agree with our independent one-loop computation of the counterterms of the SMEFT in the Green's basis~\cite{Gherardi:2020det,Carmona:2021xtq}, which we performed in the unbroken phase with the background-field method and open generation indices. With
the resulting Lagrangian $\Lag_{\rm CT}=\frac{1}{16\pi^2\epsilon}\sum_c\eta_c\,O_c$ the test is
\begin{equation}\label{eq:val:uvgreens}
  \vf^{\rm UV}=\ii\,\vf^{\rm tree}\bigl[C_c\to\eta_c(C)\bigr],
\end{equation}
so that the UV pole of an effective vertex is cancelled by the tree vertex of the counterterm Lagrangian in the broken phase and the mass basis. The
two sides were %
 compared as full Dirac and colour tensors for all generations, with complex coefficient tensors and a unitary CKM
matrix. The counterterms of the Wilson coefficients obtained from the poles with
eq.~\eqref{eq:uv:rule} also reproduce, for every class of operator, the entries of the one-loop anomalous dimensions~\cite{Jenkins:2013zja,Jenkins:2013wua,Alonso:2013hga} generated by quark, gluon and ghost loops.

The rational terms reproduce all SM coefficients of refs.~\cite{Draggiotis:2009yb,Garzelli:2009is} in the HV regularisation and those of
refs.~\cite{Demartin:2014fia,Degrande:2014tta}, whose poles differ from ours only by the redundant structures that vanish by the equations of
motion. We compared every vertex of the public \SMEFTNLO{} model with ours, as described in appendix~\ref{app:smeftatnlo}. %
We find agreement for all vertices, apart from the Levi-Civita terms of a few Goldstone-boson vertices with a
closed quark loop and a single $\gfive$. However, these vertices do not contribute in the unitary gauge, which is the choice prescribed by \SMEFTNLO{} for the processes in which they appear~\cite{SMEFTatNLO:wiki}. We also reproduce all the results of ref.~\cite{Thomas:2024dwd}. Its vertices of $ggh$ and $gghh$ at first order in ${\rm Im}\,C_{tG}$
and its $\MSbar$ counterterm of $C_{H\widetilde G}$ agree with ours, and so does every loop-induced process studied there, for all its operators.

Most of our vertices are new and have no reference in the literature. We tested their consistency %
 by computing a subset of our results a second
time with auxiliary vectors instead of projectors, or with an independent computation of the loop
integrals%
. For selected representative processes, such as the same-sign top-quark pair production of section~\ref{sec:apps:ss}, the amplitudes obtained with our rational terms agreed %
 with independent computations of the full one-loop amplitude. We also checked %
 the symmetries of
every vertex and the Ward identities of the closed quark loops with a Levi-Civita tensor.

For a representative NLO process of every class of operator, we also checked that the one-loop amplitude computed by \textsc{MadLoop} is invariant under boosts and rotations of the phase-space point and under the permutation of the external legs, satisfies
the gauge test with an external gluon and gives the same result in the Feynman and the unitary gauge of the electroweak bosons.

\label{sec:val:amp}With \textsc{MadGraph5\_aMC@NLO} we tested the example models in complete amplitudes. The first test is the cancellation of the UV
and infrared poles of the virtual corrections, which we verified %
 in processes that cover every class of operator, such as
$gg\to t\bar th$, where it also tests the chiral rotation of section~\ref{sec:uv:cpodd}. The on-shell renormalisation factors with flavour
mixing enter $t\to ug$ and $u\bar d\to t\bar bg$. In $t\to ug$ the amplitude and width agree with refs.~\cite{Zhang:2010bm,Zhang:2014rja}, since the two treatments of the
redundant operator of section~\ref{sec:uv:mix} are the same on external legs, and $u\bar d\to t\bar bg$ gives the same result in
the Feynman and the unitary gauge. Two finite parts were also %
 compared with analytic results. The virtual correction to $gg\to h$ at
$\order{C_{HG}}$ reproduces the one-loop correction of the point-like coupling of the heavy-top effective
theory~\cite{Spira:1995rr}, and the semileptonic vertices reproduce, in $\ell^-\ell'^+\to t\bar u$, the NLO correction to $e^+e^-\to tj$ of
ref.~\cite{Durieux:2014xla}. The dependence of the virtual corrections on
$\mu_{\rm eft}$ follows the one-loop anomalous dimensions.

\section{Benchmark predictions}\label{sec:apps}
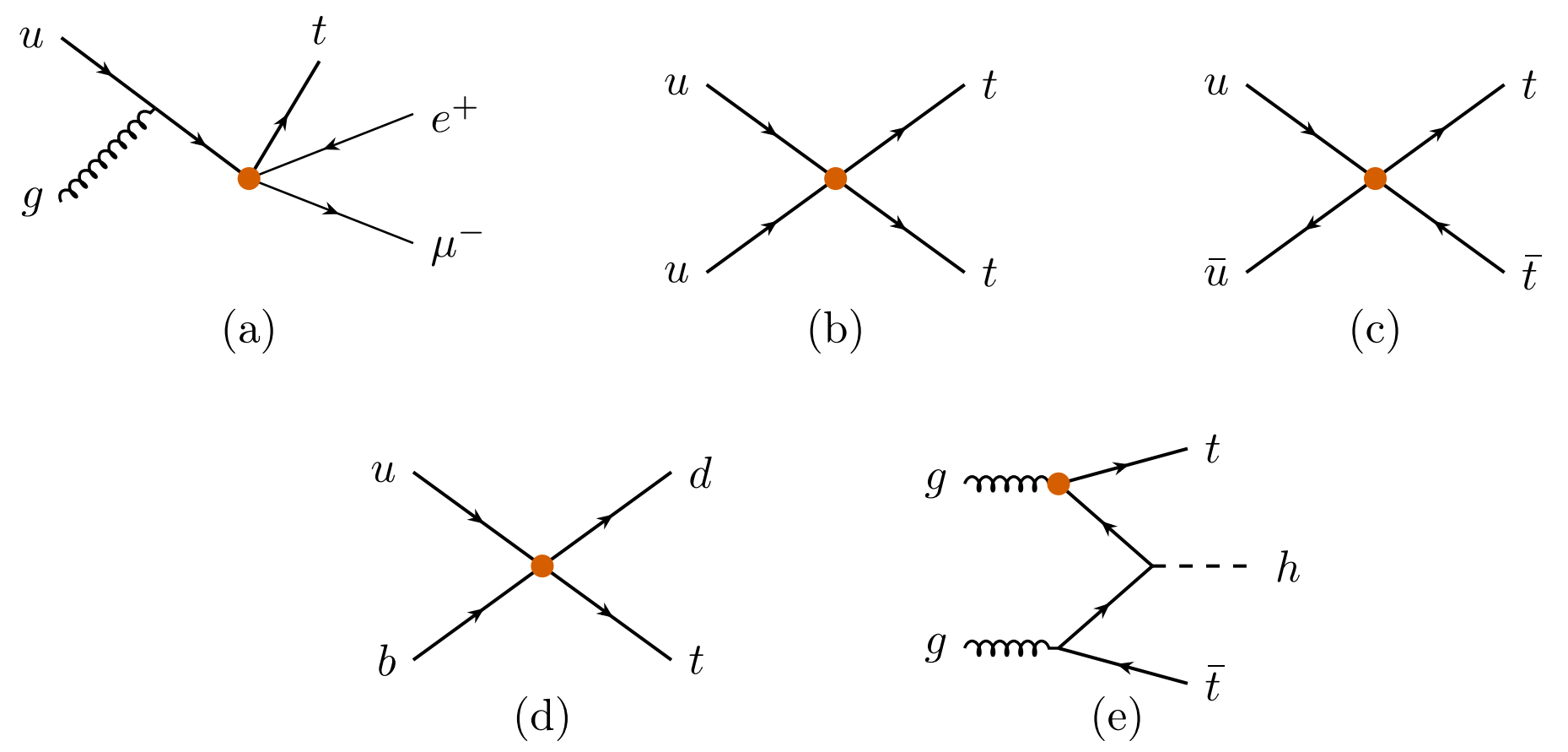
\begin{figure}[!htbp]
\centering
\begin{tikzpicture}[scale=0.92]
\begin{scope}
  \coordinate (V) at (0,0); \coordinate (Q) at (-0.8,0.6);
  \draw[sfer] (-1.6,1.2) -- (Q); \draw[sfer] (Q) -- (V); \draw[sglu] (-1.6,-0.2) -- (Q);
  \coordinate (A) at (0,0); \coordinate (B) at (0,0);
  \draw[sfer] (A) -- (0.6,1.0); \draw[slep] (1.4,0.55) -- (B); \draw[slep] (B) -- (1.4,-0.55);
  \node[sopv] at (A) {};
  \node[anchor=east] at (-1.6,1.2) {$u$}; \node[anchor=east] at (-1.6,-0.2) {$g$};
  \node[anchor=south] at (0.6,1.0) {$t$}; \node[anchor=west] at (1.4,0.55) {$e^+$}; \node[anchor=west] at (1.4,-0.55) {$\mu^-$};
  \node at (0,-1.3) {(a)};
\end{scope}
\begin{scope}[xshift=5.0cm]
  \coordinate (A) at (0,0); \coordinate (B) at (0,0);
  \draw[sfer] (-1.1,0.8) -- (A); \draw[sfer] (A) -- (1.1,0.8); \draw[sfer] (-1.1,-0.8) -- (B); \draw[sfer] (B) -- (1.1,-0.8);
  \node[sopv] at (A) {};
  \node[anchor=east] at (-1.1,0.8) {$u$}; \node[anchor=east] at (-1.1,-0.8) {$u$};
  \node[anchor=west] at (1.1,0.8) {$t$}; \node[anchor=west] at (1.1,-0.8) {$t$};
  \node at (0,-1.3) {(b)};
\end{scope}
\begin{scope}[xshift=9.6cm]
  \coordinate (A) at (0,0); \coordinate (B) at (0,0);
  \draw[sfer] (-1.1,0.8) -- (A); \draw[sfer] (A) -- (-1.1,-0.8); \draw[sfer] (1.1,-0.8) -- (B); \draw[sfer] (B) -- (1.1,0.8);
  \node[sopv] at (A) {};
  \node[anchor=east] at (-1.1,0.8) {$u$}; \node[anchor=east] at (-1.1,-0.8) {$\bar u$};
  \node[anchor=west] at (1.1,0.8) {$t$}; \node[anchor=west] at (1.1,-0.8) {$\bar t$};
  \node at (0,-1.3) {(c)};
\end{scope}
\begin{scope}[xshift=2.5cm, yshift=-3.3cm]
  \coordinate (A) at (0,0); \coordinate (B) at (0,0);
  \draw[sfer] (-1.1,0.8) -- (A); \draw[sfer] (A) -- (1.1,0.8); \draw[sfer] (-1.1,-0.8) -- (B); \draw[sfer] (B) -- (1.1,-0.8);
  \node[sopv] at (A) {};
  \node[anchor=east] at (-1.1,0.8) {$u$}; \node[anchor=east] at (-1.1,-0.8) {$b$};
  \node[anchor=west] at (1.1,0.8) {$d$}; \node[anchor=west] at (1.1,-0.8) {$t$};
  \node at (0,-1.3) {(d)};
\end{scope}
\begin{scope}[xshift=7.4cm, yshift=-3.3cm]
  \coordinate (G1) at (-0.5,0.7); \coordinate (G2) at (-0.5,-0.7); \coordinate (H) at (0.3,0);
  \draw[sglu] (-1.3,0.7) -- (G1); \draw[sglu] (-1.3,-0.7) -- (G2);
  \draw[sfer] (G2) -- (H); \draw[sfer] (H) -- (G1); \draw[sfer] (G1) -- (0.6,1.0); \draw[sfer] (0.6,-1.0) -- (G2);
  \draw[shig] (H) -- (1.2,0);
  \node[sopv] at (G1) {};
  \node[anchor=east] at (-1.3,0.7) {$g$}; \node[anchor=east] at (-1.3,-0.7) {$g$};
  \node[anchor=west] at (0.6,1.0) {$t$}; \node[anchor=west] at (0.6,-1.0) {$\bar t$}; \node[anchor=west] at (1.2,0) {$h$};
  \node at (0,-1.3) {(e)};
\end{scope}
\end{tikzpicture}
\caption{Representative leading-order diagrams of the processes of section~\ref{sec:apps}, with the operator vertex drawn as a dot. (a) $ug\to te^+\mu^-$ with a semileptonic operator. (b) Same-sign top
quarks, $uu\to tt$. (c) Top-quark pairs, $u\bar u\to t\bar t$. (d) $t$-channel single top, $ub\to dt$. (e) $gg\to t\bar th$ with the top-quark
dipole.}
\label{fig:apps:born}
\end{figure}

With the general-flavour example models of section~\ref{sec:ufo} we give NLO QCD predictions for representative processes. We show single-top production with two charged leptons of different generations, same-sign top-quark pairs,
top-quark pair and single-top production with flavour-specific four-quark operators, and a CP-odd asymmetry in $t\bar th$. Figure~\ref{fig:apps:born} shows representative
leading-order (LO) diagrams. These results should be understood only as benchmarks, and a detailed phenomenological study of these processes is outside the scope of this
work. Several of the coefficients considered below are also strongly constrained by flavour observables~\cite{UTfit:2007eik,Silvestrini:2018dos,LHCb:2017hag}.

Unless stated otherwise, we consider %
 one coefficient at a time, $C/\Lambda^2=1$~TeV$^{-2}$, defined in $\MSbar$ at $\mu_{\rm eft}=m_t$. The top quark is stable, with mass
$m_t=173$~GeV. We use massless light quarks and leptons, $M_W=80.387$~GeV, $M_Z=91.1876$~GeV, $G_F=1.166379\times10^{-5}$~GeV$^{-2}$, $m_h=125.0$~GeV, which enters only in $t\bar th$,
and the unitary CKM matrix in the standard parametrisation, with the Wolfenstein parameters $\lambda=0.2265$, $A=0.79$, $\bar\rho=0.141$ and \mbox{$\bar\eta=0.357$}
of ref.~\cite{ParticleDataGroup:2020ssz}, which we use for the matrix $V(\mu_{\rm eft})$ of section~\ref{sec:setup:ewsb}.

These values are extracted from data assuming the SM. In the SMEFT the measurements that fix them receive contributions from the Wilson
coefficients, which can be large for coefficients without a flavour symmetry~\cite{Descotes-Genon:2018foz}. A consistent treatment fits
the CKM parameters together with the Wilson coefficients~\cite{deBlas:2025xhe}, while an event generator needs numerical values of the CKM parameters as input, so that such a fit is impractical in this context. An input
scheme for the CKM matrix, analogous to that of the electroweak parameters, captures a part of these effects~\cite{Descotes-Genon:2018foz}.
Appendix~\ref{app:ckminput} gives our vertices in that scheme and its limitations, and we do not use it in this section.

We use $\sqrt s=13$~TeV and NNPDF4.0 at NLO~\cite{NNPDF:2021njg} with $\alpha_s(M_Z)=0.118$ at both
orders, and $\mu_R=\mu_F=H_T/2$, with $H_T$ the scalar sum of the transverse masses of the final-state particles. The scale uncertainty
is obtained from variations of $\mu_R$ and $\mu_F$ by factors of two, with $\mu_{\rm eft}$ fixed, so that it does not include the dependence
on $\mu_{\rm eft}$. The models are linear in the Wilson coefficients, so that their amplitudes contain at most one operator insertion. The term of order $\Lambda^{-4}$ of a cross section also receives contributions from the interference of the SM amplitude with amplitudes with two insertions and with amplitudes of dimension-eight operators, which the models do not contain. When the SM amplitude vanishes, as for the lepton-flavour-violating processes of table~\ref{tab:obs:clfv}, these interferences vanish as well, and the term of order $\Lambda^{-4}$ is the square of the amplitude with one dimension-six insertion.
The cross sections are
computed at fixed order with \textsc{MadGraph5\_aMC@NLO}.

\subsection{Top-quark flavour violation with charged leptons}\label{sec:apps:clfv}
Top-quark flavour-changing interactions with two charged leptons were described with SMEFT operators in
refs.~\cite{Durieux:2014xla,Aguilar-Saavedra:2018ksv} and studied at leading order in
refs.~\cite{Afik:2021jjh,Altmannshofer:2025lun}. At NLO in QCD the decays were computed in refs.~\cite{Zhang:2014rja,Durieux:2014xla}, and
ref.~\cite{Durieux:2014xla} obtained $pp\to t\ell^+\ell^-$ for one vector operator and one lepton generation.

Table~\ref{tab:obs:clfv} gives the cross sections of $pp\to te^+\mu^-$ from the squared dimension-six amplitude, at $\mathcal{O}(\Lambda^{-4})$. The five vector coefficients, $C^{(1)}_{lq}$, $C^{(3)}_{lq}$, $C_{lu}$, $C_{eu}$ and $C_{qe}$,
have the same cross-section contribution, as well as the same $K$-factor, $K=1.30$, and the table shows $C^{(1)}_{lq}$ for all of them, with the indices $1312$ for $C_{qe}$. The scalar coefficient $C^{(1)}_{lequ}$ and the tensor
coefficient $C^{(3)}_{lequ}$ have $K=1.25$ and $1.33$, respectively. The scale uncertainty of the NLO rates is about 6\%. No cuts are applied to the leptons, and the charge-conjugate process
$pp\to\bar te^-\mu^+$ is not included. The rates for $\mu^+\tau^-$ are the same as for $e^+\mu^-$ for massless leptons.
At NLO the real emission, for instance $gg\to t\ell^+\ell'^-\bar u$, contains diagrams in which a top-antitop pair is produced and the antitop quark decays
into $\bar u\ell^+\ell'^-$ through the operator. These diagrams describe $t\bar t$ production followed by a rare decay, which is a different process, and including them would count
$t\bar t$ production as part of $t\ell^+\ell'^-$ production.
We remove these diagrams, as done in ref.~\cite{Frixione:2008yi} for $tW^-$ production.
\begin{table}[!htbp]
\centering\small\setlength{\tabcolsep}{4pt}\renewcommand{\arraystretch}{1.25}
\begin{tabular}{@{}llcccc@{}}
\toprule
process & coefficient & $\sigma_{\rm LO}$ & $\sigma_{\rm NLO}$ & $K$ & scale [\%]\\
\midrule
$t\,e^+\mu^-$ [pb] & $C^{(1)}_{lq}$, $1213$ & $0.05915$ & $0.07687(13)$ & $1.30$ & $^{+5.0}_{-6.4}$\\
 & $C^{(1)}_{lequ}$, $1213$ & $0.02555$ & $0.03195(7)$ & $1.25$ & $^{+4.0}_{-6.0}$\\
 & $C^{(3)}_{lequ}$, $1213$ & $0.5393$ & $0.7195(15)$ & $1.33$ & $^{+5.7}_{-6.7}$\\
\midrule
$tt$ [pb] & $C^{(1,3)}_{qq}$, $C_{uu}$, $1313$ & $213.7$ & $197.1(6)$ & $0.92$ & $^{+1.2}_{-1.6}$\\
 & $C^{(1,3)}_{qq}$, $C_{uu}$, $1323$ & $30.16$ & $27.78(10)$ & $0.92$ & $^{+1.4}_{-0.9}$\\
 & $C^{(1,3)}_{qq}$, $C_{uu}$, $2323$ & $0.7735$ & $0.6969(33)$ & $0.90$ & $^{+1.6}_{-1.4}$\\
 & $C^{(1)}_{qu}$, $1313$ & $26.88$ & $28.54(11)$ & $1.06$ & $^{+1.0}_{-1.5}$\\
 & $C^{(1)}_{qu}$, $1323$ and $2313$ & $3.790$ & $4.846(21)$ & $1.28$ & $^{+4.1}_{-3.1}$\\
 & $C^{(1)}_{qu}$, $2323$ & $0.09683$ & $0.1326(6)$ & $1.37$ & $^{+4.2}_{-2.8}$\\
 & $C^{(8)}_{qu}$, $1313$ & $6.260$ & $4.613(18)$ & $0.74$ & $^{+4.4}_{-7.0}$\\
 & $C^{(8)}_{qu}$, $1323$ and $2313$ & $0.9267$ & $0.9850(43)$ & $1.06$ & $^{+2.0}_{-1.2}$\\
 & $C^{(8)}_{qu}$, $2323$ & $0.02442$ & $0.02903(13)$ & $1.19$ & $^{+3.3}_{-1.9}$\\
\bottomrule
\end{tabular}

\caption{NLO QCD cross sections at 13~TeV, in pb, for $pp\to te^+\mu^-$ and $pp\to tt$, with one real coefficient
$C/\Lambda^2=1$~TeV$^{-2}$ and the generation indices given, or two coefficients in the rows with two sets of indices. The Monte Carlo uncertainty of the NLO rates is in brackets, and the scale uncertainty of
$\sigma_{\rm NLO}$ is in the last column, in \%. The rates are of order $\Lambda^{-4}$, since these processes have no SM amplitude at this order.
$C^{(1)}_{lq}$ represents the five vector coefficients.}\label{tab:obs:clfv}\label{tab:obs:ss}
\end{table}

\subsection{Same-sign top-quark pairs}\label{sec:apps:ss}
Two top quarks of the same charge are produced by the $\Delta F=2$ four-quark operators with two top quarks~\cite{Aguilar-Saavedra:2010uur,Degrande:2011rt}. The NLO QCD corrections have been computed for the exchange
of a $Z'$ boson~\cite{Li:2012dtz}. Table~\ref{tab:obs:ss} gives the cross sections at order $\Lambda^{-4}$ for the three initial states $uu$, $uc$ and $cc$, selected by the
indices $1313$, $1323$ and $2323$. For $C^{(1)}_{qq}$, $C^{(3)}_{qq}$ and $C_{uu}$ the entries $1323$ and $2313$ are the same coefficient, by
the symmetry of the two identical currents. For $C^{(1,8)}_{qu}$ they are two different coefficients, both set to one, and the $uc$ rate is the sum of
their two contributions, which do not interfere and are equal. The coefficients $C^{(1)}_{qq}$, $C^{(3)}_{qq}$ and $C_{uu}$ give the same
results, since for four up-type quarks they reduce to the same product of currents, left-handed for $C^{(1,3)}_{qq}$ and right-handed for
$C_{uu}$. The $K$-factors
range from 0.74 to 1.37. For $C^{(1,8)}_{qu}$ the $K$-factor grows from the $uu$ to the $cc$ initial state.%

We also checked this process at the level of the amplitude. With an insertion of $O_{uu}$ we computed the one-loop amplitude of $uu\to tt$ independently in $D$ dimensions with \mbox{\textsc{FeynCalc}} and
\textsc{LoopTools}~\cite{Mertig:1990an,Shtabovenko:2023idz,Hahn:1998yk}, finding agreement with the automated computation to machine precision.

\subsection{Flavour-specific four-quark operators}\label{sec:apps:fourq}
Table~\ref{tab:ufo:rates} gives NLO interference cross sections at $\mathcal{O}(\Lambda^{-2})$ for top-quark pair and $t$-channel single-top production
with four-quark coefficients, for which we use the fixed scales $\mu_R=\mu_F=m_t$. The coefficient $(C^{(1)}_{qu})_{1133}$ has no tree-level interference with the QCD amplitude of
$q\bar q\to t\bar t$, and contributes only at NLO through the one-loop corrections and the real emission~\cite{Degrande:2020evl,Degrande:2024mbg}. Its tree-level interference with
the electroweak SM amplitude is not included.
\begin{table}[!htbp]
\centering\small\setlength{\tabcolsep}{4pt}
\begin{tabular}{@{}llccc@{}}
\toprule
process & coefficient & $\sigma_{\rm LO}$ & $\sigma_{\rm NLO}$ & $K$\\
\midrule
$pp\to t\bar t$ & $C^{(1)}_{qq}$, $1331$ & $17.33(3)$ & $17.03(3)^{+0.12}_{-0.33}$ & $0.98$\\
 & $C^{(1)}_{qu}$, $1133$ & $0$ & $-0.806(9)^{+0.15}_{-0.21}$ & --\\
 & $C^{(8)}_{qu}$, $1133$ & $6.956(9)$ & $6.29(2)^{+0.07}_{-0.28}$ & $0.90$\\
$pp\to tj$ & $C^{(3)}_{qq}$, $1133$ & $-84.65(6)$ & $-83.67(3)^{+0.7}_{-1.5}$ & $0.99$\\
\bottomrule
\end{tabular}
\caption{NLO QCD rates of top-quark pair and $t$-channel single-top production at first order in one Wilson coefficient, interference with the SM only, $\sqrt s=13$~TeV,
NNPDF4.0, in pb for $C/\Lambda^2=1$~TeV$^{-2}$, with the Monte Carlo uncertainty in brackets and the scale uncertainty in pb as superscript and subscript. The scales
are $\mu_R=\mu_F=\mu_{\rm eft}=173$~GeV. The $tj$ row has $p_T^j>20$~GeV. The SM rates at NLO at these settings are
$705.1(1.1)$~pb ($t\bar t$) and $120.4(4)$~pb ($tj$).}\label{tab:ufo:rates}
\end{table}

\subsection{CP-odd asymmetry in \texorpdfstring{$t\bar th$}{ttH}}\label{sec:apps:higgs}
CP-sensitive observables in $t\bar th$ were studied in refs.~\cite{Gunion:1996xu,Boudjema:2015nda,Miralles:2024huv}. At NLO in QCD the CP-odd
Yukawa coupling in $t\bar th$ was computed in refs.~\cite{Demartin:2014fia,Hermann:2022vit}, the CP-even operators in
refs.~\cite{Maltoni:2016yxb,Bevilacqua:2026udk}, and the CP-odd operators were added to \SMEFTNLO{} for other processes in
refs.~\cite{Thomas:2024dwd,Rossia:2024rfo}. The linear interference of a CP-odd coefficient vanishes in the total rate, so that we use the
CP-odd observable
\begin{equation}\label{eq:apps:Oodd}
  \mathcal O_{\rm odd}=\sin(\phi_t-\phi_{\bar t})\,{\rm sign}(\mathrm y_t-\mathrm y_{\bar t})
\end{equation}
in the laboratory frame, with $\phi$ the azimuthal angles and $\mathrm y$ the rapidities of the top quark and antiquark. It is the sine of the
rapidity-ordered azimuthal separation of the two top quarks, a version of the observable of ref.~\cite{Miralles:2024huv}. We quote the cross-section difference $A_{\rm odd}=\sigma(\mathcal O_{\rm odd}>0)-\sigma(\mathcal O_{\rm odd}<0)$ of the
interference. For stable top quarks it is sensitive to ${\rm Im}\,C_{tG}$ and to $C_{H\widetilde G}$, but not to the CP-odd top-quark
Yukawa coupling, whose linear interference with the SM amplitude vanishes point by point when the top-quark spins are summed. Table~\ref{tab:apps:asym} gives the results. The NLO corrections increase the asymmetry by about 30\% and 60\%
for the two coefficients. A reading of the closed top-quark loops at an SM vertex would shift $C_{H\widetilde G}$, as shown in
appendix~\ref{app:g5su3}, and change the asymmetry of ${\rm Im}\,C_{tG}$ by at most 1\%.
In the SM the asymmetry from the same runs is compatible with zero.
\begin{table}[!htbp]
\centering\small\setlength{\tabcolsep}{4pt}
\begin{tabular}{@{}llccc@{}}
\toprule
process & coefficient & $A^{\rm LO}_{\rm odd}$ & $A^{\rm NLO}_{\rm odd}$ & $K$\\
\midrule
$pp\to t\bar th$ & ${\rm Im}\,C_{tG}$ & $-0.0266(1)$ & $-0.0350(6)^{+0.0029}_{-0.0018}$ & $1.31$\\
 & $C_{H\widetilde G}$ & $0.01234(4)$ & $0.0195(3)^{+0.0023}_{-0.0025}$ & $1.58$\\
\bottomrule
\end{tabular}
\caption{The asymmetry $A_{\rm odd}$ in $pp\to t\bar th$ at NLO in QCD, at first order in one Wilson coefficient, interference with the SM only, $\sqrt s=13$~TeV,
NNPDF4.0, in pb for $C/\Lambda^2=1$~TeV$^{-2}$, with the Monte Carlo uncertainty in brackets and the scale uncertainty in pb as superscript and subscript. The scales
are $\mu_R=\mu_F=H_T/2$ with $\mu_{\rm eft}=173$~GeV. The SM rate at NLO at these settings is $0.4391(2)$~pb.}\label{tab:apps:asym}
\end{table}

\section{Conclusions}\label{sec:conclusions}
We have computed the one-loop rational terms $\Rtwo$ and the UV poles of the dimension-six SMEFT at NLO in QCD, in the broken phase and at
first order in the Wilson coefficients, for the full Warsaw basis including the CP-odd operators and without any assumption on the flavour
structure. The set of effective vertices is fixed by two diagrammatic conditions and bounded by power counting. Every vertex is computed once with
open generation indices, and its flavour dependence is given in closed form by matrix chains and traces of the coefficient tensors, the CKM matrix
and the quark masses. Together with the counterterm vertices, the on-shell renormalisation factors with flavour mixing and the
renormalisation of the CKM matrix, these vertices provide what an event generator needs to compute NLO QCD corrections with any Wilson coefficient of
the Warsaw basis. The vertices in the common flavour assumptions of \textsc{SMEFTsim} and \SMEFTNLO{} follow by substitution, except for the
colour-octet parameters, whose rational terms differ from those of their Fierz combinations by a local shift that we give explicitly.

For the closed traces with a single $\gfive$ we have adopted the reading at the operator insertion. At this order it preserves the
$SU(3)$ and electroweak Ward identities and Bose symmetry, so that no symmetry-restoring counterterm is needed. Example NLO models built from the vertices reproduce known results, and give NLO predictions for top-quark flavour violation with charged leptons, same-sign top-quark pairs, top-quark pair and single-top production with flavour-specific four-quark operators, and a CP-odd asymmetry in $t\bar th$. The models will be released in a subsequent
publication.

All results are provided in electronic form, together with a Mathematica package that returns every vertex
and its complete counterterm for any legs and flavours, also in flavour assumptions defined by the user.

The methods described in this work also apply to the electroweak corrections, which need a much larger set of vertices and renormalisation factors. We leave this and the
translation to the BMHV scheme, with its symmetry-restoring counterterms, for future work.

\section*{AI-assistance statement}
The AI models Claude Opus 5.5 and Claude Fable 5.2 by Anthropic assisted with writing the computer codes, running the numerical computations, and drafting and revising the text.

\acknowledgments
I am deeply grateful to Hesham El Faham, Victor Miralles and Marion Thomas for the long discussions that led to this work. I thank them, Fabio Maltoni and Eleni Vryonidou for reading the manuscript and for their feedback. I also thank Panagiotis Stylianou and Marion Thomas for testing some of the results.

\appendix
\section{Details about the SMEFT in the broken phase}\label{app:conventions}
Section~\ref{sec:setup} introduces the SMEFT Lagrangian, its Feynman rules after electroweak symmetry breaking and the choice of
the electroweak inputs. This appendix gives more details about the conventions, which fix the signs and the
normalisation of every vertex of the paper. Appendix~\ref{app:ops} writes the SM Lagrangian and the operators whose sign conventions
matter, appendix~\ref{app:inputs} relates the parameters of the Lagrangian to the electroweak inputs, appendix~\ref{app:opsmass}
writes the quark sector in the mass basis, where the CKM matrix enters the Yukawa and Goldstone couplings, and
appendix~\ref{app:ckminput} gives the vertices in an input scheme for the CKM matrix.

\subsection{The SM Lagrangian and the operators}\label{app:conv:summary}\label{app:ops}
The SM part of eq.~\eqref{eq:lag}, the field strengths and the QCD gauge fixing are
\begin{align}
  \Lag_{\rm SM}={}&-\frac14G^A_{\mu\nu}G^{A\mu\nu}-\frac14W^I_{\mu\nu}W^{I\mu\nu}-\frac14B_{\mu\nu}B^{\mu\nu}
  +\sum_{\psi}\bar\psi\,\ii\slashed{D}\,\psi\nonumber\\
  &+(D_\mu\varphi)^\dagger(D^\mu\varphi)+m^2\varphi^\dagger\varphi-\lambda(\varphi^\dagger\varphi)^2\nonumber\\
  &-\Big[\bar d\,Y_d\,\varphi^\dagger q+\bar u\,Y_u\,\tilde\varphi^\dagger q+\bar e\,Y_l\,\varphi^\dagger l+\text{h.c.}\Big]
  +\Lag_{\rm gf}+\Lag_{\rm gh},\label{eq:lagSM}\\
  G^A_{\mu\nu}={}&\partial_\mu G^A_\nu-\partial_\nu G^A_\mu-g_sf^{ABC}G^B_\mu G^C_\nu ,\qquad
  W^I_{\mu\nu}=\partial_\mu W^I_\nu-\partial_\nu W^I_\mu-g_2\epsilon^{IJK}W^J_\mu W^K_\nu ,\label{eq:fieldstrength}\\
  \Lag_{\rm gf}\supset{}&-\frac1{2\xi}\big(\partial^\mu G^A_\mu\big)^2 ,\qquad
  \Lag_{\rm gh}\supset-\bar c^A\,\partial^\mu\big(\partial_\mu\delta^{AC}-g_sf^{ABC}G^B_\mu\big)c^C ,\label{eq:gf}
\end{align}
with $\xi=1$, $\Tr[T^AT^B]=\frac12\delta^{AB}$, $[T^A,T^B]=\ii f^{ABC}T^C$, $\tilde\varphi_j=\epsilon_{jk}\varphi^*_k$ with $\epsilon_{12}=+1$, and
the hypercharges $\mathsf Y_q=\frac16$, $\mathsf Y_u=\frac23$, $\mathsf Y_d=-\frac13$, $\mathsf Y_l=-\frac12$, $\mathsf Y_e=-1$,
$\mathsf Y_\varphi=\frac12$. The Yukawa matrices of ref.~\cite{Grzadkowski:2010es} are $\Gamma_{u,d}=Y_{u,d}^\dagger$. No electroweak field
propagates in our loops, so that the electroweak gauge enters only through the external Goldstone bosons, for which we use the Feynman
gauge.

We use the operators of the Warsaw basis as defined in ref.~\cite{Grzadkowski:2010es}. In the names of the coefficients we label
the Higgs doublet by $H$, as in \textsc{SMEFTsim}, so that for example we write $C_{HG}$, $C_{uH}$ and $C^{(1,3)}_{Hq}$, while
ref.~\cite{Grzadkowski:2010es} writes $C_{\varphi G}$, $C_{u\varphi}$ and $C^{(1,3)}_{\varphi q}$. The signs of the vertices
follow from the covariant derivative of eq.~\eqref{eq:covd} and from the definitions of the chromomagnetic dipoles and of the purely gluonic
operators,
\begin{gather}
  \Lag_6\supset\frac{(C_{uG})_{pr}}{\Lam^2}\,(\bar q_p\sigma^{\mu\nu}T^Au_r)\,\tilde\varphi\,G^A_{\mu\nu}
  +\frac{(C_{dG})_{pr}}{\Lam^2}\,(\bar q_p\sigma^{\mu\nu}T^Ad_r)\,\varphi\,G^A_{\mu\nu}+\text{h.c.},\label{eq:dipoles}\\
  O_G=f^{ABC}G^{A\nu}_\mu G^{B\rho}_\nu G^{C\mu}_\rho ,\quad
  O_{\widetilde G}=f^{ABC}\widetilde G^{A\nu}_\mu G^{B\rho}_\nu G^{C\mu}_\rho ,\nonumber\\
  O_{HG}=\varphi^\dagger\varphi\,G^A_{\mu\nu}G^{A\mu\nu},\quad
  O_{H\widetilde G}=\varphi^\dagger\varphi\,\widetilde G^A_{\mu\nu}G^{A\mu\nu},\label{eq:gluonops}
\end{gather}
with $\sigma^{\mu\nu}=\frac\ii2[\gamma^\mu,\gamma^\nu]$ and the dual field strength of eq.~\eqref{eq:dual_fs}. The index order of $O_G$ and
$O_{\widetilde G}$ is that of \textsc{SMEFTsim}, $G_{\mu\nu}G_{\nu\rho}G_{\rho\mu}$ with $\widetilde G$ in the first slot.

\subsection{Input scheme}\label{app:inputs}
In the $\{M_W,M_Z,G_F\}$ scheme the tree-level SM values of the vacuum expectation value, the weak mixing angle and the gauge couplings are
\begin{equation}\label{eq:inputs}
  \vh=\big(\sqrt2G_F\big)^{-1/2},\quad s^2_{\hat\theta}=1-\frac{M_W^2}{M_Z^2},\quad
  \hat e=\frac{2M_Ws_{\hat\theta}}{\vh},\quad \hat g_2=\frac{\hat e}{s_{\hat\theta}},\quad
  \hat g_1=\frac{\hat e}{c_{\hat\theta}}.
\end{equation}
At first order in $1/\Lam^2$ the parameters of the Lagrangian are related to them by~\cite{Berthier:2015oma,Brivio:2017vri}
\begin{equation}\label{eq:shifts}
\begin{gathered}
  \vT=\vh\Big(1+\frac{\delta G_F}2\Big),\qquad g_2=\hat g_2(1+\delta g_2),\qquad g_1=\hat g_1(1+\delta g_1),\\[4pt]
  \delta g_2=-\frac{\delta G_F}2,\qquad
  \delta g_1=-\frac12\Big(\delta G_F+\frac{\delta M_Z^2}{s^2_{\hat\theta}}\Big),\\[4pt]
  \delta M_Z^2=\frac{\vh^2}{\Lam^2}\Big(\frac{C_{HD}}2+2c_{\hat\theta}s_{\hat\theta}C_{HWB}\Big),
\end{gathered}
\end{equation}
with $\delta G_F$ of eq.~\eqref{eq:dGFdkH}. The coefficient $C_{HWB}$ also enters through the rotation of $W^3$ and $B$ into the
photon and the $Z$ boson.

\subsection{The quark sector in the mass basis}\label{app:opsmass}
Following ref.~\cite{Brivio:2020onw}, we write the down-type Yukawa term with the doublet of eq.~\eqref{eq:warsawup},
\begin{equation}\label{eq:downyuk}
  \Lag\supset-\bar q_p\,\varphi\,(Y_d^\dagger)_{pr}\,d_{R,r}+\text{h.c.},
\end{equation}
and the bare Yukawa matrices absorb the contributions of $O_{uH}$ and $O_{dH}$ to the quark masses and the difference between $\vT$ and $\vh$,
\begin{equation}\label{eq:yukbare}
  Y_u=\hat Y_u\Big(1-\frac{\delta G_F}2\Big)+\frac{\vh^2}{2\Lam^2}C_{uH}^\dagger ,\qquad
  Y_d=\Big[\hat Y_d\Big(1-\frac{\delta G_F}2\Big)+\frac{\vh^2}{2\Lam^2}C_{dH}^\dagger V\Big]V^\dagger ,
\end{equation}
with $\hat Y_{u,d}=\sqrt2M_{u,d}/\vh$, so that the mass terms contain the physical masses. With $\varphi^-=\ii G^-$ the charged couplings are
\begin{equation}\label{eq:Gcc}
\begin{split}
  \Lag\supset{}&-\ii\,G^+\,\bar u\Big(1-\frac{\delta G_F}2\Big)\Big[\hat Y_u\,V\,P_L-V\,\hat Y_d\,P_R\Big]d\\
  &-\frac{g_2}{\sqrt2}\Big[W^+_\mu\,\bar u_L\gamma^\mu V\,d_L+W^+_\mu\,\bar\nu_L\gamma^\mu e_L\Big]+\text{h.c.}
\end{split}
\end{equation}
The Goldstone couplings carry the CKM matrix in the same way as the $W$ couplings. The
\textsc{FeynRules} implementation of ref.~\cite{Brivio:2020onw} uses the unrotated doublet in eq.~\eqref{eq:downyuk}. This changes only the
couplings of the charged Goldstone bosons to quarks and therefore matters only in a gauge with Goldstone bosons. Our vertices and the tree-level Lagrangian of the example models of section~\ref{sec:ufo} use the doublet of eq.~\eqref{eq:warsawup}, so that these couplings are the ones of eq.~\eqref{eq:Gcc}.

\subsection{\texorpdfstring{The CKM matrix in an input scheme}{The CKM matrix in an input scheme}}\label{app:ckminput}
Ref.~\cite{Descotes-Genon:2018foz} defines an input scheme for the CKM matrix $V$ of section~\ref{sec:setup:ewsb}, analogous to the
$\{M_W,M_Z,G_F\}$ scheme of appendix~\ref{app:inputs}, in which four measured observables fix the Wolfenstein parameters.
The parameters $W$ of the Lagrangian then differ from those fixed by the data, $\widetilde W$, by shifts linear in the Wilson coefficients,
\begin{equation}\label{eq:ckm:input}
  W=\widetilde W-\delta W,\qquad V(W)=V(\widetilde W)-\delta V,\qquad
  \delta V=\sum_j\frac{\partial V}{\partial W_j}\,\delta W_j ,
\end{equation}
with $\delta W=M\Delta$, where $\Delta$ contains the relative shifts of the four observables due to the Wilson coefficients and $M$ is a
matrix that depends only on the parameters $\widetilde W$, both given in ref.~\cite{Descotes-Genon:2018foz}. Since $\delta V$ is of first order in the Wilson coefficients, only the SM part of a vertex is shifted at $\mathcal{O}(\Lambda^{-2})$, and since this part is linear in $V$, an effective vertex in this input scheme is the vertex with $V(\widetilde W)$ plus its SM part with $V$ replaced by $-\delta V$, as for the electroweak input shifts. The package gives $\Delta$, $\delta W$,
$\delta V$ and the shifted vertices at tree level, as described in appendix~\ref{app:anc:ct}. The shifts are computed at first order in
$\Delta$, so that the scheme holds only when $\Delta$ is small. The Wilson coefficients enter $\Delta$ at the electroweak scale, and at NLO in QCD
the running to the hadronic scales, the one-loop matching and the hadronic inputs would need corrections of order $\alpha_s$, which we do not include.

\section{\texorpdfstring{Computation of the rational terms and UV counterterms%
}{Computation of the rational terms and UV counterterms}}\label{app:ct}\label{app:tb}
Section~\ref{sec:r2method} schematically describes the computation of the rational term $\Rtwo$ and the UV pole of the one-loop
Green's functions, while this appendix gives the details. As in section~\ref{sec:scope}, an effective vertex is the rational term and the UV pole of a one-loop Green's function
with given external fields. Appendix~\ref{app:pc} gives the power counting, appendix~\ref{app:ct:fp} the pole and $\Rtwo$ of a
diagram, appendix~\ref{app:tb:solve} the decomposition into tensor structures, and appendix~\ref{sec:method:example} the example of the
dipole quark--gluon vertex.

\subsection{Effective vertices allowed by power counting%
}\label{app:pc}
As shown in section~\ref{sec:r2:pc}, a 1PI Green's function with $f$ external fermions and $b$ external bosons can have a local
part at first order in a Wilson coefficient only if $\omega\ge0$, with $\omega$ of eq.~\eqref{eq:pc:omega}, and this local part is a polynomial of degree $\omega$ in the momenta, the masses and $v$. The allowed effective vertices are those
with $\omega\ge0$. However, when the vertex of the operator carries powers of $v$, the degree in the momenta and the masses is lower than $\omega$,
since each power of $v$ takes the place of a momentum or a mass. %
Let $d_{\min}$ be the smallest mass dimension of the coupling of
the operator vertex among the diagrams of the vertex. The degree in the momenta and the masses %
is then
\begin{equation}\label{eq:pc:Neff}
  N_{\rm eff}=4-b-\tfrac32f-d_{\min} .
\end{equation}
The chromomagnetic dipole enters the quark--gluon vertex with one power of $v$. Its coupling $Cv/\Lam^2$ has $d_{\min}=-1$, so that with $b=1$ and $f=2$ this vertex is linear in the momenta and the masses. A four-quark operator carries no power of $v$ and has $d_{\min}=-2$, and the same vertex then contains the penguin structure $q^2\gamma^\mu-q^\mu\slashed q$, of
degree two. For the same reason $O_G$ and $O_{\widetilde G}$ reach six gluons, while $O_{HG}$ and $O_{H\widetilde G}$ stop at
four. In the computation $N_{\rm eff}$ has two uses. Each diagram is expanded in the momenta and the masses only up to this degree
before the integration, and the Lorentz--Dirac structures of section~\ref{sec:tensorbasis} are limited to this degree, which limits the size of the basis of each vertex.

\subsection{Computing the UV pole and \texorpdfstring{$\Rtwo$}{R2} of a diagram%
}\label{app:ct:fp}\label{app:ct:formulae}\label{app:ct:examples}
We compute only the part $\Rtwo$ of the rational term, since the part $R_1$ is obtained by the one-loop generator from the
four-dimensional numerator, as explained in section~\ref{sec:r2:def}. A diagram with $n$ propagators $D_i=(k+p_i)^2-m_i^2$ has a
numerator $N=\bar N+\tilde N$, where $\bar N$ is built from the four-dimensional parts of the loop momentum $k$, of the Dirac matrices
and of the metric, and $\tilde N$ contains the parts in the remaining $D-4$ dimensions, such as $\tilde k^2$. After the integration
every term of $\tilde N$ carries a factor $D-4$, so that $\Rtwo=-2Q$ as in eq.~\eqref{eq:r2:A}, while the integral of $N$ has the UV pole $\mathcal P/\epsilon$. Both the pole and the rational term of a diagram come from the UV poles of scalar integrals, and no finite part of an integral is needed.
After the contraction with a projector or with the auxiliary vectors of appendix~\ref{app:tb:solve}, the denominators $D_i$ %
are combined with Feynman parameters,
\begin{equation}\label{eq:r2:fp}
  \frac1{D_0\cdots D_{n-1}}=(n-1)!\int[\dd x]\,\frac1{\big[(k+X)^2-\Delta\big]^n},
\end{equation}
with $X=\sum_ix_i\,p_i$, $\Delta=X^2-\sum_ix_i(p_i^2-m_i^2)$ and $[\dd x]=\dd x_0\cdots\dd x_{n-1}\,\delta(1-\sum_ix_i)$. Since $X$ is
four-dimensional, the shift $l=k+X$ leaves $\tilde l=\tilde k$, and the tensor integrals are reduced by Lorentz invariance, with $l^\mu l^\nu\to g^{\mu\nu}l^2/D$ and
$\tilde l^{\,2}\to(D-4)\,l^2/D$. The diagram is then a sum of the scalar integrals
\begin{equation}\label{eq:r2:SKdef}
  S_K=\int\!\frac{\dd^Dl}{\ii\pi^{D/2}}\,\frac{(l^2)^K}{(l^2-\Delta)^n}
  =(-1)^{n+K}\,\frac{\Gamma(K+\frac D2)\,\Gamma(n-K-\frac D2)}{\Gamma(\frac D2)\,\Gamma(n)}\,\Delta^{K+\frac D2-n}
\end{equation}
times coefficients $c_K(D)$, rational in $D$ and polynomial in the Feynman parameters, momenta and masses. The four-dimensional numerator
$\bar N$ gives coefficients $\bar c_K$ of the same $S_K$. Since $c_K-\bar c_K$ is of order $D-4$, only the UV pole of $S_K$ contributes to
the rational term. This pole exists for $j=K+2-n\ge0$,
\begin{equation}\label{eq:r2:pole}
  S_K\Big|_{1/\epsilon}=\frac{(K+1)!}{(n-1)!\,j!}\,\Delta^j ,
\end{equation}
and the integrals over the Feynman parameters are
\begin{equation}\label{eq:r2:simplex}
  \int[\dd x]\,\prod_ix_i^{a_i}=\frac{\prod_ia_i!}{\big(\sum_ia_i+n-1\big)!} .
\end{equation}
The numerator $N$ gives the pole $\mathcal P/\epsilon$. The difference $c_K-\bar c_K$ gives $(D-4)\,Q/\epsilon$, and hence $\Rtwo=-2Q$ of
eq.~\eqref{eq:r2:A}. For instance, with $\Delta=(1-x)m_0^2+xm_1^2-x(1-x)p^2$ for the bubble, in the measure $\int\dd^Dk/(\ii\pi^{D/2})$ of ref.~\cite{Ossola:2008xq},
\begin{equation}\label{eq:r2:basicint}
  \int\!\frac{\tilde k^2}{D_0D_1}=-\frac12\Big(m_0^2+m_1^2-\frac{p^2}{3}\Big),\quad
  \int\frac{\tilde k^2}{D_0D_1D_2}=-\frac12,\quad \int\frac{\tilde k^4}{D_0D_1D_2D_3}=-\frac16.
\end{equation}
Before the integration each vertex is truncated at its degree
$N_{\rm eff}$ of eq.~\eqref{eq:pc:Neff}, counting each external momentum with one power and each squared mass in $\Delta$ with two.

\subsection{Computing the coefficients of the tensor structures}\label{app:tb:solve}\label{app:tb:dim}\label{app:tb:cand}\label{app:tb:canon}\label{app:tb:checks}
We write the tree, the UV pole and the rational term of an effective vertex in the form of eq.~\eqref{eq:tb:vertex}, with the
structures and the projectors $P_{\alpha\beta}$ of section~\ref{sec:tensorbasis}, which give the linear system of eq.~\eqref{eq:tb:gram}.
The coefficients $c_{\alpha\beta}$ are obtained in three steps, by fixing the basis of structures, projecting every diagram onto it,
and solving eq.~\eqref{eq:tb:gram}.%

The first step is the list of structures. For a given set of external legs we write all products
$\mathcal C_\alpha\otimes\mathcal T_\beta$ allowed by the indices of the legs, with $\mathcal T_\beta$ of degree at most $N_{\rm eff}$, as in
eq.~\eqref{eq:ex1:cand} for the example below. Some of these candidates are related in four dimensions, by the Dirac algebra or by the
Schouten identity. We compute their Gram matrices of eq.~\eqref{eq:tb:gram} and keep a maximal set of independent candidates, which is the
basis of the vertex.

With the basis fixed, the %
second step is the projection. Each diagram is contracted with the projector $P_{\alpha\beta}$ of every candidate and integrated as
in appendix~\ref{app:ct:fp}. This gives the left-hand side of eq.~\eqref{eq:tb:gram} as a rational function of $D$, once for the numerator
$N$ and once for $\bar N$.

The third step is the solution of eq.~\eqref{eq:tb:gram}, once for $N$ and once for $\bar N$, for which we use the \textsc{FLINT}
library~\cite{Hart:2010flint}. The two solutions have the same pole, and their difference is $(D-4)\,Q/\epsilon$ of
eq.~\eqref{eq:r2:A}, which gives $\Rtwo=-2Q$.

The projectors carry four-dimensional metric tensors on the external indices, since the external momenta, polarisations and spinors are four-dimensional%
. With $D$-dimensional
projectors, a Levi-Civita structure removed from the basis by the Schouten identity would contribute $D-4$ times its pole to the others.
This spurious rational term would show up as a violation of Bose symmetry, for instance in the three-gluon vertex of $O_{\widetilde G}$.

The route of auxiliary vectors described in section~\ref{sec:tensorbasis} does not need the basis construction or the solution of the linear system. We used it for the five- and six-gluon
vertices, and for a subset of the other vertices as an independent check of the projection.

\begin{figure}[!htbp]
\centering
\begin{tikzpicture}[scale=0.9]
\begin{scope}
  \coordinate (O) at (0.6,0); \coordinate (Q) at (-0.25,-0.55);
  \draw[sfer] (-0.9,-1.0) -- (Q); \draw[sfer] (Q) -- (O); \draw[sfer] (O) -- (-0.9,1.0);
  \draw[sglu] (Q) to[bend right=55] (O); \draw[sglu] (O) -- (1.5,0);
  \node[sopv] at (O) {};
  \node[anchor=east] at (-0.9,-1.0) {$u_i$}; \node[anchor=east] at (-0.9,1.0) {$\bar u_j$};
  \node at (0.3,-1.6) {(1)};
\end{scope}
\begin{scope}[xshift=3.2cm]
  \coordinate (O) at (0.6,0); \coordinate (Q) at (-0.25,0.55);
  \draw[sfer] (-0.9,-1.0) -- (O); \draw[sfer] (O) -- (Q); \draw[sfer] (Q) -- (-0.9,1.0);
  \draw[sglu] (Q) to[bend left=55] (O); \draw[sglu] (O) -- (1.5,0);
  \node[sopv] at (O) {};
  \node at (0.3,-1.6) {(2)};
\end{scope}
\begin{scope}[xshift=6.3cm]
  \coordinate (O) at (-0.1,0); \coordinate (T) at (1.0,0);
  \draw[sfer] (-0.9,-1.0) -- (O); \draw[sfer] (O) -- (-0.9,1.0);
  \draw[sglu] (O) arc (180:0:0.55); \draw[sglu] (O) arc (180:360:0.55); \draw[sglu] (T) -- (1.7,0);
  \node[sopv] at (O) {};
  \node at (0.4,-1.6) {(3)};
\end{scope}
\begin{scope}[xshift=9.6cm]
  \coordinate (A) at (0,-0.6); \coordinate (B) at (0,0.6); \coordinate (O) at (0.9,0);
  \draw[sfer] (-0.7,-1.0) -- (A); \draw[sfer] (A) -- (O); \draw[sfer] (O) -- (B); \draw[sfer] (B) -- (-0.7,1.0);
  \draw[sglu] (A) -- (B); \draw[sglu] (O) -- (1.7,0);
  \node[sopv] at (O) {};
  \node at (0.5,-1.6) {(4)};
\end{scope}
\begin{scope}[xshift=12.9cm]
  \coordinate (A) at (0,-0.6); \coordinate (B) at (0,0.6); \coordinate (O) at (0.9,0);
  \draw[sfer] (-0.7,-1.0) -- (A); \draw[sfer] (A) -- (B); \draw[sfer] (B) -- (-0.7,1.0);
  \draw[sglu] (A) -- (O); \draw[sglu] (B) -- (O); \draw[sglu] (O) -- (1.7,0);
  \node[sopv] at (A) {};
  \node at (0.5,-1.6) {(5)};
\end{scope}
\end{tikzpicture}
\caption{The five one-loop topologies of the quark--gluon vertex $u_i\bar u_jg$ at $\order{C_{uG}}$, numbered as in
table~\ref{tab:ex1:diag}. In (1)--(3) the operator enters through its contact vertex $q\bar qgg$, in (4) and (5) through the dipole
vertex $q\bar qg$. The internal quarks
carry the generation of the external line to which they are attached.}
\label{fig:method:qqg}
\end{figure}
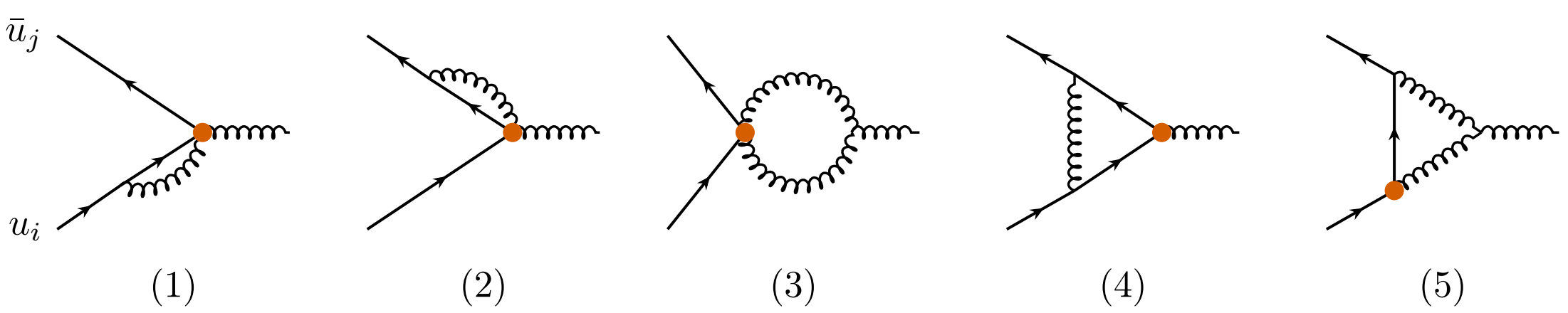%
\subsection{The dipole quark--gluon vertex, from the diagrams to the counterterm}\label{sec:method:example}
As an example of the whole computation, from the diagrams to the counterterm, we show %
the effective vertex of two up-type quarks and a gluon with an insertion of the dipole operator $O_{uG}$. Its
legs are an incoming quark $u_i$ with momentum $p_1$, an incoming antiquark $\bar u_j$ with momentum $p_2$ and an outgoing gluon with momentum $q=p_1+p_2$ and Lorentz and colour indices $\mu$ and $A$, respectively.
The tree-level vertex is
\begin{equation}\label{eq:ex1:tree}
\begin{gathered}
  \Gamma^{\mu}_{\rm tree}=-\ii g_s\,\delta_{ij}\,T^A\gamma^\mu
  +\frac{\sqrt2\,v}{\Lambda^2}\,T^A\sigma^{\mu\nu}q_\nu\bigl(C\,P_R+\Cb\,P_L\bigr),\\
  C\equiv (C_{uG})_{ji},\qquad \Cb\equiv (C^*_{uG})_{ij}.
\end{gathered}
\end{equation}
The first term is the SM vertex and the second is the dipole vertex. The operator also gives a contact vertex with two gluons, so that
there are the five diagrams of figure~\ref{fig:method:qqg}. By the power counting of appendix~\ref{app:pc} the vertex is linear in the
momenta and the masses.

The colour structure is $T^A$. The Dirac structures that can appear are, with $p$ equal to $p_1$ or $p_2$,
\begin{equation}\label{eq:ex1:cand}
  p^\mu,\qquad p^\mu\gfive,\qquad \gamma^\mu,\qquad \gamma^\mu\gfive,\qquad \gamma^\mu\slashed p,\qquad \slashed p\,\gamma^\mu,\qquad
  -\ii\,\epsilon^{\rho\sigma\mu p}\gamma_\rho\gamma_\sigma .
\end{equation}
These twelve structures are related by
$\gamma^\mu\slashed p+\slashed p\gamma^\mu=2p^\mu$, so that we drop $\slashed p\gamma^\mu$ and keep the other ten as the basis. The vertex has no
closed quark loop, so that the reading point of $\gfive$ plays no role. Table~\ref{tab:ex1:diag} gives the contribution of each diagram
to each structure, for the rational term and for the pole. The coefficients of $\gamma^\mu$ and $\gamma^\mu\gfive$ are linear in the
quark masses, and the table gives their $m_{u,i}$ terms.
\begin{table}[!htbp]
\centering\small
\begin{tabular}{@{}l@{\hspace{3mm}}rrrrr@{\hspace{3mm}}r@{\hspace{6mm}}rrrrr@{\hspace{3mm}}r@{}}\toprule
& \multicolumn{6}{c}{$\Rtwo$} & \multicolumn{6}{c}{UV pole}\\
structure & 1 & 2 & 3 & 4 & 5 & sum & 1 & 2 & 3 & 4 & 5 & sum\\\midrule
$p_1^\mu$ & $0$ & $0$ & $0$ & $\frac14$ & $\frac{21}8$ & $\frac{23}8$ & $\frac34$ & $0$ & $\frac94$ & $-\frac13$ & $-\frac92$ & $-\frac{11}6$\\[2pt]
$p_2^\mu$ & $0$ & $\frac32$ & $0$ & $-\frac5{12}$ & $\frac{39}8$ & $\frac{143}{24}$ & $0$ & $-\frac32$ & $\frac94$ & $\frac16$ & $-\frac{45}8$ & $-\frac{113}{24}$\\[2pt]
$p_1^\mu\gfive$ & $-\frac34$ & $0$ & $0$ & $\frac13$ & $-\frac98$ & $-\frac{37}{24}$ & $\frac98$ & $0$ & $0$ & $-\frac14$ & $\frac9{16}$ & $\frac{23}{16}$\\[2pt]
$p_2^\mu\gfive$ & $0$ & $\frac34$ & $0$ & $-\frac13$ & $\frac98$ & $\frac{37}{24}$ & $0$ & $-\frac98$ & $0$ & $\frac14$ & $-\frac9{16}$ & $-\frac{23}{16}$\\[2pt]
$\gamma^\mu\,m_{u,i}$ & $\frac32$ & $0$ & $0$ & $-\frac1{12}$ & $-\frac98$ & $\frac7{24}$ & $-\frac94$ & $0$ & $0$ & $\frac18$ & $\frac{27}{16}$ & $-\frac7{16}$\\[2pt]
$\gamma^\mu\gfive\,m_{u,i}$ & $-\frac32$ & $0$ & $0$ & $\frac1{12}$ & $\frac98$ & $-\frac7{24}$ & $\frac94$ & $0$ & $0$ & $-\frac18$ & $-\frac{27}{16}$ & $\frac7{16}$\\[2pt]
$\gamma^\mu\slashed p_1$ & $-\frac34$ & $0$ & $0$ & $\frac1{12}$ & $-\frac{15}4$ & $-\frac{53}{12}$ & $\frac38$ & $0$ & $-\frac94$ & $\frac1{12}$ & $\frac{81}{16}$ & $\frac{157}{48}$\\[2pt]
$\gamma^\mu\slashed p_2$ & $0$ & $-\frac34$ & $0$ & $\frac1{12}$ & $-\frac{15}4$ & $-\frac{53}{12}$ & $0$ & $\frac38$ & $-\frac94$ & $\frac1{12}$ & $\frac{81}{16}$ & $\frac{157}{48}$\\[2pt]
$-\ii\,\epsilon^{\rho\sigma\mu p_1}\gamma_\rho\gamma_\sigma$ & $\frac38$ & $0$ & $0$ & $-\frac1{24}$ & $\frac{15}8$ & $\frac{53}{24}$ & $-\frac3{16}$ & $0$ & $\frac98$ & $-\frac1{24}$ & $-\frac{81}{32}$ & $-\frac{157}{96}$\\[2pt]
$-\ii\,\epsilon^{\rho\sigma\mu p_2}\gamma_\rho\gamma_\sigma$ & $0$ & $\frac38$ & $0$ & $-\frac1{24}$ & $\frac{15}8$ & $\frac{53}{24}$ & $0$ & $-\frac3{16}$ & $\frac98$ & $-\frac1{24}$ & $-\frac{81}{32}$ & $-\frac{157}{96}$\\\bottomrule
\end{tabular}
\caption{Contributions of the five diagrams of the $u_i\bar u_jg$ vertex to the coefficient of $C=(C_{uG})_{ji}$, in units of $\sqrt2\,g_s^2v/\Lambda^2$
and of eq.~\eqref{eq:units}, colour factor $T^A$. The coefficient of $\Cb$ is the same
for the structures without $\gfive$ and without the Levi-Civita tensor, and opposite for the others, and the $m_{u,j}$ entries follow from the $m_{u,i}$ entries by exchanging the quark and the
antiquark, which exchanges diagrams (1) and (2).}\label{tab:ex1:diag}
\end{table}

We combine the entries with $\gfive$ and with the Levi-Civita tensor into chiral
projectors, with $\gamma^\mu\slashed p=p^\mu-\ii\sigma^{\mu p}$ and
$\epsilon^{\rho\sigma\mu\nu}\gamma_\rho\gamma_\sigma=-2\sigma^{\mu\nu}\gfive$. The rational term is
\begin{equation}\label{eq:res:qqgR2}%
\begin{split}
  \vf^{\Rtwo,\mu}_{u_i\bar u_jg}=\;&\tfrac83\,g_s^3\,\delta_{ij}\,T^A\gamma^\mu\\
  &+\frac{\sqrt2\,g_s^2v}{\Lambda^2}\,T^A\Bigl\{\tfrac{7}{12}\,\gamma^\mu\bigl[C\,(m_{u,i}P_L+m_{u,j}P_R)
  +\Cb\,(m_{u,i}P_R+m_{u,j}P_L)\bigr]\\
  &\qquad+\Bigl[\tfrac{23}{4}\,p_1^\mu+\tfrac{143}{12}\,p_2^\mu-\tfrac{53}{6}\,\gamma^\mu(\slashed p_1+\slashed p_2)\Bigr]
  \bigl(C\,P_R+\Cb\,P_L\bigr)\Bigr\},
\end{split}
\end{equation}
and the coefficient of the pole is
\begin{equation}\label{eq:res:qqgUV}
\begin{split}
  \vf^{\rm UV,\mu}_{u_i\bar u_jg}=\;&-\tfrac{13}3\,g_s^3\,\delta_{ij}\,T^A\gamma^\mu\\
  &+\frac{\sqrt2\,g_s^2v}{\Lambda^2}\,T^A\Bigl\{-\tfrac{7}{8}\,\gamma^\mu\bigl[C\,(m_{u,i}P_L+m_{u,j}P_R)
  +\Cb\,(m_{u,i}P_R+m_{u,j}P_L)\bigr]\\
  &\qquad+\Bigl[-\tfrac{11}{3}\,p_1^\mu-\tfrac{113}{12}\,p_2^\mu+\tfrac{157}{24}\,\gamma^\mu(\slashed p_1+\slashed p_2)\Bigr]
  \bigl(C\,P_R+\Cb\,P_L\bigr)\Bigr\}.
\end{split}
\end{equation}
The SM part of the pole is $(C_F+C_A)\,g_s^2/(16\pi^2\epsilon)$ times the SM tree vertex, with $C_F=4/3$ and $C_A=3$.

The counterterm vertex follows from the pole with eq.~\eqref{eq:uv:rule}. We write it as
\begin{equation}\label{eq:uvex:dc}
  \delta c=\delta c_Z-\Gamma^{\rm UV}_C\Big(\frac1\epsilon-L_{\rm eft}\Big),
\end{equation}
where $\delta c_Z$ is the sum of the first two terms of eq.~\eqref{eq:uv:rule}, which renormalise the SM fields and parameters, while
the last term renormalises the Wilson coefficients.

In $\delta c_Z$ the SM vertex is multiplied by the renormalisation factors of first order in $C_{uG}$, and the dipole vertex by the
SM factors,
\begin{equation}\label{eq:uvex:first}
\begin{split}
  \delta c_Z={}&-\frac{\ii g_s}2\,T^A\gamma^\mu\Big[\big(\delta\bar Z^L+\delta Z^L\big)_{ji}P_L+\big(\delta\bar Z^R+\delta Z^R\big)_{ji}P_R\Big]\\
  &+\frac{\sqrt2\,v}{2\Lambda^2}\,T^A\sigma^{\mu\nu}q_\nu\bigl(C\,P_R+\Cb\,P_L\bigr)\big(\delta Z_{u_i}+\delta Z_{u_j}+\delta Z_g\big).
\end{split}
\end{equation}
Here $\delta Z^{L,R}$ and $\delta\bar Z^{L,R}$ are the field factors of the incoming and the outgoing quark of
section~\ref{sec:uv:mix}, at first order in $C_{uG}$, and $\delta Z_{u_i}$, $\delta Z_{u_j}$ and $\delta Z_g$ are the SM field renormalisation factors
of the two quarks and of the gluon. The counterterm of $g_s$ proportional to the dipole coefficient does not appear, because it cancels against
the gluon field factor of the same order.

The coefficient $\Gamma^{\rm UV}_C$ is what is left of the UV pole of the loop once the UV poles of $\delta c_Z$ are added to it, and reads
\begin{equation}\label{eq:uvex:gammac}
\begin{split}
  \Gamma^{\rm UV}_C=\frac{\ii}{16\pi^2}\,\frac{\sqrt2\,g_s^2v}{\Lambda^2}\,T^A\Bigl\{&\tfrac98\,\gamma^\mu\bigl[C\,(m_{u,i}P_L+m_{u,j}P_R)
  +\Cb\,(m_{u,i}P_R+m_{u,j}P_L)\bigr]\\
  &+\Bigl[-\tfrac{17}{6}\,p_1^\mu-\tfrac{103}{12}\,p_2^\mu+\tfrac{137}{24}\,\gamma^\mu(\slashed p_1+\slashed p_2)\Bigr]
  \bigl(C\,P_R+\Cb\,P_L\bigr)\Bigr\}.
\end{split}
\end{equation}
Unlike its SM part, $\Gamma^{\rm UV}_C$ is not proportional to the tree-level vertex. It also contains the vertices of redundant
operators, specifically $\bar qD^2u\,\tilde\varphi$ of section~\ref{sec:uv:mix}, and the
gauge-variant unphysical operator~\mbox{\cite{Naterop:2023dek,Duhr:2025zqw}}
\begin{equation}\label{eq:ex1:gv}
  -\frac98\,\frac{\sqrt2\,g_s^2v}{\Lambda^2}\,C\,\Bigl[\bar u_j\,T^A\slashed G^A P_L\,\bigl(\ii\slashed\partial-m_{u,i}\bigr)u_i
  +\bar u_j\bigl(-\ii\overleftarrow{\slashed\partial}-m_{u,j}\bigr)T^A\slashed G^A P_R\,u_i\Bigr]
\end{equation}
and the corresponding operator for $\Cb$. This operator is proportional to the equations of motion of the quarks. It vanishes on shell and does not
contribute to amplitudes. Between on-shell spinors $\Gamma^{\rm UV}_C$ becomes
\begin{equation}\label{eq:uvex:os}
\begin{split}
  \Gamma^{\rm UV}_C\to\frac{\alpha_s}{4\pi}\,\frac{\sqrt2\,v}{\Lambda^2}\,T^A\Bigl\{&\tfrac{17}6\,\sigma^{\mu\nu}q_\nu\bigl(C\,P_R+\Cb\,P_L\bigr)\\
  &+4\ii\,\gamma^\mu\bigl[C\,(m_{u,i}P_L+m_{u,j}P_R)+\Cb\,(m_{u,i}P_R+m_{u,j}P_L)\bigr]\Bigr\}.
\end{split}
\end{equation}
The first term is $\frac{17}6\,\alpha_s/(4\pi)$ times the dipole vertex of the tree. It is the counterterm of $C_{uG}$, with the
anomalous dimension of refs.~\cite{Jenkins:2013zja,Jenkins:2013wua,Alonso:2013hga} for six flavours. The second term contains no
momentum. It comes from the redundant operators, which on shell act as a shift of the SM fields and parameters, as discussed in
section~\ref{sec:uv:os}.

The package function \texttt{UVCounterterm} of appendix~\ref{app:anc} returns $\delta c$ for any choice of generations. The down-type vertex
$d_r\bar d_sg$ at first order in $C_{dG}$ follows with $m_u\to m_d$, $C\to(V^\dagger C_{dG})_{sr}$ and
$\Cb\to(V^\dagger C_{dG})^*_{rs}$, since the down-type quark of the doublet carries the CKM matrix.

\section{The reading point and the evanescent schemes}\label{app:scheme}

Section~\ref{sec:gamma5} fixes the three choices of our scheme that change the rational terms but not the UV poles. These are
the reading point of a closed quark loop with a single $\gfive$, the Levi-Civita tensor of an operator on an open quark line and the
reduction of the four-fermion Dirac structures. Appendix~\ref{app:g5list}
lists the vertices that depend on the reading point and discusses which readings respect gauge invariance, and appendix~\ref{app:evtests}
gives the relations that define our evanescent scheme.

\subsection{Reading-point dependence and gauge invariance of the readings}\label{app:g5list}\label{app:g5su3}\label{app:ward}
Table~\ref{tab:app:g5classes} lists, operator by operator, every effective vertex whose rational term depends on the reading point of a
closed quark loop with a single $\gfive$. These are the three classes of section~\ref{sec:g5:affected}. For the operators marked ``none'' no vertex depends on the reading.
\begin{table}[!htbp]
\centering\small
\renewcommand{\arraystretch}{1.15}
\begin{tabularx}{\textwidth}{@{}>{\raggedright\arraybackslash}p{0.23\textwidth}>{\raggedright\arraybackslash}p{0.2\textwidth}>{\raggedright\arraybackslash}X@{}}
\toprule
operators & source of $\gfive$ & vertices whose $\Rtwo$ depends on the reading point\\
\midrule
\multicolumn{3}{@{}l}{\emph{Vertices $gg\,X$ and $ggg\,X$, listed by the bosons $X$}}\\[2pt]
$C_{Hq}^{(1,3)}$, $C_{Hu}$, $C_{Hd}$ & operator (axial) vertex &
  $G^0$,\; $G^0h$,\; $Z$,\; $Zh$,\; $Zhh$,\; $ZG^0G^0$,\; $ZG^+G^-$,\; $G^+G^-$,\; $\gamma G^+G^-$\newline
  for $C^{(1)}_{Hq}$, $C_{Hu}$, $C_{Hd}$ also\; $W^\pm G^\mp$,\; $W^\pm G^\mp h$,\; $W^\pm G^0G^\mp$\\
\addlinespace
$\Re C_{uG}$, $\Re C_{dG}$ & operator and SM & $G^0$,\; $G^0h$,\; $Z$,\; $Zh$,\; $Zhh$,\; $ZG^0G^0$\\
$\Im C_{uG}$, $\Im C_{dG}$ & operator and SM & $h$,\; $hh$,\; $G^0G^0$\\
$C_{uG}$, $C_{dG}$ (both parts) & operator and SM & $G^+G^-$,\; $\gamma G^+G^-$,\; $ZG^+G^-$,\; $W^\pm G^\mp$,\; $W^\pm G^\mp h$,\; $W^\pm G^0G^\mp$\\
\addlinespace
$C_{uW}$, $C_{uB}$, $C_{dW}$, $C_{dB}$ & operator and SM & none (every closed-loop vertex vanishes at $\order{C}$)\\
$C_{uH}$, $C_{dH}$ & $G^0$, $G^\pm$ couplings & none\\
$C_{Hud}$ & operator vertex & none\\
\midrule
\multicolumn{3}{@{}l}{\emph{Vertices with two external fermions}}\\[2pt]
four-fermion & chiral projector of the four-fermion Feynman rule &
  $\ell\bar\ell gg(g)$,\; $\nu\bar\nu gg(g)$,\; $q\bar qgg(g)$\newline
  colour singlets\; $q\bar q\gamma\gamma$,\; $q\bar q\gamma Z$,\; $q\bar qZZ$\newline
  colour octets\; $q\bar q\gamma g$,\; $q\bar qgZ$,\; $q\bar q\gamma gg$,\; $q\bar qggZ$\newline
  vertices with $W^\pm$ for left-handed currents\newline
  none for $C^{(1,8)}_{quqd}$, $C^{(1,3)}_{lequ}$, $C_{ledq}$\\
\bottomrule
\end{tabularx}
\caption{Effective vertices whose rational term depends on the reading point. In the upper part each entry $X$ stands for the two vertices $gg\,X$ and
$ggg\,X$, and in the lower part $gg(g)$ stands for two and for three gluons. Only the Levi-Civita parts of $\Rtwo$ are affected. The insertions of $C_{uH}$ and $C_{dH}$ are scalar vertices without
gamma matrices, and the closed loops that contain them do not have enough gamma matrices in their divergent parts for a Levi-Civita term to depend on the reading. For $\Im C_{uG}$ and $\Im C_{dG}$ the
Levi-Civita part of $ggZG^0$ and $gggZG^0$ vanishes in every reading.}\label{tab:app:g5classes}
\end{table}

Different readings of a closed quark loop give different coefficients of the Levi-Civita tensor in $\Rtwo$%
. Whether a reading other than the one at the operator respects the gauge
symmetries depends on the vertex. For some vertices every reading does, and the readings differ only by a finite shift of a Wilson coefficient,
as for $\Im C_{tG}$ discussed later. For other vertices, such readings can violate gauge invariance. We show an
example of the latter case, the closed top loop of $ggG^0$ and $gggG^0$ at $\order{\Re\,(C_{uG})_{33}}$ of figure~\ref{fig:g5:ggG0}, and
then discuss why the reading at the operator respects the identities in all cases.

Gauge invariance imposes two conditions on these vertices. The first condition comes from $SU(3)$ invariance, which forces the Levi-Civita couplings of $G^0$ to two and to three gluons to come from the same local term of the form
\begin{equation}\label{eq:g5:kappa}
  \Lag\supset\kappa\,G^0\,\epsilon^{\mu\nu\rho\sigma}G^A_{\mu\nu}G^A_{\rho\sigma}.
\end{equation}
The two Green's functions then contain
\begin{equation}\label{eq:g5:c2c3}
  \Gamma^{\mu_1\mu_2}_{ggG^0}\supset c_2\,\delta^{AB}\,\epsilon^{\mu_1\mu_2\nu\rho}p_{1\nu} p_{2\rho},\qquad
  \Gamma^{\mu_1\mu_2\mu_3}_{gggG^0}\supset c_3\,f^{ABC}\,\epsilon^{\mu_1\mu_2\mu_3\nu}(p_1+p_2+p_3)_\nu,
\end{equation}
and their coefficients are related by
\begin{equation}\label{eq:g5:su3}
  c_3=-\ii\,g_s\,c_2 .
\end{equation}
A reading respects invariance under $SU(3)$ only if its two coefficients satisfy this relation. For
$\Re C_{tG}$ the readings at the operator, at the $G^0$ vertex and at one gluon vertex satisfy it, each with a different value of
$\kappa$, while the average over the gluons does not, as shown in table~\ref{tab:app:ward}.
\begin{table}[!htbp]
\centering\small\setlength{\tabcolsep}{5pt}
\begin{tabular}{@{}lcccccc@{}}
\toprule
 & \multicolumn{3}{c}{$ggG^0$, $gggG^0$} & \multicolumn{3}{c}{$ggG^+G^-$, $gggG^+G^-$}\\
\cmidrule(lr){2-4}\cmidrule(l){5-7}
 & \multicolumn{2}{c}{$\Re C_{tG}$} & $C_{Hu}$ & $\Re C_{tG}$ & $\Im C_{tG}$ & $C_{Hu}$\\
reading & $SU(3)$ & electroweak & $SU(3)$ & $SU(3)$, CP & $SU(3)$, CP & $SU(3)$, CP\\
\midrule
operator (ours) & \ding{51} & \ding{51} & \ding{51} & \ding{51} & \ding{51} & \ding{51}\\
$G^0$ / $G^+$ vertex & \ding{51} & \ding{55} & \ding{51} & \ding{55} (CP) & \ding{51} & \ding{51}\\
$G^+$, $G^-$ averaged & -- & -- & -- & \ding{51} & \ding{51} & \ding{51}\\
gluon leg 1 & \ding{51} & \ding{55} & \ding{51} & \ding{55} ($SU(3)$) & \ding{51} & \ding{55} ($SU(3)$)\\
gluons averaged & \ding{55} & \ding{55} & \ding{51} & \ding{55} ($SU(3)$) & \ding{55} ($SU(3)$) & \ding{55} ($SU(3)$)\\
\bottomrule
\end{tabular}
\caption{The identities satisfied (\ding{51}) or violated (\ding{55}) by the closed top loop for different reading points. The identities are the $SU(3)$
relation of eq.~\eqref{eq:g5:su3}, for the charged vertices also CP, and for $ggG^0$ at $\order{\Re\,C_{tG}}$ the electroweak identity of
eq.~\eqref{eq:g5:ewward}. A dash marks a reading that does not exist for the vertex%
. For $C_{Hu}$ the
$G^0$ and $G^\pm$ vertices are operator vertices. %
}\label{tab:app:ward}\label{tab:g5:ward}
\end{table}

The second condition is the electroweak Slavnov--Taylor identity, which for an outgoing $Z$ boson reads
\begin{equation}\label{eq:g5:ewward}
  (p_1+p_2)_\mu\,\Gamma^{\mu\mu_1\mu_2}_{ggZ}=\ii M_Z\,\Gamma^{\mu_1\mu_2}_{ggG^0}.
\end{equation}
This identity relates the vertex with a $Z$ boson to the vertex with a $G^0$, and in this case fixes
$\kappa$. The reading at the operator gives this value. The readings at the $G^0$ vertex and at a gluon vertex give a
different one. They respect $SU(3)$ but violate eq.~\eqref{eq:g5:ewward}, and would need a counterterm of the form $-\delta\kappa\,G^0\epsilon GG$ to
restore it, with $\delta\kappa$ the difference of their $\kappa$ from the value read at the operator.

\begin{figure}[!htbp]
\centering
\begin{tikzpicture}[scale=0.95]
\begin{scope}
  \coordinate (A) at (0,0.6); \coordinate (B) at (1.2,0.6); \coordinate (C) at (1.2,-0.6); \coordinate (D) at (0,-0.6);
  \draw[sfer] (A) -- (B); \draw[sfer] (B) -- (C); \draw[sfer] (C) -- (D); \draw[sfer] (D) -- (A);
  \draw[sglu] (-0.8,1.1) -- (A); \draw[sglu] (-0.8,-1.1) -- (D);
  \draw[spho] (B) -- (2.0,1.1); \draw[shig] (C) -- (2.0,-1.1);
  \node[sopv] at (A) {};
  \node[anchor=east] at (-0.8,1.1) {$g$}; \node[anchor=east] at (-0.8,-1.1) {$g$};
  \node[anchor=west] at (2.0,1.1) {$Z$}; \node[anchor=west] at (2.0,-1.1) {$h$};
  \node[anchor=south] at (0.6,0.62) {$t$};
  \node at (0.6,-1.75) {(a)};
\end{scope}
\begin{scope}[xshift=3.7cm]
  \coordinate (G1) at (0,0.6); \coordinate (G2) at (0,-0.6); \coordinate (V) at (1.0,0); \coordinate (W) at (1.75,0);
  \draw[sfer] (G1) -- (V); \draw[sfer] (V) -- (G2); \draw[sfer] (G2) -- (G1);
  \draw[sglu] (-0.8,1.1) -- (G1); \draw[sglu] (-0.8,-1.1) -- (G2);
  \draw[spho] (V) -- (W); \draw[spho] (W) -- (2.35,0.8); \draw[shig] (W) -- (2.35,-0.8);
  \node[sopv] at (G1) {};
  \node[anchor=south] at (1.38,0.12) {$Z$};
  \node[anchor=west] at (2.35,0.8) {$Z$}; \node[anchor=west] at (2.35,-0.8) {$h$};
  \node at (1.0,-1.75) {(b)};
\end{scope}
\begin{scope}[xshift=7.4cm]
  \coordinate (G1) at (0,0.6); \coordinate (G2) at (0,-0.6); \coordinate (V) at (1.0,0); \coordinate (W) at (1.75,0);
  \draw[sfer] (G1) -- (V); \draw[sfer] (V) -- (G2); \draw[sfer] (G2) -- (G1);
  \draw[sglu] (-0.8,1.1) -- (G1); \draw[sglu] (-0.8,-1.1) -- (G2);
  \draw[shig] (V) -- (W); \draw[spho] (W) -- (2.35,0.8); \draw[shig] (W) -- (2.35,-0.8);
  \node[sopv] at (G1) {};
  \node[anchor=south] at (1.38,0.05) {$G^0$};
  \node[anchor=west] at (2.35,0.8) {$Z$}; \node[anchor=west] at (2.35,-0.8) {$h$};
  \node at (1.0,-1.75) {(c)};
\end{scope}
\begin{scope}[xshift=11.3cm]
  \coordinate (V) at (0,0); \coordinate (W) at (1.1,0);
  \draw[sglu] (-0.8,1.0) -- (V); \draw[sglu] (-0.8,-1.0) -- (V);
  \draw[shig] (V) -- (W); \draw[spho] (W) -- (1.8,0.8); \draw[shig] (W) -- (1.8,-0.8);
  \node[sr2] at (V) {};
  \node[anchor=south] at (0.55,0.05) {$G^0$}; \node[anchor=north west] at (0.02,-0.08) {$\Rtwo$};
  \node[anchor=west] at (1.8,0.8) {$Z$}; \node[anchor=west] at (1.8,-0.8) {$h$};
  \node at (0.5,-1.75) {(d)};
\end{scope}
\end{tikzpicture}
\caption{Classes of one-loop diagrams of $gg\to Zh$ at $\order{\Re C_{tG}}$ in the Feynman gauge, with the dipole vertex (dot) at one of
its possible places on the top loop. (a) Boxes. (b) Triangles with an $s$-channel $Z$ boson. (c) Triangles with an $s$-channel $G^0$.
(d) The rational term of $ggG^0$, eq.~\eqref{eq:g5:ggG0}, as an effective vertex (square), followed by $G^0$ exchange. In the unitary
gauge (c) and (d) are absent and the longitudinal part of the $Z$ propagator in (b) takes their place, so that the two gauges agree only
if the electroweak Slavnov--Taylor identities are satisfied.}
\label{fig:g5:zh}
\end{figure}
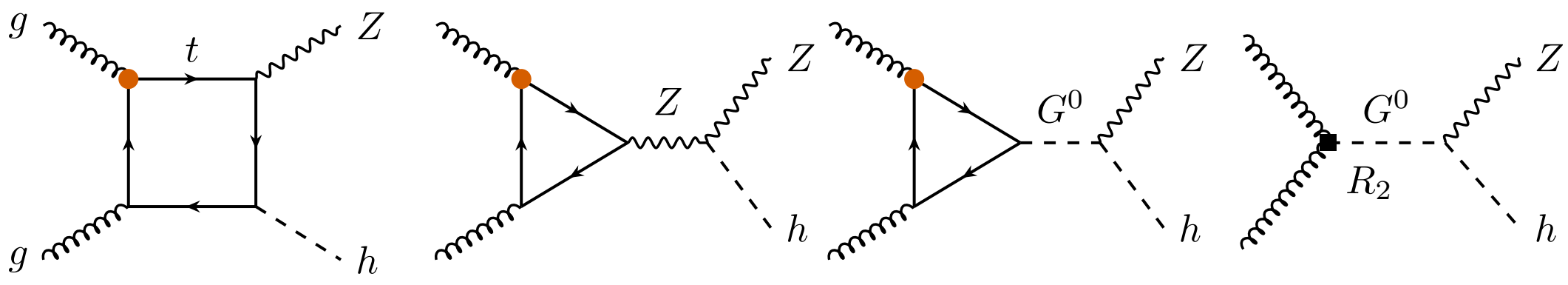

The violation can be tested in a physical amplitude. In $gg\to Zh$ at $\order{\Re\,(C_{uG})_{33}}$ the Feynman gauge contains the exchange of
a $G^0$ through the $\Rtwo$ vertex $ggG^0$ and the unitary gauge does not, as shown in figure~\ref{fig:g5:zh}. With the reading at the
operator the two gauges agree. With the reading at the $G^0$ vertex they differ, as shown in table~\ref{tab:g5:zh}. Adding the counterterm $-\delta\kappa\,G^0\epsilon GG$ restores the
agreement.
\begin{table}[!htbp]
\centering\small
\begin{tabular}{@{}lccc@{}}
\toprule
$\sqrt s$ [GeV] & 500 & 800 & 1500\\
\midrule
reading at the operator & 1 & 1 & 1\\
reading at the $G^0$ vertex & 1.44--1.45 & 1.05--1.07 & 0.47--0.51\\
\bottomrule
\end{tabular}
\caption{The interference of the amplitude of $gg\to Zh$ at $\order{\Re\,(C_{uG})_{33}}$ with the SM amplitude in the Feynman gauge,
divided by its value in the unitary gauge, for two scattering angles at each energy.}\label{tab:g5:zh}
\end{table}

We now turn to $\Im C_{tG}$, for which %
every reading at a single vertex satisfies the identities without additional counterterms, and the
dependence on the reading can be absorbed by a finite shift of $C_{H\widetilde G}$. In $ggh$, $gghh$, $ggG^0G^0$ and $ggG^+G^-$ with two and three gluons, a reading at a gluon vertex
differs from the one at the operator by $\frac12$ times the structure of $O_{H\widetilde G}$, and a reading at the vertex of the scalar by one times this structure. The difference is
\begin{equation}\label{eq:g5:imctg}
\begin{gathered}
  \rho=-\frac{2g_s}{16\pi^2}\Bigl[\sum_ay_{u,a}\Im(C_{uG})_{aa}+\sum_{a,b}y_{d,b}\Im\bigl(V^*_{ab}(C_{dG})_{ab}\bigr)\Bigr],\\
  \bar{C}_{H\widetilde G}=C_{H\widetilde G}+\lambda_{\rm rp}\,\rho ,
\end{gathered}
\end{equation}
with $\lambda_{\rm rp}=\frac12$ for a reading at a gluon vertex and $\lambda_{\rm rp}=1$ for a reading at the vertex of the scalar, so that a computation read at a gluon vertex or at the vertex of the scalar with the coefficient $C_{H\widetilde G}$ equals ours with $\bar{C}_{H\widetilde G}$.

\begin{figure}[!htbp]
\centering
\begin{tikzpicture}[scale=0.85, every node/.append style={transform shape}]
\begin{scope}
  \coordinate (A) at (0,0.75); \coordinate (B) at (0,-0.75); \coordinate (V) at (1.3,0);
  \draw[sfer] (A) -- (V); \draw[sfer] (V) -- (B); \draw[sfer] (B) -- (A);
  \draw[sglu] (-0.9,1.2) -- (A); \draw[sglu] (-0.9,-1.2) -- (B); \draw[spho] (V) -- (2.3,0);
  \node[sopv] at (V) {}; \node[sreadpt] at (V) {};
  \node[anchor=east] at (-0.9,1.2) {$g$}; \node[anchor=east] at (-0.9,-1.2) {$g$}; \node[anchor=west] at (2.3,0) {$Z,\,G^0$};
  \node at (0.7,-1.9) {(a)};
\end{scope}
\begin{scope}[xshift=5.2cm]
  \coordinate (A) at (0,0.75); \coordinate (B) at (0,-0.75); \coordinate (V) at (1.3,0);
  \draw[sfer] (A) -- (V); \draw[sfer] (V) -- (B); \draw[sfer] (B) -- (A);
  \draw[sglu] (-0.9,1.2) -- (A); \draw[sglu] (-0.9,-1.2) -- (B); \draw[shig] (V) -- (2.3,0);
  \node[sopv] at (A) {}; \node[sreadpt] at (A) {};
  \node[anchor=east] at (-0.9,1.2) {$g$}; \node[anchor=east] at (-0.9,-1.2) {$g$}; \node[anchor=west] at (2.3,0) {$G^0,\,h$};
  \node at (0.7,-1.9) {(b)};
\end{scope}
\begin{scope}[xshift=10.6cm]
  \coordinate (O) at (0,0);
  \draw[sfer] (-0.9,0.9) -- (O); \draw[sfer] (O) -- (-0.9,-0.9);
  \draw[sfer] (O) arc (180:0:0.7); \draw[sfer] (1.4,0) arc (0:-180:0.7);
  \draw[sglu] (1.195,0.495) -- (2.15,1.05); \draw[sglu] (1.195,-0.495) -- (2.15,-1.05);
  \node[sopv] at (O) {}; \node[sreadpt] at (O) {};
  \node[anchor=east] at (-0.9,0.9) {$q$}; \node[anchor=east] at (-0.9,-0.9) {$q$};
  \node[anchor=west] at (2.15,1.05) {$g$}; \node[anchor=west] at (2.15,-1.05) {$g$};
  \node at (0.7,-1.9) {(c)};
\end{scope}
\end{tikzpicture}
\caption{One diagram for each of the three classes of vertices that can depend on the reading point, with the operator marked by a dot and the
reading at the operator vertex marked by the blue ring. (a) A Higgs-current operator emits the $Z$ boson or the Goldstone boson. (b) A
gluon dipole emits one of the gluons. (c) One current of a four-fermion operator closes into the loop, and no gauge boson is attached to
the operator. All other bosons are attached to SM vertices.}
\label{fig:g5:classes}
\end{figure}
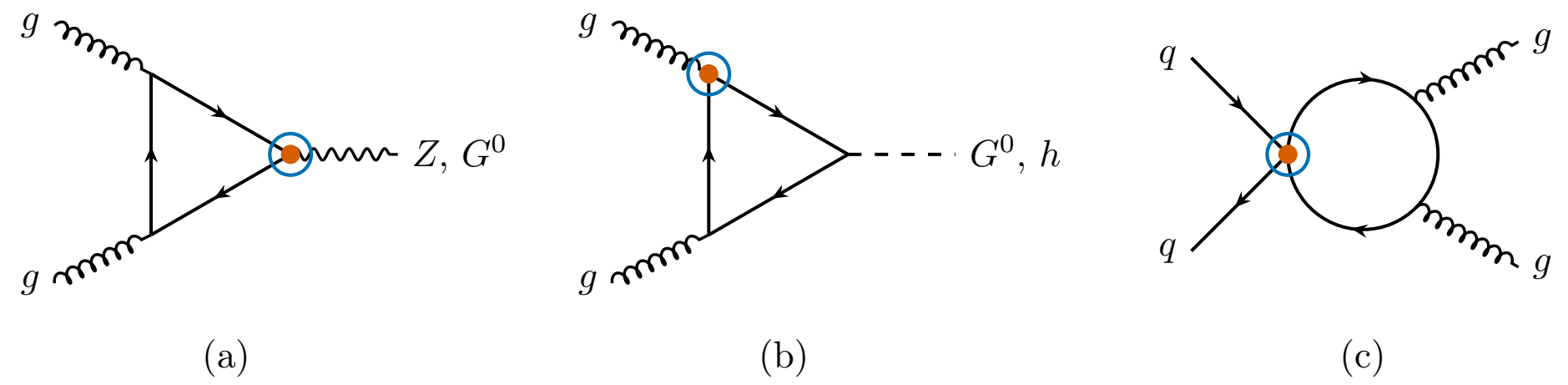
The reading at the operator does not need any symmetry-restoring counterterm in any of the three classes of vertices of
section~\ref{sec:g5:affected} that depend on the reading point, shown in figure~\ref{fig:g5:classes}. The Ward identity is tested by
contracting the leg of a gauge boson with its momentum. With a fixed reading point it holds at every vertex of the loop except the one at
which the trace is read~\cite{Korner:1991sx,Born:2026xkr}, and for our reading this is the operator vertex. Therefore, only the gauge bosons attached to the operator vertex have to be tested, and for them the identity follows
from the form of the operator itself. Figure~\ref{fig:g5:classes} shows one diagram for each class. In the first class, diagram (a), the
operator is a Higgs current. It couples the $Z$ boson and the derivative of $G^0$ through the same matrix, $\gamma^\mu P_R$ for $O_{Hu}$,
so that its $Z$ vertex contracted with the momentum is directly its $G^0$ vertex, as eq.~\eqref{eq:g5:ewward} requires. In the second class, diagram (b), the operator is a gluon dipole. Its gluon couples through $\sigma^{\mu\nu}q_\nu$, which always vanishes when contracted with $q_\mu$. In the third class, diagram (c), the current of the four-fermion operator that closes the loop
carries no gauge boson at all. All other gauge bosons are attached to SM vertices, which are not the reading
vertex and satisfy the Ward identities.

\subsection{The evanescent four-fermion structures}\label{app:evtests}
At one loop a gluon exchanged between the two currents of a four-quark operator produces products of three Dirac matrices on each
quark line. As discussed in section~\ref{sec:schemes}, their reduction to the Dirac structures of the tree-level Feynman rules is unique only in four
dimensions, and the terms of order $\epsilon$ are a choice of scheme, which changes $\Rtwo$. %

A four-fermion Dirac structure is written as $X_1\otimes X_2$, with $X_1$ and $X_2$ the products of Dirac matrices on the two fermion lines, and the
four-dimensional basis is
\begin{equation}\label{eq:sch:basis}
\begin{aligned}
  \{\Gamma_i\otimes\tilde\Gamma_i\}=\{&P_L\otimes P_L,\;P_R\otimes P_R,\;P_L\otimes P_R,\;P_R\otimes P_L,\\
  &\gamma^\mu P_L\otimes\gamma_\mu P_L,\;\gamma^\mu P_R\otimes\gamma_\mu P_R,\;
  \gamma^\mu P_L\otimes\gamma_\mu P_R,\;\gamma^\mu P_R\otimes\gamma_\mu P_L,\\
  &\sigma^{\mu\nu}P_L\otimes\sigma_{\mu\nu}P_L,\;\sigma^{\mu\nu}P_R\otimes\sigma_{\mu\nu}P_R\}.
\end{aligned}
\end{equation}
In $D$ dimensions every structure is split as
\begin{equation}\label{eq:sch:split}
  X_1\otimes X_2=\sum_{i=1}^{10}b_i(X_1,X_2;D)\;\Gamma_i\otimes\tilde\Gamma_i+E(X_1,X_2),\qquad
  b_i=b_i^{(0)}+\epsilon\,b_i^{(1)}+\order{\epsilon^2},
\end{equation}
which defines the evanescent structure $E$. The poles and the one-loop anomalous dimensions use only $b^{(0)}$. We fix the coefficients $b_i$ as in refs.~\cite{Buras:1989xd,Buras:2000if}, which for three gamma matrices on each line gives
\begin{equation}\label{eq:sch:rank3}
\begin{aligned}
  \gamma^\mu\gamma^\nu\gamma^\rho P_L\otimes\gamma_\mu\gamma_\nu\gamma_\rho P_L&=(16-4\epsilon)\,\gamma^\mu P_L\otimes\gamma_\mu P_L+E^{LL}_{3},\\
  \gamma^\mu\gamma^\nu\gamma^\rho P_L\otimes\gamma_\mu\gamma_\nu\gamma_\rho P_R&=(4+4\epsilon)\,\gamma^\mu P_L\otimes\gamma_\mu P_R+E^{LR}_{3},\\
  \gamma^\mu\gamma^\nu\gamma^\rho P_L\otimes\gamma_\rho\gamma_\nu\gamma_\mu P_L&=(4-8\epsilon)\,\gamma^\mu P_L\otimes\gamma_\mu P_L+\tilde E^{LL}_{3},\\
  \gamma^\mu\gamma^\nu\gamma^\rho P_L\otimes\gamma_\rho\gamma_\nu\gamma_\mu P_R&=(16-16\epsilon)\,\gamma^\mu P_L\otimes\gamma_\mu P_R+\tilde E^{LR}_{3},
\end{aligned}
\end{equation}
and the same with $L\leftrightarrow R$. The first relation is eq.~\eqref{eq:sch:ex} with
$b_{\rm ev}=1$. Similar relations for two gamma matrices and for the mixed structures with a tensor are given in
ref.~\cite{Fuentes-Martin:2022vvu}.

\section{On-shell renormalisation factors}\label{app:os}

Section~\ref{sec:uv} gives the on-shell conditions and shows some examples of the renormalisation factors. This appendix collects the relations that are not shown
there. These are the remaining renormalisation factors, the self-energies in closed form, the factors of the top quark, its CP-odd self-energy and the counterterm of the CKM matrix. All counterterms of this appendix carry the factor $1/(16\pi^2)$ explicitly, unlike the vertices $\mathcal V$ of
eq.~\eqref{eq:units}.

\subsection{Remaining renormalisation factors}\label{app:os:lsz}
Eq.~\eqref{eq:mix:dZ} gives the off-diagonal factor of an incoming quark. The factor of an outgoing quark is
\begin{equation}\label{eq:mix:dZbar}
  \delta\bar Z^L_{ij}=\frac{-2}{m_j^2-m_i^2}\Big[m_i^2\Sigma^L_{ij}+m_im_j\Sigma^R_{ij}+m_i\Sigma^{SL}_{ij}+m_j\Sigma^{SR}_{ij}\Big](m_i^2)
  \quad(L\leftrightarrow R,\ SL\leftrightarrow SR),
\end{equation}
as in refs.~\cite{Espriu:2002xv,Endo:2020kie}. On the diagonal, the field renormalisation factor is the real part of
\begin{equation}\label{eq:mix:dZdiag}
  \delta Z^{L,R}_{ii}=\Sigma^{L,R}+m_i^2(\Sigma^{L\prime}+\Sigma^{R\prime})+m_i(\Sigma^{SL\prime}+\Sigma^{SR\prime})
\end{equation}
at $p^2=m_i^2$, with $\Sigma'=\dd\Sigma/\dd p^2$, and the mass counterterm is
\begin{equation}\label{eq:mix:diag}
  {\rm Re}\,\delta m_{ii}=-\tfrac12\big[m_i(\Sigma^L+\Sigma^R)+\Sigma^{SL}+\Sigma^{SR}\big](m_i^2),\qquad
  {\rm Im}\,\delta m_{ii}=-{\rm Im}\,\Sigma^{SL}_{ii}(m_i^2),
\end{equation}
with ${\rm Im}\,\Sigma^{SL}_{ii}=-\Sigma_5$ of section~\ref{sec:uv:cpodd}.

Two massless quarks are mixed only by the four-quark tadpoles of
appendix~\ref{app:os:selfenergies}, which do not depend on the momentum. The mixing is removed by the mass counterterm
$\delta m_{ij}=-\Sigma^{SL}_{ij}(0)$, the off-diagonal analogue of eq.~\eqref{eq:mix:mq}.

For a massless quark the diagonal on-shell field renormalisation factor vanishes as it only contains scaleless integrals.

\subsection{Dipole and four-quark self-energies}\label{app:os:selfenergies}
The dipole self-energy of up-type quarks, diagram (a) of figure~\ref{fig:uv:mix}, is
\begin{equation}\label{eq:mix:dip}
\begin{split}
  \Sigma^{\rm dip}_{ij}=-\frac{C_F\,g_s\,(D-1)\,v}{16\sqrt2\,\pi^2\Lam^2}\Big\{&(F_i+F_j)\big[(C_{uG})_{ij}P_R+(C_{uG})^*_{ji}P_L\big]\\
  &+\frac{\slashed p}{p^2}\,P_L\Big[m_iH_i\,(C_{uG})^*_{ji}+m_jH_j\,(C_{uG})_{ij}\Big]\\
  &+\frac{\slashed p}{p^2}\,P_R\Big[m_iH_i\,(C_{uG})_{ij}+m_jH_j\,(C_{uG})^*_{ji}\Big]\Big\},
\end{split}
\end{equation}
with the two loop functions
\begin{equation}\label{eq:mix:dipKFH}
  F_a=A_0(m_a)+(m_a^2-p^2)B_0(p^2;0,m_a),\quad
  H_a=A_0(m_a)+(p^2-m_a^2)B_0(p^2;0,m_a).
\end{equation}
Here $A_0$ and $B_0$ are the scalar one-point and two-point functions of ref.~\cite{Passarino:1978jh}, in the normalisation of ref.~\cite{Denner:1991kt}.
For down-type quarks $(C_{uG})_{ij}$ is replaced by $\sum_aV^*_{ai}(C_{dG})_{aj}$. The function
$H_a$ vanishes at $p^2=0$, so that $H_a/p^2$ is regular. The pole of $F_a$ is $(2m_a^2-p^2)/\epsilon$, and its $p^2$ term is the pole of
the redundant operator of eq.~\eqref{eq:mix:p2}.

The four-quark tadpoles, diagrams (b) and (c) of figure~\ref{fig:uv:mix}, do not depend on the momentum and have only a scalar part,
\begin{equation}\label{eq:mix:tad}
  \Sigma_{ij}=P_RM^R_{ij}+P_LM^L_{ij},\qquad M^L=M^{R\dagger} ,
\end{equation}
with, in the up sector,
\begin{equation}\label{eq:mix:4q}
\begin{split}
  M^{R}_{ij}={}&\frac{D\,m_{u,a}A_0(m_{u,a})}{16\pi^2\Lam^2}\big(C^{(1)}_{qu}+C_FC^{(8)}_{qu}\big)_{iaaj}\\
  &-\frac{2\,m_{d,b}A_0(m_{d,b})}{16\pi^2\Lam^2}\Big[3\,(C^{(1)}_{quqd})_{ijab}+\tfrac12\big(C^{(1)}_{quqd}+C_FC^{(8)}_{quqd}\big)_{ajib}\Big]V^*_{ab},
\end{split}
\end{equation}
summed over $a$ and $b$, and analogously in the down sector. The diagonal entries give the mass counterterms through
eq.~\eqref{eq:mix:diag}. For the third generation and $V=1$ they agree with the one-loop mass counterterm of the top quark of
ref.~\cite{Duhr:2026btw} and with that of the bottom quark of ref.~\cite{Gauld:2015lmb}. For the top quark and a massless up-type quark
the tadpoles give $\delta\bar Z^{L,R}_{3i}=-\delta Z^{L,R}_{3i}$, which is a rotation of the two fields and enters $\delta V$ through
eq.~\eqref{eq:ckm:dV}.

\subsection{Top-quark renormalisation factors of the chromomagnetic operator}\label{app:os:top}
For the top quark, $i=j=3$, the dipole self-energy of eq.~\eqref{eq:mix:dip} becomes
\begin{equation}\label{eq:mix:diptt}
  \Sigma^{\rm dip}_{tt}=-\frac{\sqrt2\,C_F\,g_s\,(D-1)\,v}{16\pi^2\Lam^2}\Big\{F_t\big[{\rm Re}\,C_{tG}+\ii\,{\rm Im}\,C_{tG}\,\gfive\big]
  +\frac{m_t}{p^2}\,H_t\,{\rm Re}\,C_{tG}\,\slashed p\Big\} ,
\end{equation}
with $C_{tG}=(C_{uG})_{33}$. Its terms with ${\rm Re}\,C_{tG}$, inserted in eqs.~\eqref{eq:mix:dZdiag} and~\eqref{eq:mix:diag}, give the
field and mass renormalisation factors of the top quark,
\begin{equation}\label{eq:uv:ctG}
\begin{aligned}
  \delta Z_t&=-\frac{\sqrt2\,g_sm_tv\,{\rm Re}\,C_{tG}}{2\pi^2\Lam^2}\Big(\frac1{2\epsilon}-\frac16+\frac{L_t}2-L_{\rm eft}\Big),\\
  \frac{\delta m_t}{m_t}&=\frac{\sqrt2\,g_sm_tv\,{\rm Re}\,C_{tG}}{2\pi^2\Lam^2}\Big(\frac3{2\epsilon}+\frac13-L_t-\frac{L_{\rm eft}}2\Big).
\end{aligned}
\end{equation}
The terms with $L_{\rm eft}$ come from the counterterm of the redundant operator of eq.~\eqref{eq:mix:p2} with $i=j=3$. The top-quark
loop with one dipole vertex in the gluon self-energy gives a gluon field factor and a counterterm of $g_s$,
\begin{equation}\label{eq:uv:ctGg}
  \delta Z_g=\frac{\sqrt2\,g_sm_tv\,{\rm Re}\,C_{tG}}{4\pi^2\Lam^2}\Big(\frac1\epsilon-L_t\Big),\qquad
  \frac{\delta g_s}{g_s}=-\frac12\,\delta Z_g .
\end{equation}
These factors multiply the SM vertices. The dipole vertices are multiplied by the SM factors,
\begin{equation}\label{eq:uv:sm0}
\begin{gathered}
  \delta Z_t=\frac{\delta m_t}{m_t}=-\frac{\alpha_s}{4\pi}\,C_F\Big(\frac3\epsilon+4-3L_t\Big),\qquad
  \delta Z_g=-\frac{\alpha_s}{6\pi}\Big(\frac1\epsilon-L_t\Big),\\
  \frac{\delta g_s}{g_s}=-\frac{\alpha_s}{4\pi}\,\frac{23}{6\,\epsilon}+\frac{\alpha_s}{12\pi}\Big(\frac1\epsilon-L_t\Big),
\end{gathered}
\end{equation}
where $23/6$ is half the first coefficient of the beta function of QCD with five flavours. The four-quark tadpole of
eq.~\eqref{eq:mix:4q} with $i=j=a=3$ gives in addition
\begin{equation}\label{eq:uv:dmt4q}
  \frac{\delta m_t}{m_t}=-\frac{m_t^2}{4\pi^2\Lam^2}\,\big(C^{(1)}_{qu}+C_FC^{(8)}_{qu}\big)_{3333}\Big(\frac1\epsilon+\frac12-L_t\Big).
\end{equation}
All the factors of this subsection agree with those of \SMEFTNLO{}, once its parameters are written in our notation as
described in appendix~\ref{app:maps}%
.

\subsection{The CP-odd top self-energy}\label{app:os:cpodd}
The term with $\gfive$ of eq.~\eqref{eq:mix:diptt} is the CP-odd self-energy $\Sigma_5$ of eq.~\eqref{eq:cp:sigma5},
\begin{equation}\label{eq:cp:sigma5loop}
  \Sigma_5(p^2)=-\frac{C_Fg_sv\,{\rm Im}\,(C_{uG})_{33}}{8\sqrt2\,\pi^2\Lam^2}\,(D-1)\,F_t(p^2),
\end{equation}
whose pole is
\begin{equation}\label{eq:cp:sigma5pole}
  \Sigma_5(p^2)\Big|_{\rm pole}=-\frac{g_sv\,{\rm Im}\,(C_{uG})_{33}}{2\sqrt2\,\pi^2\Lam^2}\,\frac{2m_t^2-p^2}{\epsilon}.
\end{equation}
The term proportional to $p^2$ is removed by the counterterm $\delta K_5$ of the redundant operator, and the condition of eq.~\eqref{eq:cp:os},
with $(D-1)F_t(m_t^2)=3m_t^2/\epsilon+m_t^2(1-3L_t)$, gives $\delta m_5$,
\begin{equation}\label{eq:cp:dm5}
\begin{aligned}
  \delta K_5&=\frac{g_sv\,{\rm Im}\,(C_{uG})_{33}}{2\sqrt2\,\pi^2\Lam^2}\Big(\frac1\epsilon-L_{\rm eft}\Big),\\
  \delta m_5&=-\frac{\sqrt2\,g_sm_t^2v\,{\rm Im}\,(C_{uG})_{33}}{2\pi^2\Lam^2}
  \Big(\frac1\epsilon+\frac16-\frac{L_t}2-\frac{L_{\rm eft}}2\Big).
\end{aligned}
\end{equation}
The factors of eq.~\eqref{eq:uv:ctG} and the gluon field factor receive no contribution from ${\rm Im}\,(C_{uG})_{33}$.

\subsection{The counterterm of the CKM matrix}\label{app:os:ckm}
As discussed in section~\ref{sec:uv:ckm}, the off-diagonal field renormalisation factors of section~\ref{sec:uv:mix} contain a rotation
of the quarks, which changes the vertex of the $W$ boson and gives a counterterm of the CKM matrix. We give it in the two schemes of
section~\ref{sec:uv:ckm}, for quarks of different masses. Between two massless quarks the field renormalisation factors have no off-diagonal part, since their mixing is removed by the mass counterterm of appendix~\ref{app:os:lsz}, and the corresponding entries of $A^\psi$ and terms of the sum in eq.~\eqref{eq:ckm:dV} below are zero.

In the on-shell scheme the counterterm is built from the field renormalisation factors of the left-handed quarks as in
ref.~\cite{Denner:1990yz},
\begin{equation}\label{eq:ckm:dV}
  \delta V_{ij}=\frac14\sum_k\Big[\big(\delta Z^{u,L}_{ik}-\delta Z^{u,L*}_{ki}\big)V_{kj}-V_{ik}\big(\delta Z^{d,L}_{kj}-\delta Z^{d,L*}_{jk}\big)\Big].
\end{equation}
It is used with its pole and its finite part, without the imaginary parts of the loop functions, such as the term $-\ii\pi$ of
eq.~\eqref{eq:mix:tq}.

In the $\MSbar$ scheme the counterterm is a pole subtracted at $\mu_{\rm eft}$,
\begin{equation}\label{eq:ckm:dVMS}
  \delta V^{\MSbar}=\frac{\dd V}{\dd\ln\mu_{\rm eft}^2}\,\Big(\frac1\epsilon-L_{\rm eft}\Big),\qquad
  \frac{\dd V}{\dd\ln\mu_{\rm eft}^2}=-A^u\,V+V\,A^d ,
\end{equation}
where the second equation is the running of $V(\mu_{\rm eft})$ at linear order in the SMEFT. For $i\neq j$ and $\psi=u,d$,
\begin{equation}\label{eq:ckm:A}
  A^\psi_{ij}=\frac{1}{m_{\psi,j}^2-m_{\psi,i}^2}\bigg[m_{\psi,j}\,\frac{\dd M^\psi_{ij}}{\dd\ln\mu_{\rm eft}^2}
  +m_{\psi,i}\,\frac{\dd M^{\psi*}_{ji}}{\dd\ln\mu_{\rm eft}^2}\bigg].
\end{equation}
Here $M^\psi$ is the mass matrix of the quarks $\psi$ in the basis where it is diagonal at $\mu_{\rm eft}$. Its derivative is not
diagonal, because the dipoles and the four-quark operators mix into $C_{uH}$ and $C_{dH}$~\cite{Jenkins:2013wua,Alonso:2013hga}. For the up-type quarks it reads
\begin{equation}\label{eq:ckm:Du}
\begin{split}
  \frac{\dd M^u_{ij}}{\dd\ln\mu_{\rm eft}^2}=\frac{1}{16\pi^2\Lam^2}\bigg\{&4\sqrt2g_sv\,\big(m_{u,i}^2+m_{u,j}^2\big)(C_{uG})_{ij}\\
  &-4\sum_k m_{u,k}^3\big(C^{(1)}_{qu}+C_FC^{(8)}_{qu}\big)_{ikkj}\\
  &+\sum_{a,k}m_{d,k}^3V^*_{ak}\Big[6\,(C^{(1)}_{quqd})_{ijak}+\big(C^{(1)}_{quqd}+C_FC^{(8)}_{quqd}\big)_{ajik}\Big]\bigg\},
\end{split}
\end{equation}
and similarly for the down-type quarks
\begin{equation}\label{eq:ckm:Dd}
\begin{split}
  \frac{\dd M^d_{ij}}{\dd\ln\mu_{\rm eft}^2}=\frac{1}{16\pi^2\Lam^2}\sum_aV^*_{ai}\bigg\{&4\sqrt2g_sv\,\big(m_{d,i}^2+m_{d,j}^2\big)(C_{dG})_{aj}\\
  &-4\sum_{b,k}m_{d,k}^3V_{bk}\big(C^{(1)}_{qd}+C_FC^{(8)}_{qd}\big)_{abkj}\\
  &+\sum_k m_{u,k}^3\Big[6\,(C^{(1)}_{quqd})_{kkaj}+\big(C^{(1)}_{quqd}+C_FC^{(8)}_{quqd}\big)_{akkj}\Big]\bigg\}.
\end{split}
\end{equation}
These derivatives are minus the coefficients of the poles of the scalar parts of the self-energies at $p^2=0$, with the dipole terms from eq.~\eqref{eq:mix:dip} and the four-quark terms from eq.~\eqref{eq:mix:4q}. The diagonal elements of $A^\psi$, which eq.~\eqref{eq:ckm:A} does not fix, only change the phases of the quark fields. The matrices $A^\psi$ are anti-Hermitian, and the counterterm of eq.~\eqref{eq:ckm:dV} has the same form, so that the renormalised CKM matrix stays unitary in both schemes. 

The package of appendix~\ref{app:anc} gives both counterterms for any choice of massless quarks. The function
\texttt{OnShellConstants} returns the pole and the finite part of eq.~\eqref{eq:ckm:dV}, and \texttt{UVCounterterm} returns the counterterm vertices with either matrix.

\section{Comparison with \texorpdfstring{\SMEFTNLO{}}{SMEFT@NLO}}\label{app:smeftatnlo}\label{app:smeftnlo:conv}
The \SMEFTNLO{} model~\cite{Degrande:2020evl} is the public model with the rational terms and UV counterterms of the SMEFT at
NLO in QCD, for a flavour assumption with only the top quark massive and a unit CKM matrix. This appendix discusses the comparison. Appendix~\ref{app:maps} writes the parameters of \SMEFTNLO{} in our notation, and appendix~\ref{app:smeftnlo:comp} compares the rational terms and UV counterterms.

\subsection{The parameters of \texorpdfstring{\SMEFTNLO{}}{SMEFT@NLO}}\label{app:maps}
The parameters of \SMEFTNLO{}~\cite{Degrande:2020evl} are defined in terms of the Warsaw coefficients in
ref.~\cite{SMEFTatNLO:definitions}, following ref.~\cite{Aguilar-Saavedra:2018ksv}. Three differences from our conventions change the
vertices. The covariant derivative has the opposite sign, which is equivalent to ours under $G^A_\mu\to-G^A_\mu$. The fields $\gamma$,
$W^\pm$, $G^\pm$ and $G^0$ of \SMEFTNLO{} are minus ours, so that a vertex with $n$ such legs differs from ours by $(-1)^n$. The chromomagnetic operator
contains a factor $g_s$, so that, with $g_s>0$ in both models,
\begin{equation}\label{eq:app:ctG}
  \mathrm{Re}\,(C_{uG})_{33}=-g_s\,c_{tG},\qquad \mathrm{Im}\,(C_{uG})_{33}=-g_s\,c^I_{tG}.
\end{equation}
The operators $O_G$ and $O_{\widetilde G}$ also contain a factor $g_s$, with $C_G=-g_s\,c_G$ and $C_{\widetilde G}=-g_s\,c_{\widetilde G}$.
The Higgs operators of \SMEFTNLO{}, written with $\varphi^\dagger\varphi-v^2/2$, give $c_{\varphi G}=C_{HG}$ and $c_{t\varphi}=\mathrm{Re}\,(C_{uH})_{33}$
after the redefinitions of section~\ref{sec:setup:ewsb}. The sums of eq.~\eqref{eq:lag} run over all indices, so that a heavy--light coefficient of
an operator with two identical currents appears twice, as $C_{ii33}$ and $C_{33ii}$. The four-quark parameters of \SMEFTNLO{} are also
Fierzed combinations of the Warsaw operators. With $i=1,2$ a light generation, three of them read%
\begin{equation}\label{eq:app:nlo4f}
  c^{(1)}_{tu}=2(C_{uu})_{ii33}+\tfrac23(C_{uu})_{i33i},\qquad c^{(8)}_{tu}=4(C_{uu})_{i33i},\qquad
  c^{(8)}_{QQ}=8(C^{(3)}_{qq})_{3333},
\end{equation}
where the entries $i33i$ enter through the colour and four-dimensional Dirac Fierz identities. The consequences of these identities at one
loop are discussed below.

\subsection{Comparison of the rational terms and UV counterterms}\label{app:smeftnlo:comp}
We compared every rational term and UV counterterm of \SMEFTNLO{} with ours. For this we restrict our general-flavour vertices
to its flavour assumption, with only the top quark massive and a unit CKM matrix, and write its parameters %
as in appendix~\ref{app:maps}.

With these maps the effective vertices agree, except for the Goldstone vertices with a closed quark loop and a single $\gfive$. The terms that differ are %
the Levi-Civita terms of the vertices of $O^{(1,3)}_{Hq}$, $O_{Hu}$, $O_{Hd}$ and the chromomagnetic dipole. We
read these loops at the operator, as in eq.~\eqref{eq:g5:string}, and the $Z$ vertices of \SMEFTNLO{} %
agree with this reading. %
 We reproduce its Goldstone vertices %
by changing %
the reading point vertex by vertex%
. For instance, most vertices are reproduced when the trace is read at the vertex of an external gluon leg, and $ggG^0$ of the dipole when it is read at the $G^0$ vertex. For the current operators $ggG^0$ is read at the axial vertex, as in our reading, but $gggG^0$ at a gluon vertex, so that the two vertices do not satisfy the relation of eq.~\eqref{eq:g5:su3}.

With these readings the vertices do not satisfy the $SU(3)$ and electroweak identities of appendix~\ref{app:g5su3}. They enter only in the Feynman gauge, in loop-induced processes in which a $G^0$ or $G^\pm$
connects the gluons to the rest of the process, such as $gg\to Zh$ and $gg\to W^+W^-$~\cite{Rossia:2023hen}. The documentation of \SMEFTNLO{} prescribes the unitary gauge for the anomalous quark-loop amplitudes, such as $gg\to Zh$~\cite{SMEFTatNLO:wiki}, in which these vertices do not contribute, and
in that gauge the model is consistent. In $gg\to Zh$ linear in $c_{tG}$ at 13~TeV the interference with the SM agrees with ours in the
unitary gauge. If the model is instead used in the Feynman gauge, outside its prescription, the interference is 80\% larger.

The agreement includes the colour-octet Wilson coefficients. For $c^{(8)}_{QQ}$, for instance, the \SMEFTNLO{} vertices are those of the Fierz
combination of Warsaw operators on the right-hand side of eq.~\eqref{eq:res:fierzQQ8}, and not those of the octet operator
$O^{(8)}_{QQ}$. The two differ by the shift $\Delta\Lag$ of eq.~\eqref{eq:anc:octshift}. When we write our effective vertices in the flavour assumption of \SMEFTNLO{}, the octet coefficients multiply by default the same
Fierz combination, as in \SMEFTNLO{}. When we write them instead for the octet operators, as in section~\ref{sec:res:octet}, they differ
from those of \SMEFTNLO{} by $\Delta\Lag$.%

\section{Results in electronic format and the Mathematica package}\label{app:anc}
The results are attached to this paper in two forms, a set of files in a machine-readable form (JSON) and the Mathematica package
\texttt{SMEFTR2}, which returns every vertex and its counterterm for any choice of legs. Appendix~\ref{app:anc:files} lists the files and gives the format and the conventions of an entry. The remaining subsections describe the package.

\subsection{Files and format}\label{app:anc:files}\label{app:anc:format}
Table~\ref{tab:anc:files} lists the electronic material. The vertices of section~\ref{sec:res:content} are in the set \texttt{main},
with one file for each contribution of an operator to a vertex. The 4743 contributions are stored in 4814 files, since 71 contributions are also given with the legs in another order. Of these files, 59 are obtained from their conjugates by hermiticity. The 25 contributions of the input shifts are in the set
\texttt{input\_shifts}, since they come from the shifts of the electroweak inputs and of the normalisation of $h$ and $G^0$ of appendix~\ref{app:inputs}, which combine
several operators, and not from the Feynman rules of a single operator. Both sets are in \texttt{functions.zip}, where the file \texttt{FORMAT.txt} describes the format of the files. Table~\ref{tab:res:families} lists the operators with their names in the files and in the package.
\begin{table}[!htbp]
\centering\small
\renewcommand{\arraystretch}{1.2}
\begin{tabularx}{\textwidth}{@{}>{\raggedright\arraybackslash}p{0.2\textwidth}>{\raggedright\arraybackslash}X@{}}
\toprule
folder & content\\
\midrule
\texttt{mathematica} & the package \texttt{SMEFTR2}, with the usage notebook and checks\\
\texttt{functions.zip} & the vertices as data files, one for each contribution of an operator to a vertex, with the list of the files and the description of their format, and the renormalisation factors of appendix~\ref{app:os} as a data file\\
\texttt{documents} & in readable form, the effective vertices of the SMEFT with general flavour, the flavour assumptions of table~\ref{tab:res:assumptions} with their Wilson coefficients in terms of the Warsaw basis, and the renormalisation factors of appendix~\ref{app:os}\\
\bottomrule
\end{tabularx}
\caption{The electronic material attached to the paper.}\label{tab:anc:files}
\end{table}
\begin{table}[!htbp]
\centering\small
\renewcommand{\arraystretch}{1.2}
\begin{tabular}{@{}llrr@{\hspace{22pt}}llrr@{}}
\toprule
name & coefficient & all & $\neq0$ & name & coefficient & all & $\neq0$\\
\midrule
\multicolumn{4}{@{}l}{\emph{Bosonic}} & \multicolumn{4}{l@{}}{\emph{Four-quark}}\\
\texttt{G}       & $C_G$               & 35 & 35  & \texttt{qq1}   & $C^{(1)}_{qq}$     & 259 & 145\\
\texttt{Gtilde}  & $C_{\widetilde G}$  & 35 & 34  & \texttt{qq3}   & $C^{(3)}_{qq}$     & 259 & 145\\
\texttt{HG}      & $C_{HG}$            & 79 & 74  & \texttt{uu}    & $C_{uu}$           & 162 & 41\\
\texttt{HGtilde} & $C_{H\widetilde G}$ & 79 & 16  & \texttt{dd}    & $C_{dd}$           & 162 & 41\\
\multicolumn{4}{@{}l}{\emph{Two quarks}}             & \texttt{ud1}   & $C^{(1)}_{ud}$     & 257 & 95\\
\texttt{uH}      & $C_{uH}$            & 55 & 36  & \texttt{ud8}   & $C^{(8)}_{ud}$     & 257 & 63\\
\texttt{dH}      & $C_{dH}$            & 55 & 36  & \texttt{qu1}   & $C^{(1)}_{qu}$     & 259 & 158\\
\texttt{uG}      & $C_{uG}$            & 297 & 198 & \texttt{qu8}   & $C^{(8)}_{qu}$     & 259 & 133\\
\texttt{dG}      & $C_{dG}$            & 297 & 198 & \texttt{qd1}   & $C^{(1)}_{qd}$     & 259 & 158\\
\texttt{uW}      & $C_{uW}$            & 110 & 0   & \texttt{qd8}   & $C^{(8)}_{qd}$     & 259 & 133\\
\texttt{dW}      & $C_{dW}$            & 110 & 0   & \texttt{quqd1} & $C^{(1)}_{quqd}$   & 257 & 159\\
\texttt{uB}      & $C_{uB}$            & 49 & 0   & \texttt{quqd8} & $C^{(8)}_{quqd}$   & 257 & 159\\
\texttt{dB}      & $C_{dB}$            & 49 & 0   & \multicolumn{4}{l@{}}{\emph{Semileptonic}}\\
\texttt{Hq1}     & $C^{(1)}_{Hq}$      & 90 & 87  & \texttt{lq1}   & $C^{(1)}_{lq}$     & 16 & 12\\
\texttt{Hq3}     & $C^{(3)}_{Hq}$      & 119 & 102 & \texttt{lq3}   & $C^{(3)}_{lq}$     & 22 & 16\\
\texttt{Hu}      & $C_{Hu}$            & 90 & 72  & \texttt{lu}    & $C_{lu}$           & 16 & 10\\
\texttt{Hd}      & $C_{Hd}$            & 90 & 72  & \texttt{ld}    & $C_{ld}$           & 16 & 10\\
\texttt{Hud}     & $C_{Hud}$           & 62 & 32  & \texttt{qe}    & $C_{qe}$           & 8 & 6\\
                 &                     &     &     & \texttt{eu}    & $C_{eu}$           & 8 & 5\\
                 &                     &     &     & \texttt{ed}    & $C_{ed}$           & 8 & 5\\
                 &                     &     &     & \texttt{ledq}  & $C_{ledq}$         & 14 & 8\\
                 &                     &     &     & \texttt{lequ1} & $C^{(1)}_{lequ}$   & 14 & 8\\
                 &                     &     &     & \texttt{lequ3} & $C^{(3)}_{lequ}$   & 14 & 0\\
\midrule
\multicolumn{2}{@{}l}{total} & 4743 & 2502 & & & & \\
\bottomrule
\end{tabular}
\caption{The 39 operators with their names in the electronic files, the number of their contributions and of contributions with a
non-vanishing $\order{C}$ part in $\Rtwo$ or in the pole. The 25 contributions of the input shifts, which combine several operators,
are not included in the counts and are in the separate set \texttt{input\_shifts}%
.}\label{tab:res:families}
\end{table}

A JSON file gives, for each structure $\mathcal C_\alpha\otimes\mathcal T_\beta$ of its vertex, the tree, $\Rtwo$ and the pole as lists of
flavour factors with their coefficients, as in eq.~\eqref{eq:anc:vertex}. Table~\ref{tab:anc:conv} collects the conventions.
\begin{table}[!htbp]
\centering\small
\renewcommand{\arraystretch}{1.1}
\begin{tabularx}{\textwidth}{@{}>{\raggedright\arraybackslash}p{0.27\textwidth}>{\raggedright\arraybackslash}X@{}}
\toprule
 & convention of the files\\
\midrule
generations of the external legs & $i,j,k,l$ for up-type quarks, $r,s,t,u$ for down-type quarks, $\alpha,\beta,\gamma,\delta$ for leptons\\
other flavour indices & summed over the three generations\\
momenta & $p_1,p_2,\dots$ incoming and $q_1,q_2,\dots$ outgoing, the last one eliminated\\
Levi-Civita symbol & $e$, with $\epsilon^{\mu\nu\rho\sigma}=\ii\,e^{\mu\nu\rho\sigma}$\\
SM part & in every file, to be included only once\\
\bottomrule
\end{tabularx}
\caption{Conventions of the vertex files.}\label{tab:anc:conv}
\end{table}

The Wilson coefficients are given by default in the up basis of eq.~\eqref{eq:warsawup}, the WCxf basis \texttt{Warsaw up}~\cite{Aebischer:2017ugx}, with $C=\Lam^2$ times the WCxf value
$c$. The only exceptions are the coefficients of the operators with two identical currents, $C_{ll}$, $C^{(1,3)}_{qq}$, $C_{uu}$ and $C_{dd}$, for which
\begin{equation}\label{eq:anc:wcxf}
  C_{prst}=\frac{\Lam^2}2\,c_{prst}\quad\text{for }(p,r)\ne(s,t),\qquad C_{prst}=\Lam^2\,c_{prst}\quad\text{for }(p,r)=(s,t),
\end{equation}
since the sums in eq.~\eqref{eq:lag} run over all indices. The function \texttt{WCxfName} gives the WCxf name of a coefficient, and \texttt{FromWCxf} converts WCxf values into our coefficients.

\subsection{An example vertex}\label{app:anc:first}
The Mathematica package is loaded from its folder. A vertex is specified by the list of its external legs, with the names
of table~\ref{tab:anc:fields}. A leg is incoming by default, and \texttt{OutLeg} makes it outgoing. The functions \texttt{R2},
\texttt{UVPole} and \texttt{TreeVertex} return $\vf^{\Rtwo}$, $\vf^{\rm UV}$ and the tree of eq.~\eqref{eq:anc:vertex}. The result
contains all operators, unless the option \texttt{"Operators"} selects some of them by their names in table~\ref{tab:res:families}.
\begin{table}[!htbp]
\centering\small
\begin{tabular}{@{}ll@{}}
\toprule
fields & names\\
\midrule
quarks and leptons of open generation & \texttt{U}, \texttt{D}, \texttt{L}, \texttt{N}\\
their antiparticles & \texttt{Ubar}, \texttt{Dbar}, \texttt{Lbar}, \texttt{Nbar}\\
quarks of fixed generation & \texttt{u}, \texttt{c}, \texttt{t}, \texttt{d}, \texttt{s}, \texttt{b}\\
leptons of fixed generation & \texttt{e}, \texttt{mu}, \texttt{ta}, \texttt{ve}, \texttt{vm}, \texttt{vt}\\
gauge bosons & \texttt{g}, \texttt{a}, \texttt{Z}, \texttt{Wp}, \texttt{Wm}\\
Higgs and Goldstone bosons & \texttt{H}, \texttt{G0}, \texttt{Gp}, \texttt{Gm}\\
\bottomrule
\end{tabular}
\caption{Names of the fields in the package. Here \texttt{U}, \texttt{D}, \texttt{L} and \texttt{N} are an up-type quark, a down-type
quark, a charged lepton and a neutrino.}\label{tab:anc:fields}
\end{table}

As an example we take the two-point vertex of eq.~\eqref{eq:sigma}, with an incoming up-type quark of generation $i$ and an
outgoing one of generation $j$.

{\footnotesize
\begin{verbatim}
In[1]:= Get["SMEFTR2.wl"]
In[2]:= v = R2[{"U", OutLeg["U"]}, "Operators" -> {"uG"}, "Flavour" -> {i, j}];
In[3]:= WilsonCoefficients[v]
Out[3]= {Conjugate[WC["uG", {i, j}]], WC["uG", {j, i}]}
In[4]:= Collect[v[WC["uG", {j, i}]], _Str, Factor]
Out[4]= -Sqrt[2] gs vev (7 FMass["u", i]^2 + 7 FMass["u", j]^2 - 6 SProd[p2, p2])
          (Str[1] + Str[2])/(3 Lam^2)
        -2 Sqrt[2] gs vev (FMass["u", i] + FMass["u", j]) Str[3]/(3 Lam^2)
        +2 Sqrt[2] gs vev (FMass["u", i] - FMass["u", j]) Str[4]/(3 Lam^2)
In[5]:= v["Structures"]
Out[5]= <|Str[1] -> Dl[Col[c1], Col[c2]] Gam[Spn[s2], Spn[s1]],
          Str[2] -> Dl[Col[c1], Col[c2]] Gam[5, Spn[s2], Spn[s1]],
          Str[3] -> Dl[Col[c1], Col[c2]] Gam[Lor[$nu1], Spn[s2], Spn[s1]]
                      MomV[p2, Lor[$nu1]],
          Str[4] -> Dl[Col[c1], Col[c2]] Gam[Lor[$nu1], 5, Spn[s2], Spn[s1]]
                      MomV[p2, Lor[$nu1]]|>
\end{verbatim}
}
\noindent
A vertex is a sum of coefficients times structures. Here, \texttt{p2} is the momentum of the outgoing quark, and the output is the part of eq.~\eqref{eq:sigma} proportional
to $(C_{uG})_{ji}$. The vertex also contains its SM part, given
by \texttt{v["SM"]}. The numbering of the structures \texttt{Str[n]} is specific to each vertex. The input shifts
appear as the symbols \texttt{dGF}, \texttt{dgW}, \texttt{dg1}, \texttt{dkH}, \texttt{dkG0} and \texttt{WC["HWB"]}. They are selected
with the name \texttt{"shifts"} in the option \texttt{"Operators"}, and \texttt{"Convention" -> "Shifts"} gives their contributions alone. Table~\ref{tab:anc:symbols} lists the symbols
of the output.
\begin{table}[!htbp]
\centering\small
\renewcommand{\arraystretch}{1.1}
\begin{tabularx}{\textwidth}{@{}>{\raggedright\arraybackslash}p{0.36\textwidth}>{\raggedright\arraybackslash}X@{}}
\toprule
symbol & meaning\\
\midrule
\texttt{Str[1]}, \texttt{Str[2]}, \dots & the structures $\mathcal C_\alpha\otimes\mathcal T_\beta$ of the vertex\\
\texttt{WC["uG", \{i, j\}]} & the Wilson coefficient $(C_{uG})_{ij}$, and likewise for the other operators\\
\texttt{Gam} & a product of Dirac matrices with its two spinor indices, with \texttt{5} for $\gfive$\\
\texttt{Tg}, \texttt{Dl} & the colour generator, the Kronecker delta or the metric tensor\\
\texttt{MomV}, \texttt{SProd} & a momentum component, a scalar product\\
\texttt{LCTensor} & the Levi-Civita tensor with $\epsilon^{0123}=+1$\\
\texttt{gs}, \texttt{ee}, \texttt{sw}, \texttt{cw}, \texttt{vev}, \texttt{Lam} & $g_s$, $e$, $s_\theta$, $c_\theta$, $v$ and $\Lam$\\
\texttt{FMass["u", i]}, \texttt{Yuk["u", i]} & the mass $m_{u,i}$ and the Yukawa coupling $y_{u,i}$\\
\texttt{VCKM[i, j]} & the element $V_{ij}$ of the CKM matrix\\
\texttt{FlavourSum} & a sum over the internal generations listed in its second argument\\
\texttt{OSLog[m]}, \texttt{Leff} & $\ln(m^2/\mu_R^2)$ and $L_{\rm eft}$ in the renormalisation factors\\
\texttt{Par["cQQ8"]} & a parameter of a flavour assumption\\
\texttt{WCDown}, \texttt{WCRot} & the Wilson coefficients in the down basis and in a basis chosen with \texttt{RotateDoublets}\\
\texttt{dVCKM[i, j]} & the shift $\delta V_{ij}$ of the CKM input scheme\\
\texttt{Lor}, \texttt{Col}, \texttt{Adj}, \texttt{Spn} & the type of an index, Lorentz, colour, adjoint or spinor\\
\texttt{epsilonDR} & the regulator $\epsilon$ in the counterterms\\
\texttt{NLight} & the number of massless quark flavours in the renormalisation factor of $g_s$\\
\texttt{dGF}, \texttt{dgW}, \texttt{dg1}, \texttt{dkH}, \texttt{dkG0} & the input shifts $\delta G_F$, $\delta g_2$, $\delta g_1$, $\delta k_H$ and $\delta k_{G^0}$\\
\texttt{GreenOp}, \texttt{UVConst} & an operator of $\Delta\Lag$ and a renormalisation factor with open generations\\
\texttt{p1}, \texttt{s1}, \texttt{c1}, \texttt{a1}, \texttt{mu1}, \texttt{f1} & the default names of the momentum and of the spinor, colour, adjoint, Lorentz and generation indices of leg 1, and likewise for the other legs, which are protected\\
\bottomrule
\end{tabularx}
\caption{Symbols in the output of the package.}\label{tab:anc:symbols}
\end{table}

The sign of a vertex
with fermions follows the order in which its fermion legs are listed. With the option \texttt{"FeynmanRule"} the functions include the factors of eq.~\eqref{eq:anc:norm}, $\ii/(16\pi^2)$ for the rational term and $\ii/(16\pi^2\epsilon)$ for the pole.

\subsection{Generations and flavour sums}\label{app:anc:flavour}
With open generations the flavour factors of section~\ref{sec:flavour} appear as sums over internal generations. The function
\texttt{CKMUnitarity} uses the unitarity of the CKM matrix in these sums, and \texttt{EvaluateFlavour} fixes the generations and carries
out the sums. As an example we take the pole of the penguin vertex $b\bar sg$ of eq.~\eqref{eq:res:penguin}, in which all the coefficients are inside flavour sums.

{\footnotesize
\begin{verbatim}
In[6]:= bsg = UVPole[{"b", OutLeg["s"], OutLeg["g"]}, "Operators" -> {"qq1"}];
In[7]:= ev = EvaluateFlavour[CKMUnitarity[bsg]];
In[8]:= {Length[WilsonCoefficients[bsg]], Length[WilsonCoefficients[ev]]}
Out[8]= {0, 27}
In[9]:= Factor[Coefficient[ev[WC["qq1", {2, 3, 3, 3}]], Str[5]]]
Out[9]= 2 gs Conjugate[VCKM[2, 2]] VCKM[3, 3]
          (SProd[p1, p1] - 2 SProd[p1, p2] + SProd[p2, p2])/(3 Lam^2)
\end{verbatim}
}
\noindent
Here \texttt{Str[5]} is $\gamma^\mu T^A$, and \texttt{Out[9]} is proportional to $(p_1-p_2)^2$, the square of the momentum of the gluon.

The function \texttt{RotateDoublets} writes a vertex in the coefficients \texttt{WCDown[...]} of the WCxf basis \texttt{Warsaw}
of ref.~\cite{Aebischer:2017ugx}, in which the quark doublets are aligned with the down-type quarks, or back in the coefficients of the
paper, or in those of any other basis obtained by a unitary rotation of the doublets%
. After \texttt{SetMassBasis["Warsaw"]} every function
returns its results in the basis \texttt{Warsaw}, and \texttt{FromWCxf} reads values given in that basis.

\subsection{Counterterms and renormalisation factors}\label{app:anc:ct}
The function \texttt{OnShellConstants} returns the renormalisation factors of appendix~\ref{app:os} for given generations. Its option
\texttt{"Massless"} sets the massless quarks. In the five-flavour scheme the factor $\delta Z^L_{13}$ of eq.~\eqref{eq:mix:tq} and the
rotation angle of eq.~\eqref{eq:cp:rot} are obtained as follows.

{\footnotesize
\begin{verbatim}
In[10]:= fs5 = "Massless" -> {"u" -> {1, 2}, "d" -> {1, 2, 3}};
In[11]:= OnShellConstants["dZL_uG_offdiag", "u", {1, 3}, fs5]
Out[11]= <|"Pole" -> -3 gs vev FMass["u", 3] WC["uG", {1, 3}]/(2 Sqrt[2] Lam^2 Pi^2),
           "Finite" -> gs vev FMass["u", 3] (2 + 6 Leff + 3 I Pi + 3 OSLog[FMass["u", 3]])
                         WC["uG", {1, 3}]/(6 Sqrt[2] Lam^2 Pi^2),
           "DefinedAt" -> "incoming quark j on shell, p^2 = m[j]^2"|>
In[12]:= OnShellConstants["theta_uG_diag", "u", {3, 3}, fs5][[{"Pole", "Finite"}]]
Out[12]= <|"Pole" -> gs vev FMass["u", 3] Im[WC["uG", {3, 3}]]/(Sqrt[2] Lam^2 Pi^2),
           "Finite" -> -gs vev FMass["u", 3] Im[WC["uG", {3, 3}]]
                         (-1 + 3 Leff + 3 OSLog[FMass["u", 3]])/(6 Sqrt[2] Lam^2 Pi^2)|>
In[13]:= ct = UVCounterterm[{"t", OutLeg["t"], OutLeg["g"]}, fs5, "Operators" -> {"uG"}];
\end{verbatim}
}
\noindent
The function \texttt{UVCounterterm} returns the complete counterterm vertex $\delta c$ of eq.~\eqref{eq:uv:rule} for any legs, with the
factor $1/(16\pi^2)$ included. The option \texttt{"CKM"} chooses the CKM matrix in $\MSbar$ or on shell. The pole and the finite
part are \texttt{ct["Pole"]} and \texttt{ct["Finite"]}, and \texttt{ct["GammaC"]} is $\Gamma^{\rm UV}_C$.

The function \texttt{CKMInputShift} returns $\Delta$, $\delta W$ and $\delta V$ of eq.~\eqref{eq:ckm:input} in terms of the Wilson
coefficients, with the inputs of ref.~\cite{Descotes-Genon:2018foz} as defaults, and \texttt{CKMShiftVertex} returns, for any vertex,
its SM part with $V$ replaced by $-\delta V$. The vertex in the input scheme of
appendix~\ref{app:ckminput} is the vertex plus this shift.

\subsection{Flavour assumptions and colour octets}\label{app:anc:schemes}
The function \texttt{ApplyFlavour} writes a vertex in the parameters of one of the flavour assumptions of
table~\ref{tab:res:assumptions}, and \texttt{DefineFlavourScheme} defines a new assumption. The function \texttt{OctetShift} returns
$\Delta\Lag$ of section~\ref{sec:res:octet} for an octet parameter.

{\footnotesize
\begin{verbatim}
In[14]:= FlavourSchemes[]
Out[14]= {"SMEFTsim_U35", "SMEFTsim_MFV", "SMEFTsim_top", "SMEFTsim_topU3l",
         "SMEFTatNLO_CPV"}
In[15]:= top = ApplyFlavour[R2[{"t", OutLeg["t"], OutLeg["g"]}, "Operators" -> {"uG"}],
           "SMEFTsim_top"];
In[16]:= top[Par["ctGIm"]]
Out[16]= -37 I gs^2 vev (Str[3] + Str[4])/(6 Sqrt[2] Lam^2)
           - 53 gs^2 vev (Str[9] - Str[10])/(6 Sqrt[2] Lam^2)
In[17]:= OctetShift["SMEFTsim_top", "cQQ8"]
Out[17]= Par["cQQ8"] (ee GreenOp["B", "q", 3]/(9 cw) - 4 gs GreenOp["G", "q", 3]/3
           + ee GreenOp["W", "q", 3]/sw + GreenOp["H1", "q", 3] Yuk["u", 3]^2/2
           - 3 GreenOp["H3", "q", 3] Yuk["u", 3]^2/2)/(16 Lam^2 Pi^2)
In[18]:= DefineFlavourScheme["third",
           {WC[n_String, idx : {(3) ..}] :> WC[n, idx], WC[n_String] :> WC[n]}];
In[19]:= WilsonCoefficients[
           ApplyFlavour[EvaluateFlavour[R2[{"g", "g", OutLeg["H"]}]], "third"]]
Out[19]= {WC["HG"], WC["HGtilde"], Conjugate[WC["dG", {3, 3}]], WC["dG", {3, 3}],
          Conjugate[WC["dH", {3, 3}]], WC["dH", {3, 3}], Conjugate[WC["uG", {3, 3}]],
          WC["uG", {3, 3}], Conjugate[WC["uH", {3, 3}]], WC["uH", {3, 3}]}
\end{verbatim}
}
\noindent
With $g_1=e/c_\theta$ and $g_2=e/s_\theta$, \texttt{Out[17]} is eq.~\eqref{eq:anc:octshift}. The function \texttt{OctetShiftVertex}
gives the Feynman rules of $\Delta\Lag$ for any legs, as a vertex object like that of \texttt{R2}. The assumption defined in \texttt{In[18]} keeps the coefficients without indices
and those with third-generation indices only.

\subsection{Interfaces}\label{app:anc:interfaces}
The function \texttt{ToFeynCalc} writes a vertex with \textsc{FeynCalc}
objects~\cite{Mertig:1990an,Shtabovenko:2023idz}, \texttt{ToLaTeX} writes it as \LaTeX{} source in the layout
of the catalogue, and \texttt{FromWCxf} translates coefficients given in WCxf. We
use the rational term of $t\bar tZ$ at first order in $(C^{(3)}_{Hq})_{33}$ and the loop-induced vertex $ggh$ as examples.

{\footnotesize
\begin{verbatim}
In[20]:= Needs["FeynCalc`"]
In[21]:= Simplify[ToFeynCalc[R2[{"t", OutLeg["t"], OutLeg["Z"]}, "Operators" -> {"Hq3"}],
           WC["Hq3", {3, 3}]]]
Out[21]= 2 ee gs^2 vev^2 (DiracChain[DiracGamma[LorentzIndex[mu3]],
           DiracIndex[s2], DiracIndex[s1]] - DiracChain[DiracGamma[LorentzIndex[mu3]]
           . DiracGamma[5], DiracIndex[s2], DiracIndex[s1]])
           SUNFDelta[SUNFIndex[c1], SUNFIndex[c2]]/(3 cw Lam^2 sw)
In[22]:= ggh = EvaluateFlavour[R2[{"g", "g", OutLeg["H"]}, "Operators" -> {"uG"}]];
In[23]:= Simplify[ggh[WC["uG", {3, 3}]] - ggh[Conjugate[WC["uG", {3, 3}]]]]
Out[23]= -(16 I/3) gs vev Str[6] Yuk["u", 3]/Lam^2
In[24]:= ggh["Structures"][Str[6]]
Out[24]= Dl[Adj[a1], Adj[a2]] LCTensor[p1, p2, Lor[mu1], Lor[mu2]]
In[25]:= FromWCxf[{"phiq1_13" -> x, "qq1_1122" -> y}]
Out[25]= {WC["Hq1", {1, 3}] -> Lam^2 x, WC["Hq1", {3, 1}] -> Lam^2 Conjugate[x],
          WC["qq1", {1, 1, 2, 2}] -> Lam^2 y/2, WC["qq1", {2, 2, 1, 1}] -> Lam^2 y/2}
\end{verbatim}
}
\noindent
With $y_t=\sqrt2\,m_t/v$, \texttt{Out[23]} times $\ii$ is the coefficient of $\Im(C_{uG})_{33}$ in the top-quark term of eq.~\eqref{eq:res:ggh}. The translation of
\texttt{Out[25]} includes the factor $\frac12$ of eq.~\eqref{eq:anc:wcxf}.

\subsection{Summary of the functions}\label{app:anc:summary}
Table~\ref{tab:anc:functions} lists the functions of the package. Each has a usage message, most of them with an example. The package comes with a
usage notebook, and its function \texttt{RunChecks} runs all the checks of the package.
\begin{table}[!htbp]
\centering\small
\renewcommand{\arraystretch}{1.1}
\begin{tabularx}{\textwidth}{@{}>{\raggedright\arraybackslash}p{0.3\textwidth}>{\raggedright\arraybackslash}X@{}}
\toprule
function & returns\\
\midrule
\multicolumn{2}{@{}l}{\emph{Vertices}}\\
\texttt{TreeVertex} & the tree-level Feynman rule\\
\texttt{R2}, \texttt{UVPole} & $\vf^{\Rtwo}$ and $\vf^{\rm UV}$ of a vertex\\
\texttt{UVct} & the pole with the opposite sign, $-\ii\,\vf^{\rm UV}/(16\pi^2\epsilon)$\\
\texttt{OutLeg}, \texttt{InLeg}, \texttt{Leg} & an outgoing leg, an incoming leg, a leg with its own index names\\
\texttt{ListVertices} & the 396 sets of external fields\\
\texttt{Operators}, \texttt{ListOperators} & the operators that contribute to a vertex, all operators\\
\texttt{WilsonCoefficients} & the Wilson coefficients of a vertex\\
\texttt{VertexTable}, \texttt{R2UVTable} & the structures against chosen coefficients, and the tree, $\Rtwo$ and pole side by side\\
\texttt{EliminateMomentum} & a vertex without the momentum of a chosen leg\\
\addlinespace
\multicolumn{2}{@{}l}{\emph{Flavour}}\\
\texttt{EvaluateFlavour} & a vertex with fixed generations and the flavour sums carried out\\
\texttt{CKMUnitarity} & a vertex with the unitarity of $V$ used in the flavour sums\\
\texttt{FlavourSchemes} & the flavour assumptions and their maps\\
\texttt{ApplyFlavour} & a vertex in the parameters of an assumption\\
\texttt{DefineFlavourScheme}, \texttt{ClearFlavourScheme} & the definition and the removal of an assumption of the user\\
\texttt{OctetShift}, \texttt{OctetShiftVertex} & $\Delta\Lag$ of section~\ref{sec:res:octet} and its Feynman rules\\
\texttt{OctetOperator} & a colour-octet four-quark operator\\
\addlinespace
\multicolumn{2}{@{}l}{\emph{Renormalisation}}\\
\texttt{UVCounterterm} & the complete counterterm vertex $\delta c$ of eq.~\eqref{eq:uv:rule}\\
\texttt{OnShellConstants} & the renormalisation factors of appendix~\ref{app:os}\\
\texttt{CKMInputShift} & the shifts of the CKM input scheme of appendix~\ref{app:ckminput}\\
\texttt{CKMShiftVertex} & the shift of a vertex in that scheme\\
\texttt{RotateDoublets} & a result in another basis %
of the quark doublets\\
\texttt{SetMassBasis}, \texttt{MassBasis} & all results in the basis \texttt{Warsaw} or \texttt{Warsaw up}, and the basis in use\\
\texttt{EvaluateUVConstants} & the values of the renormalisation factors in a counterterm with open generations\\
\addlinespace
\multicolumn{2}{@{}l}{\emph{Interfaces and information}}\\
\texttt{ToFeynCalc} & a vertex or an expression with \textsc{FeynCalc} objects\\
\texttt{ToLaTeX} & a vertex or an expression as \LaTeX{} source, in the layout of the catalogue\\
\texttt{WCxfName}, \texttt{FromWCxf} & the WCxf name of a coefficient, rules for the coefficients from WCxf values\\
\texttt{SMEFTR2Info} & a summary of names and conventions\\
\texttt{RunChecks} & the result of the checks of the package\\
\bottomrule
\end{tabularx}
\caption{The functions of the package \texttt{SMEFTR2}.}\label{tab:anc:functions}
\end{table}

The package released with this paper contains the vertices of this paper only, that is, QCD corrections at first order in the Wilson coefficients for $N_c=3$ and for
vertices with at least one coloured leg.

\bibliographystyle{JHEP}
\bibliography{refs}

\providecommand{\href}[2]{#2}\begingroup\raggedright\begin{thebibliography}{100}

\bibitem{Grzadkowski:2010es}
B.~Grzadkowski, M.~Iskrzynski, M.~Misiak and J.~Rosiek, \emph{{Dimension-Six
  Terms in the Standard Model Lagrangian}},
  \href{https://doi.org/10.1007/JHEP10(2010)085}{\emph{JHEP} {\bfseries 10}
  (2010) 085} [\href{https://arxiv.org/abs/1008.4884}{{\ttfamily 1008.4884}}].

\bibitem{Jenkins:2013zja}
E.E.~Jenkins, A.V.~Manohar and M.~Trott, \emph{{Renormalization Group Evolution
  of the Standard Model Dimension Six Operators I: Formalism and lambda
  Dependence}}, \href{https://doi.org/10.1007/JHEP10(2013)087}{\emph{JHEP}
  {\bfseries 10} (2013) 087} [\href{https://arxiv.org/abs/1308.2627}{{\ttfamily
  1308.2627}}].

\bibitem{Jenkins:2013wua}
E.E.~Jenkins, A.V.~Manohar and M.~Trott, \emph{{Renormalization Group Evolution
  of the Standard Model Dimension Six Operators II: Yukawa Dependence}},
  \href{https://doi.org/10.1007/JHEP01(2014)035}{\emph{JHEP} {\bfseries 01}
  (2014) 035} [\href{https://arxiv.org/abs/1310.4838}{{\ttfamily 1310.4838}}].

\bibitem{Alonso:2013hga}
R.~Alonso, E.E.~Jenkins, A.V.~Manohar and M.~Trott, \emph{{Renormalization
  Group Evolution of the Standard Model Dimension Six Operators III: Gauge
  Coupling Dependence and Phenomenology}},
  \href{https://doi.org/10.1007/JHEP04(2014)159}{\emph{JHEP} {\bfseries 04}
  (2014) 159} [\href{https://arxiv.org/abs/1312.2014}{{\ttfamily 1312.2014}}].

\bibitem{Bern:2020ikv}
Z.~Bern, J.~Parra-Martinez and E.~Sawyer, \emph{{Structure of two-loop SMEFT
  anomalous dimensions via on-shell methods}},
  \href{https://doi.org/10.1007/JHEP10(2020)211}{\emph{JHEP} {\bfseries 10}
  (2020) 211} [\href{https://arxiv.org/abs/2005.12917}{{\ttfamily
  2005.12917}}].

\bibitem{Born:2024mgz}
L.~Born, J.~Fuentes-Mart{\'\i}n, S.~Kvedarait{\.{e}} and A.E.~Thomsen,
  \emph{{Two-loop running in the bosonic SMEFT using functional methods}},
  \href{https://doi.org/10.1007/JHEP05(2025)121}{\emph{JHEP} {\bfseries 05}
  (2025) 121} [\href{https://arxiv.org/abs/2410.07320}{{\ttfamily
  2410.07320}}].

\bibitem{DiNoi:2024ajj}
S.~Di~Noi, R.~Gr{\"o}ber and M.K.~Mandal, \emph{{Two-loop running effects in
  Higgs physics in Standard Model Effective Field Theory}},
  \href{https://doi.org/10.1007/JHEP12(2024)220}{\emph{JHEP} {\bfseries 12}
  (2024) 220} [\href{https://arxiv.org/abs/2408.03252}{{\ttfamily 2408.03252}}]
  [Erratum: JHEP 04, 037 (2026)].

\bibitem{DiNoi:2025tka}
S.~Di~Noi, B.A.~Erdelyi and R.~Gr{\"o}ber, \emph{{Complete two-loop
  Yukawa-induced running of the Higgs-gluon coupling in SMEFT}},
  \href{https://doi.org/10.1007/JHEP06(2026)177}{\emph{JHEP} {\bfseries 06}
  (2026) 177} [\href{https://arxiv.org/abs/2510.14680}{{\ttfamily
  2510.14680}}].

\bibitem{Thomsen:2025kka}
A.E.~Thomsen, \emph{{On the renormalization group in EFTs: on-shell bases,
  flavor ambiguities, and divergences}},
  \href{https://doi.org/10.1007/JHEP06(2026)011}{\emph{JHEP} {\bfseries 06}
  (2026) 011} [\href{https://arxiv.org/abs/2512.14260}{{\ttfamily
  2512.14260}}].

\bibitem{Duhr:2025zqw}
C.~Duhr, A.~Vasquez, G.~Ventura and E.~Vryonidou, \emph{{Two-loop
  renormalisation of quark and gluon fields in the SMEFT}},
  \href{https://doi.org/10.1007/JHEP07(2025)160}{\emph{JHEP} {\bfseries 07}
  (2025) 160} [\href{https://arxiv.org/abs/2503.01954}{{\ttfamily
  2503.01954}}].

\bibitem{Duhr:2025yor}
C.~Duhr, G.~Ventura and E.~Vryonidou, \emph{{Two-loop renormalisation of quark
  and gluon fields in the SMEFT in the on-shell scheme}},
  \href{https://doi.org/10.1007/JHEP11(2025)046}{\emph{JHEP} {\bfseries 11}
  (2025) 046} [\href{https://arxiv.org/abs/2508.04500}{{\ttfamily
  2508.04500}}].

\bibitem{Duhr:2026btw}
C.~Duhr, G.~Ventura and E.~Vryonidou, \emph{{Two-loop renormalisation of the
  quark fields in the presence of four-fermion SMEFT operators}},
  \href{https://doi.org/10.1007/JHEP08(2026)151}{\emph{JHEP} {\bfseries 08}
  (2026) 151} [\href{https://arxiv.org/abs/2602.01250}{{\ttfamily
  2602.01250}}].

\bibitem{DiNoi:2025arz}
S.~Di~Noi and R.~Gr{\"o}ber, \emph{{Two loops, four tops and two $\gamma_5$
  schemes: A renormalization story}},
  \href{https://doi.org/10.1016/j.physletb.2025.139878}{\emph{Phys. Lett. B}
  {\bfseries 869} (2025) 139878}
  [\href{https://arxiv.org/abs/2507.10295}{{\ttfamily 2507.10295}}].

\bibitem{DiNoi:2023ygk}
S.~Di~Noi, R.~Gr{\"o}ber, G.~Heinrich, J.~Lang and M.~Vitti, \emph{{$\gamma_5$
  schemes and the interplay of SMEFT operators in the Higgs-gluon coupling}},
  \href{https://doi.org/10.1103/PhysRevD.109.095024}{\emph{Phys. Rev. D}
  {\bfseries 109} (2024) 095024}
  [\href{https://arxiv.org/abs/2310.18221}{{\ttfamily 2310.18221}}].

\bibitem{Haisch:2024wnw}
U.~Haisch and L.~Schnell, \emph{{Precision tests of third-generation four-quark
  operators: one- and two-loop matching}},
  \href{https://doi.org/10.1007/JHEP02(2025)038}{\emph{JHEP} {\bfseries 02}
  (2025) 038} [\href{https://arxiv.org/abs/2410.13304}{{\ttfamily
  2410.13304}}].

\bibitem{Haisch:2025vqj}
U.~Haisch and M.~Niggetiedt, \emph{{Precision tests of third-generation
  four-quark operators: $gg \to h$ and $h \to \gamma\gamma$}},
  \href{https://arxiv.org/abs/2507.20803}{{\ttfamily 2507.20803}}.

\bibitem{Haisch:2025lvd}
U.~Haisch, \emph{{Higgs production from anomalous gluon dynamics}},
  \href{https://doi.org/10.1007/JHEP06(2025)004}{\emph{JHEP} {\bfseries 06}
  (2025) 004} [\href{https://arxiv.org/abs/2503.06249}{{\ttfamily
  2503.06249}}].

\bibitem{Born:2026xkr}
L.~Born, J.~Fuentes-Mart{\'\i}n and A.E.~Thomsen, \emph{{Next-to-leading order
  running in the SMEFT}},
  \href{https://doi.org/10.1007/JHEP09(2026)106}{\emph{JHEP} {\bfseries 09}
  (2026) 106} [\href{https://arxiv.org/abs/2601.19974}{{\ttfamily
  2601.19974}}].

\bibitem{Ethier:2021bye}
{\scshape SMEFiT} collaboration, \emph{{Combined SMEFT interpretation of Higgs,
  diboson, and top quark data from the LHC}},
  \href{https://doi.org/10.1007/JHEP11(2021)089}{\emph{JHEP} {\bfseries 11}
  (2021) 089} [\href{https://arxiv.org/abs/2105.00006}{{\ttfamily
  2105.00006}}].

\bibitem{Ellis:2020unq}
J.~Ellis, M.~Madigan, K.~Mimasu, V.~Sanz and T.~You, \emph{{Top, Higgs, Diboson
  and Electroweak Fit to the Standard Model Effective Field Theory}},
  \href{https://doi.org/10.1007/JHEP04(2021)279}{\emph{JHEP} {\bfseries 04}
  (2021) 279} [\href{https://arxiv.org/abs/2012.02779}{{\ttfamily
  2012.02779}}].

\bibitem{deBlas:2025gyz}
J.~de~Blas et~al., \emph{{Physics Briefing Book: Input for the 2026 update of
  the European Strategy for Particle Physics}},
  \href{https://arxiv.org/abs/2511.03883}{{\ttfamily 2511.03883}}.

\bibitem{Armadillo:2026mvp}
T.~Armadillo, E.~Celada, J.~ter Hoeve, F.~Maltoni, L.~Mantani, J.~Rojo et~al.,
  \emph{{New Physics Reach through Precision at Future Colliders: a
  Multi-Pronged Approach}},  \href{https://arxiv.org/abs/2604.16596}{{\ttfamily
  2604.16596}}.

\bibitem{Aoude:2022aro}
R.~Aoude, F.~Maltoni, O.~Mattelaer, C.~Severi and E.~Vryonidou,
  \emph{{Renormalisation group effects on SMEFT interpretations of LHC data}},
  \href{https://doi.org/10.1007/JHEP09(2023)191}{\emph{JHEP} {\bfseries 09}
  (2023) 191} [\href{https://arxiv.org/abs/2212.05067}{{\ttfamily
  2212.05067}}].

\bibitem{DiNoi:2023onw}
S.~Di~Noi and R.~Gr{\"o}ber, \emph{{Renormalisation group running effects in
  $pp\rightarrow t{\bar{t}}h$ in the Standard Model Effective Field Theory}},
  \href{https://doi.org/10.1140/epjc/s10052-024-12661-5}{\emph{Eur. Phys. J. C}
  {\bfseries 84} (2024) 403}
  [\href{https://arxiv.org/abs/2312.11327}{{\ttfamily 2312.11327}}].

\bibitem{Maltoni:2024dpn}
F.~Maltoni, G.~Ventura and E.~Vryonidou, \emph{{Impact of SMEFT renormalisation
  group running on Higgs production at the LHC}},
  \href{https://doi.org/10.1007/JHEP12(2024)183}{\emph{JHEP} {\bfseries 12}
  (2024) 183} [\href{https://arxiv.org/abs/2406.06670}{{\ttfamily
  2406.06670}}].

\bibitem{terHoeve:2025gey}
J.~ter Hoeve, L.~Mantani, J.~Rojo, A.N.~Rossia and E.~Vryonidou,
  \emph{{Connecting scales: RGE effects in the SMEFT at the LHC and future
  colliders}}, \href{https://doi.org/10.1007/JHEP06(2025)125}{\emph{JHEP}
  {\bfseries 06} (2025) 125}
  [\href{https://arxiv.org/abs/2502.20453}{{\ttfamily 2502.20453}}].

\bibitem{Biekotter:2026dlb}
A.~Biek{\"o}tter, L.E.G.~Maskos and B.D.~Pecjak, \emph{{Threshold corrections
  in SMEFT}}, \href{https://doi.org/10.1007/JHEP07(2026)100}{\emph{JHEP}
  {\bfseries 07} (2026) 100}
  [\href{https://arxiv.org/abs/2601.15901}{{\ttfamily 2601.15901}}].

\bibitem{Mantani:2026fao}
L.~Mantani, P.~Olgoso and A.N.~Rossia, \emph{{The effect of the two-loop SMEFT
  RGEs at future colliders}},
  \href{https://arxiv.org/abs/2604.16599}{{\ttfamily 2604.16599}}.

\bibitem{Maltoni:2016yxb}
F.~Maltoni, E.~Vryonidou and C.~Zhang, \emph{{Higgs production in association
  with a top-antitop pair in the Standard Model Effective Field Theory at NLO
  in QCD}}, \href{https://doi.org/10.1007/JHEP10(2016)123}{\emph{JHEP}
  {\bfseries 10} (2016) 123}
  [\href{https://arxiv.org/abs/1607.05330}{{\ttfamily 1607.05330}}].

\bibitem{Degrande:2020evl}
C.~Degrande, G.~Durieux, F.~Maltoni, K.~Mimasu, E.~Vryonidou and C.~Zhang,
  \emph{{Automated one-loop computations in the standard model effective field
  theory}}, \href{https://doi.org/10.1103/PhysRevD.103.096024}{\emph{Phys. Rev.
  D} {\bfseries 103} (2021) 096024}
  [\href{https://arxiv.org/abs/2008.11743}{{\ttfamily 2008.11743}}].

\bibitem{Haisch:2025jqr}
U.~Haisch, J.~Linder, G.~Pelliccioli, E.~Re and G.~Zanderighi,
  \emph{{Polarized-boson pairs at NLO in the SMEFT}},
  \href{https://doi.org/10.1007/JHEP11(2025)080}{\emph{JHEP} {\bfseries 11}
  (2025) 080} [\href{https://arxiv.org/abs/2507.21768}{{\ttfamily
  2507.21768}}].

\bibitem{Bevilacqua:2026udk}
G.~Bevilacqua, M.~Reinartz and M.~Worek, \emph{{SMEFT everywhere: a NLO study
  of $pp \to t\bar{t}H$ with decaying tops}},
  \href{https://arxiv.org/abs/2605.11747}{{\ttfamily 2605.11747}}.

\bibitem{Alwall:2014hca}
J.~Alwall, R.~Frederix, S.~Frixione, V.~Hirschi, F.~Maltoni, O.~Mattelaer
  et~al., \emph{{The automated computation of tree-level and next-to-leading
  order differential cross sections, and their matching to parton shower
  simulations}}, \href{https://doi.org/10.1007/JHEP07(2014)079}{\emph{JHEP}
  {\bfseries 07} (2014) 079} [\href{https://arxiv.org/abs/1405.0301}{{\ttfamily
  1405.0301}}].

\bibitem{Hirschi:2011pa}
V.~Hirschi, R.~Frederix, S.~Frixione, M.V.~Garzelli, F.~Maltoni and R.~Pittau,
  \emph{{Automation of one-loop QCD corrections}},
  \href{https://doi.org/10.1007/JHEP05(2011)044}{\emph{JHEP} {\bfseries 05}
  (2011) 044} [\href{https://arxiv.org/abs/1103.0621}{{\ttfamily 1103.0621}}].

\bibitem{Cascioli:2011va}
F.~Cascioli, P.~Maierhofer and S.~Pozzorini, \emph{{Scattering Amplitudes with
  Open Loops}},
  \href{https://doi.org/10.1103/PhysRevLett.108.111601}{\emph{Phys. Rev. Lett.}
  {\bfseries 108} (2012) 111601}
  [\href{https://arxiv.org/abs/1111.5206}{{\ttfamily 1111.5206}}].

\bibitem{Buccioni:2019sur}
F.~Buccioni, J.-N.~Lang, J.M.~Lindert, P.~Maierh{\"o}fer, S.~Pozzorini,
  H.~Zhang et~al., \emph{{OpenLoops 2}},
  \href{https://doi.org/10.1140/epjc/s10052-019-7306-2}{\emph{Eur. Phys. J. C}
  {\bfseries 79} (2019) 866}
  [\href{https://arxiv.org/abs/1907.13071}{{\ttfamily 1907.13071}}].

\bibitem{Ossola:2006us}
G.~Ossola, C.G.~Papadopoulos and R.~Pittau, \emph{{Reducing full one-loop
  amplitudes to scalar integrals at the integrand level}},
  \href{https://doi.org/10.1016/j.nuclphysb.2006.11.012}{\emph{Nucl. Phys. B}
  {\bfseries 763} (2007) 147}
  [\href{https://arxiv.org/abs/hep-ph/0609007}{{\ttfamily hep-ph/0609007}}].

\bibitem{Ossola:2008xq}
G.~Ossola, C.G.~Papadopoulos and R.~Pittau, \emph{{On the Rational Terms of the
  one-loop amplitudes}},
  \href{https://doi.org/10.1088/1126-6708/2008/05/004}{\emph{JHEP} {\bfseries
  05} (2008) 004} [\href{https://arxiv.org/abs/0802.1876}{{\ttfamily
  0802.1876}}].

\bibitem{Degrande:2011ua}
C.~Degrande, C.~Duhr, B.~Fuks, D.~Grellscheid, O.~Mattelaer and T.~Reiter,
  \emph{{UFO - The Universal FeynRules Output}},
  \href{https://doi.org/10.1016/j.cpc.2012.01.022}{\emph{Comput. Phys. Commun.}
  {\bfseries 183} (2012) 1201}
  [\href{https://arxiv.org/abs/1108.2040}{{\ttfamily 1108.2040}}].

\bibitem{Darme:2023jdn}
L.~Darm{\'e} et~al., \emph{{UFO 2.0: the {\textquoteleft}Universal Feynman
  Output{\textquoteright} format}},
  \href{https://doi.org/10.1140/epjc/s10052-023-11780-9}{\emph{Eur. Phys. J. C}
  {\bfseries 83} (2023) 631}
  [\href{https://arxiv.org/abs/2304.09883}{{\ttfamily 2304.09883}}].

\bibitem{Draggiotis:2009yb}
P.~Draggiotis, M.V.~Garzelli, C.G.~Papadopoulos and R.~Pittau, \emph{{Feynman
  Rules for the Rational Part of the QCD 1-loop amplitudes}},
  \href{https://doi.org/10.1088/1126-6708/2009/04/072}{\emph{JHEP} {\bfseries
  04} (2009) 072} [\href{https://arxiv.org/abs/0903.0356}{{\ttfamily
  0903.0356}}].

\bibitem{Garzelli:2009is}
M.V.~Garzelli, I.~Malamos and R.~Pittau, \emph{{Feynman rules for the rational
  part of the Electroweak 1-loop amplitudes}},
  \href{https://doi.org/10.1007/JHEP10(2010)097}{\emph{JHEP} {\bfseries 01}
  (2010) 040} [\href{https://arxiv.org/abs/0910.3130}{{\ttfamily 0910.3130}}]
  [Erratum: JHEP 10, 097 (2010)].

\bibitem{Pittau:2011qp}
R.~Pittau, \emph{{Primary Feynman rules to calculate the epsilon-dimensional
  integrand of any 1-loop amplitude}},
  \href{https://doi.org/10.1007/JHEP02(2012)029}{\emph{JHEP} {\bfseries 02}
  (2012) 029} [\href{https://arxiv.org/abs/1111.4965}{{\ttfamily 1111.4965}}].

\bibitem{Degrande:2014vpa}
C.~Degrande, \emph{{Automatic evaluation of UV and $R_2$ terms for beyond the
  Standard Model Lagrangians: a proof-of-principle}},
  \href{https://doi.org/10.1016/j.cpc.2015.08.015}{\emph{Comput. Phys. Commun.}
  {\bfseries 197} (2015) 239}
  [\href{https://arxiv.org/abs/1406.3030}{{\ttfamily 1406.3030}}].

\bibitem{Alloul:2013bka}
A.~Alloul, N.D.~Christensen, C.~Degrande, C.~Duhr and B.~Fuks, \emph{{FeynRules
  2.0 - A complete toolbox for tree-level phenomenology}},
  \href{https://doi.org/10.1016/j.cpc.2014.04.012}{\emph{Comput. Phys. Commun.}
  {\bfseries 185} (2014) 2250}
  [\href{https://arxiv.org/abs/1310.1921}{{\ttfamily 1310.1921}}].

\bibitem{Pozzorini:2020hkx}
S.~Pozzorini, H.~Zhang and M.F.~Zoller, \emph{{Rational Terms of UV Origin at
  Two Loops}}, \href{https://doi.org/10.1007/JHEP05(2020)077}{\emph{JHEP}
  {\bfseries 05} (2020) 077}
  [\href{https://arxiv.org/abs/2001.11388}{{\ttfamily 2001.11388}}].

\bibitem{Duhr:2023wsr}
C.~Duhr and P.~Thakkar, \emph{{Rational terms of UV origin to all loop
  orders}}, \href{https://doi.org/10.1007/JHEP07(2024)210}{\emph{JHEP}
  {\bfseries 07} (2024) 210}
  [\href{https://arxiv.org/abs/2310.14551}{{\ttfamily 2310.14551}}].

\bibitem{Degrande:2014tta}
C.~Degrande, F.~Maltoni, J.~Wang and C.~Zhang, \emph{{Automatic computations at
  next-to-leading order in QCD for top-quark flavor-changing neutral
  processes}}, \href{https://doi.org/10.1103/PhysRevD.91.034024}{\emph{Phys.
  Rev. D} {\bfseries 91} (2015) 034024}
  [\href{https://arxiv.org/abs/1412.5594}{{\ttfamily 1412.5594}}].

\bibitem{Degrande:2016dqg}
C.~Degrande, B.~Fuks, K.~Mawatari, K.~Mimasu and V.~Sanz, \emph{{Electroweak
  Higgs boson production in the standard model effective field theory beyond
  leading order in QCD}},
  \href{https://doi.org/10.1140/epjc/s10052-017-4793-x}{\emph{Eur. Phys. J. C}
  {\bfseries 77} (2017) 262}
  [\href{https://arxiv.org/abs/1609.04833}{{\ttfamily 1609.04833}}].

\bibitem{BuarqueFranzosi:2015jrv}
D.~Buarque~Franzosi and C.~Zhang, \emph{{Probing the top-quark chromomagnetic
  dipole moment at next-to-leading order in QCD}},
  \href{https://doi.org/10.1103/PhysRevD.91.114010}{\emph{Phys. Rev. D}
  {\bfseries 91} (2015) 114010}
  [\href{https://arxiv.org/abs/1503.08841}{{\ttfamily 1503.08841}}].

\bibitem{Zhang:2016omx}
C.~Zhang, \emph{{Single Top Production at Next-to-Leading Order in the Standard
  Model Effective Field Theory}},
  \href{https://doi.org/10.1103/PhysRevLett.116.162002}{\emph{Phys. Rev. Lett.}
  {\bfseries 116} (2016) 162002}
  [\href{https://arxiv.org/abs/1601.06163}{{\ttfamily 1601.06163}}].

\bibitem{BessidskaiaBylund:2016jvp}
O.~Bessidskaia~Bylund, F.~Maltoni, I.~Tsinikos, E.~Vryonidou and C.~Zhang,
  \emph{{Probing top quark neutral couplings in the Standard Model Effective
  Field Theory at NLO in QCD}},
  \href{https://doi.org/10.1007/JHEP05(2016)052}{\emph{JHEP} {\bfseries 05}
  (2016) 052} [\href{https://arxiv.org/abs/1601.08193}{{\ttfamily
  1601.08193}}].

\bibitem{Degrande:2018fog}
C.~Degrande, F.~Maltoni, K.~Mimasu, E.~Vryonidou and C.~Zhang,
  \emph{{Single-top associated production with a $Z$ or $H$ boson at the LHC:
  the SMEFT interpretation}},
  \href{https://doi.org/10.1007/JHEP10(2018)005}{\emph{JHEP} {\bfseries 10}
  (2018) 005} [\href{https://arxiv.org/abs/1804.07773}{{\ttfamily
  1804.07773}}].

\bibitem{Artoisenet:2013puc}
P.~Artoisenet et~al., \emph{{A framework for Higgs characterisation}},
  \href{https://doi.org/10.1007/JHEP11(2013)043}{\emph{JHEP} {\bfseries 11}
  (2013) 043} [\href{https://arxiv.org/abs/1306.6464}{{\ttfamily 1306.6464}}].

\bibitem{Demartin:2014fia}
F.~Demartin, F.~Maltoni, K.~Mawatari, B.~Page and M.~Zaro, \emph{{Higgs
  characterisation at NLO in QCD: CP properties of the top-quark Yukawa
  interaction}},
  \href{https://doi.org/10.1140/epjc/s10052-014-3065-2}{\emph{Eur. Phys. J. C}
  {\bfseries 74} (2014) 3065}
  [\href{https://arxiv.org/abs/1407.5089}{{\ttfamily 1407.5089}}].

\bibitem{Page:2013xla}
B.~Page and R.~Pittau, \emph{{$R_{2}$ vertices for the effective ggH theory}},
  \href{https://doi.org/10.1007/JHEP09(2013)078}{\emph{JHEP} {\bfseries 09}
  (2013) 078} [\href{https://arxiv.org/abs/1307.6142}{{\ttfamily 1307.6142}}].

\bibitem{Aguilar-Saavedra:2018ksv}
J.A.~Aguilar-Saavedra et~al., \emph{{Interpreting top-quark LHC measurements in
  the standard-model effective field theory}},
  \href{https://arxiv.org/abs/1802.07237}{{\ttfamily 1802.07237}}.

\bibitem{Thomas:2024dwd}
M.O.A.~Thomas and E.~Vryonidou, \emph{{CP violation in loop-induced diboson
  production}}, \href{https://doi.org/10.1007/JHEP03(2025)038}{\emph{JHEP}
  {\bfseries 03} (2025) 038}
  [\href{https://arxiv.org/abs/2411.00959}{{\ttfamily 2411.00959}}].

\bibitem{Rossia:2024rfo}
A.N.~Rossia and E.~Vryonidou, \emph{{CP-odd effects at NLO in SMEFT WH and ZH
  production}}, \href{https://doi.org/10.1007/JHEP11(2024)142}{\emph{JHEP}
  {\bfseries 11} (2024) 142}
  [\href{https://arxiv.org/abs/2409.00168}{{\ttfamily 2409.00168}}].

\bibitem{Allwicher:2023shc}
L.~Allwicher, C.~Cornella, G.~Isidori and B.A.~Stefanek, \emph{{New physics in
  the third generation. A comprehensive SMEFT analysis and future prospects}},
  \href{https://doi.org/10.1007/JHEP03(2024)049}{\emph{JHEP} {\bfseries 03}
  (2024) 049} [\href{https://arxiv.org/abs/2311.00020}{{\ttfamily
  2311.00020}}].

\bibitem{Bartocci:2023nvp}
R.~Bartocci, A.~Biek{\"o}tter and T.~Hurth, \emph{{A global analysis of the
  SMEFT under the minimal MFV assumption}},
  \href{https://doi.org/10.1007/JHEP05(2024)074}{\emph{JHEP} {\bfseries 05}
  (2024) 074} [\href{https://arxiv.org/abs/2311.04963}{{\ttfamily
  2311.04963}}].

\bibitem{Brivio:2017btx}
I.~Brivio, Y.~Jiang and M.~Trott, \emph{{The SMEFTsim package, theory and
  tools}}, \href{https://doi.org/10.1007/JHEP12(2017)070}{\emph{JHEP}
  {\bfseries 12} (2017) 070}
  [\href{https://arxiv.org/abs/1709.06492}{{\ttfamily 1709.06492}}].

\bibitem{Brivio:2020onw}
I.~Brivio, \emph{{SMEFTsim 3.0 {\textemdash} a practical guide}},
  \href{https://doi.org/10.1007/JHEP04(2021)073}{\emph{JHEP} {\bfseries 04}
  (2021) 073} [\href{https://arxiv.org/abs/2012.11343}{{\ttfamily
  2012.11343}}].

\bibitem{Dedes:2023zws}
A.~Dedes, J.~Rosiek, M.~Ryczkowski, K.~Suxho and L.~Trifyllis, \emph{{SmeftFR
  v3 {\textendash} Feynman rules generator for the Standard Model Effective
  Field Theory}},
  \href{https://doi.org/10.1016/j.cpc.2023.108943}{\emph{Comput. Phys. Commun.}
  {\bfseries 294} (2024) 108943}
  [\href{https://arxiv.org/abs/2302.01353}{{\ttfamily 2302.01353}}].

\bibitem{Hourtz:2026ijm}
L.~Hourtz, T.~Ketheeswaran, M.~Kr{\"a}mer, A.E.~Thomsen, P.~Weber and
  F.~Wilsch, \emph{{UFO Sightings with Matchete: Connecting BSM Lagrangians
  with Event Generators}},  \href{https://arxiv.org/abs/2609.28619}{{\ttfamily
  2609.28619}}.

\bibitem{Fuentes-Martin:2022jrf}
J.~Fuentes-Mart{\'\i}n, M.~K{\"o}nig, J.~Pag{\`e}s, A.E.~Thomsen and F.~Wilsch,
  \emph{{A proof of concept for matchete: an automated tool for matching
  effective theories}},
  \href{https://doi.org/10.1140/epjc/s10052-023-11726-1}{\emph{Eur. Phys. J. C}
  {\bfseries 83} (2023) 662}
  [\href{https://arxiv.org/abs/2212.04510}{{\ttfamily 2212.04510}}].

\bibitem{Carmona:2021xtq}
A.~Carmona, A.~Lazopoulos, P.~Olgoso and J.~Santiago, \emph{{Matchmakereft:
  automated tree-level and one-loop matching}},
  \href{https://doi.org/10.21468/SciPostPhys.12.6.198}{\emph{SciPost Phys.}
  {\bfseries 12} (2022) 198}
  [\href{https://arxiv.org/abs/2112.10787}{{\ttfamily 2112.10787}}].

\bibitem{Kreimer:1989ke}
D.~Kreimer, \emph{{The $\gamma$(5) Problem and Anomalies: A Clifford Algebra
  Approach}}, \href{https://doi.org/10.1016/0370-2693(90)90461-E}{\emph{Phys.
  Lett. B} {\bfseries 237} (1990) 59}.

\bibitem{Korner:1991sx}
J.G.~Korner, D.~Kreimer and K.~Schilcher, \emph{{A Practicable $\gamma_5$
  scheme in dimensional regularization}},
  \href{https://doi.org/10.1007/BF01559471}{\emph{Z. Phys. C} {\bfseries 54}
  (1992) 503}.

\bibitem{Fuentes-Martin:2022vvu}
J.~Fuentes-Mart{\'\i}n, M.~K{\"o}nig, J.~Pag{\`e}s, A.E.~Thomsen and F.~Wilsch,
  \emph{{Evanescent operators in one-loop matching computations}},
  \href{https://doi.org/10.1007/JHEP02(2023)031}{\emph{JHEP} {\bfseries 02}
  (2023) 031} [\href{https://arxiv.org/abs/2211.09144}{{\ttfamily
  2211.09144}}].

\bibitem{Buras:1989xd}
A.J.~Buras and P.H.~Weisz, \emph{{QCD Nonleading Corrections to Weak Decays in
  Dimensional Regularization and 't Hooft-Veltman Schemes}},
  \href{https://doi.org/10.1016/0550-3213(90)90223-Z}{\emph{Nucl. Phys. B}
  {\bfseries 333} (1990) 66}.

\bibitem{Dugan:1990df}
M.J.~Dugan and B.~Grinstein, \emph{{On the vanishing of evanescent operators}},
  \href{https://doi.org/10.1016/0370-2693(91)90680-O}{\emph{Phys. Lett. B}
  {\bfseries 256} (1991) 239}.

\bibitem{Herrlich:1994kh}
S.~Herrlich and U.~Nierste, \emph{{Evanescent operators, scheme dependences and
  double insertions}},
  \href{https://doi.org/10.1016/0550-3213(95)00474-7}{\emph{Nucl. Phys. B}
  {\bfseries 455} (1995) 39}
  [\href{https://arxiv.org/abs/hep-ph/9412375}{{\ttfamily hep-ph/9412375}}].

\bibitem{Dekens:2019ept}
W.~Dekens and P.~Stoffer, \emph{{Low-energy effective field theory below the
  electroweak scale: matching at one loop}},
  \href{https://doi.org/10.1007/JHEP10(2019)197}{\emph{JHEP} {\bfseries 10}
  (2019) 197} [\href{https://arxiv.org/abs/1908.05295}{{\ttfamily 1908.05295}}]
  [Erratum: JHEP 11, 148 (2022)].

\bibitem{Denner:1990yz}
A.~Denner and T.~Sack, \emph{{Renormalization of the quark mixing matrix}},
  \href{https://doi.org/10.1016/0550-3213(90)90557-T}{\emph{Nucl. Phys. B}
  {\bfseries 347} (1990) 203}.

\bibitem{Kniehl:1996bd}
B.A.~Kniehl and A.~Pilaftsis, \emph{{Mixing renormalization in Majorana
  neutrino theories}},
  \href{https://doi.org/10.1016/0550-3213(96)00280-5}{\emph{Nucl. Phys. B}
  {\bfseries 474} (1996) 286}
  [\href{https://arxiv.org/abs/hep-ph/9601390}{{\ttfamily hep-ph/9601390}}].

\bibitem{Espriu:2002xv}
D.~Espriu, J.~Manzano and P.~Talavera, \emph{{Flavor mixing, gauge invariance
  and wave function renormalization}},
  \href{https://doi.org/10.1103/PhysRevD.66.076002}{\emph{Phys. Rev. D}
  {\bfseries 66} (2002) 076002}
  [\href{https://arxiv.org/abs/hep-ph/0204085}{{\ttfamily hep-ph/0204085}}].

\bibitem{Zhang:2014rja}
C.~Zhang, \emph{{Effective field theory approach to top-quark decay at
  next-to-leading order in QCD}},
  \href{https://doi.org/10.1103/PhysRevD.90.014008}{\emph{Phys. Rev. D}
  {\bfseries 90} (2014) 014008}
  [\href{https://arxiv.org/abs/1404.1264}{{\ttfamily 1404.1264}}].

\bibitem{tHooft:1972tcz}
G.~'t~Hooft and M.J.G.~Veltman, \emph{{Regularization and Renormalization of
  Gauge Fields}},
  \href{https://doi.org/10.1016/0550-3213(72)90279-9}{\emph{Nucl. Phys. B}
  {\bfseries 44} (1972) 189}.

\bibitem{Gnendiger:2017pys}
C.~Gnendiger et~al., \emph{{To ${d}$, or not to ${d}$: recent developments and
  comparisons of regularization schemes}},
  \href{https://doi.org/10.1140/epjc/s10052-017-5023-2}{\emph{Eur. Phys. J. C}
  {\bfseries 77} (2017) 471}
  [\href{https://arxiv.org/abs/1705.01827}{{\ttfamily 1705.01827}}].

\bibitem{Breitenlohner:1977hr}
P.~Breitenlohner and D.~Maison, \emph{{Dimensional Renormalization and the
  Action Principle}}, \href{https://doi.org/10.1007/BF01609069}{\emph{Commun.
  Math. Phys.} {\bfseries 52} (1977) 11}.

\bibitem{Aebischer:2017ugx}
J.~Aebischer et~al., \emph{{WCxf: an exchange format for Wilson coefficients
  beyond the Standard Model}},
  \href{https://doi.org/10.1016/j.cpc.2018.05.022}{\emph{Comput. Phys. Commun.}
  {\bfseries 232} (2018) 71}
  [\href{https://arxiv.org/abs/1712.05298}{{\ttfamily 1712.05298}}].

\bibitem{Dedes:2017zog}
A.~Dedes, W.~Materkowska, M.~Paraskevas, J.~Rosiek and K.~Suxho, \emph{{Feynman
  rules for the Standard Model Effective Field Theory in R$_{\xi}$-gauges}},
  \href{https://doi.org/10.1007/JHEP06(2017)143}{\emph{JHEP} {\bfseries 06}
  (2017) 143} [\href{https://arxiv.org/abs/1704.03888}{{\ttfamily
  1704.03888}}].

\bibitem{Descotes-Genon:2018foz}
S.~Descotes-Genon, A.~Falkowski, M.~Fedele, M.~Gonz{\'a}lez-Alonso and
  J.~Virto, \emph{{The CKM parameters in the SMEFT}},
  \href{https://doi.org/10.1007/JHEP05(2019)172}{\emph{JHEP} {\bfseries 05}
  (2019) 172} [\href{https://arxiv.org/abs/1812.08163}{{\ttfamily
  1812.08163}}].

\bibitem{deBlas:2025xhe}
J.~de~Blas, A.~Goncalves, V.~Miralles, L.~Reina, L.~Silvestrini and M.~Valli,
  \emph{{Constraining new physics effective interactions via a global fit of
  electroweak, Drell-Yan, Higgs, top, and flavour observables}},
  \href{https://doi.org/10.1007/JHEP03(2026)013}{\emph{JHEP} {\bfseries 03}
  (2026) 013} [\href{https://arxiv.org/abs/2507.06191}{{\ttfamily
  2507.06191}}].

\bibitem{Berthier:2015oma}
L.~Berthier and M.~Trott, \emph{{Towards consistent Electroweak Precision Data
  constraints in the SMEFT}},
  \href{https://doi.org/10.1007/JHEP05(2015)024}{\emph{JHEP} {\bfseries 05}
  (2015) 024} [\href{https://arxiv.org/abs/1502.02570}{{\ttfamily
  1502.02570}}].

\bibitem{Brivio:2017vri}
I.~Brivio and M.~Trott, \emph{{The Standard Model as an Effective Field
  Theory}}, \href{https://doi.org/10.1016/j.physrep.2018.11.002}{\emph{Phys.
  Rept.} {\bfseries 793} (2019) 1}
  [\href{https://arxiv.org/abs/1706.08945}{{\ttfamily 1706.08945}}].

\bibitem{Weinberg:1959nj}
S.~Weinberg, \emph{{High-energy behavior in quantum field theory}},
  \href{https://doi.org/10.1103/PhysRev.118.838}{\emph{Phys. Rev.} {\bfseries
  118} (1960) 838}.

\bibitem{Nogueira:1991ex}
P.~Nogueira, \emph{{Automatic Feynman Graph Generation}},
  \href{https://doi.org/10.1006/jcph.1993.1074}{\emph{J. Comput. Phys.}
  {\bfseries 105} (1993) 279}.

\bibitem{Vermaseren:2000nd}
J.A.M.~Vermaseren, \emph{{New features of FORM}},
  \href{https://arxiv.org/abs/math-ph/0010025}{{\ttfamily math-ph/0010025}}.

\bibitem{Ruijl:2017dtg}
B.~Ruijl, T.~Ueda and J.~Vermaseren, \emph{{FORM version 4.2}},
  \href{https://arxiv.org/abs/1707.06453}{{\ttfamily 1707.06453}}.

\bibitem{tHooft:1978jhc}
G.~'t~Hooft and M.J.G.~Veltman, \emph{{Scalar One Loop Integrals}},
  \href{https://doi.org/10.1016/0550-3213(79)90605-9}{\emph{Nucl. Phys. B}
  {\bfseries 153} (1979) 365}.

\bibitem{Passarino:1978jh}
G.~Passarino and M.J.G.~Veltman, \emph{{One Loop Corrections for $e^+e^-$
  Annihilation Into $\mu^+\mu^-$ in the Weinberg Model}},
  \href{https://doi.org/10.1016/0550-3213(79)90234-7}{\emph{Nucl. Phys. B}
  {\bfseries 160} (1979) 151}.

\bibitem{Belusca-Maito:2020ala}
H.~B{\'e}lusca-Ma{\"\i}to, A.~Ilakovac, M.~Ma{\dj}or-Bo{\v{z}}inovi{\'c} and
  D.~St{\"o}ckinger, \emph{{Dimensional regularization and
  Breitenlohner-Maison/{\textquoteright}t Hooft-Veltman scheme for $\gamma_5$
  applied to chiral YM theories: full one-loop counterterm and RGE structure}},
  \href{https://doi.org/10.1007/JHEP08(2020)024}{\emph{JHEP} {\bfseries 08}
  (2020) 024} [\href{https://arxiv.org/abs/2004.14398}{{\ttfamily
  2004.14398}}].

\bibitem{Ossola:2007ax}
G.~Ossola, C.G.~Papadopoulos and R.~Pittau, \emph{{CutTools: A Program
  implementing the OPP reduction method to compute one-loop amplitudes}},
  \href{https://doi.org/10.1088/1126-6708/2008/03/042}{\emph{JHEP} {\bfseries
  03} (2008) 042} [\href{https://arxiv.org/abs/0711.3596}{{\ttfamily
  0711.3596}}].

\bibitem{Denner:2016kdg}
A.~Denner, S.~Dittmaier and L.~Hofer, \emph{{COLLIER: a fortran-based Complex
  One-Loop LIbrary in Extended Regularizations}},
  \href{https://doi.org/10.1016/j.cpc.2016.10.013}{\emph{Comput. Phys. Commun.}
  {\bfseries 212} (2017) 220}
  [\href{https://arxiv.org/abs/1604.06792}{{\ttfamily 1604.06792}}].

\bibitem{Binoth:2006hk}
T.~Binoth, J.P.~Guillet and G.~Heinrich, \emph{{Algebraic evaluation of
  rational polynomials in one-loop amplitudes}},
  \href{https://doi.org/10.1088/1126-6708/2007/02/013}{\emph{JHEP} {\bfseries
  02} (2007) 013} [\href{https://arxiv.org/abs/hep-ph/0609054}{{\ttfamily
  hep-ph/0609054}}].

\bibitem{Bredenstein:2008zb}
A.~Bredenstein, A.~Denner, S.~Dittmaier and S.~Pozzorini, \emph{{NLO QCD
  corrections to $t\bar tb\bar b$ production at the LHC: 1. Quark-antiquark
  annihilation}},
  \href{https://doi.org/10.1088/1126-6708/2008/08/108}{\emph{JHEP} {\bfseries
  08} (2008) 108} [\href{https://arxiv.org/abs/0807.1248}{{\ttfamily
  0807.1248}}].

\bibitem{Misiak:1994zw}
M.~Misiak and M.~Munz, \emph{{Two loop mixing of dimension five flavor changing
  operators}}, \href{https://doi.org/10.1016/0370-2693(94)01553-O}{\emph{Phys.
  Lett. B} {\bfseries 344} (1995) 308}
  [\href{https://arxiv.org/abs/hep-ph/9409454}{{\ttfamily hep-ph/9409454}}].

\bibitem{Chetyrkin:1997fm}
K.G.~Chetyrkin, M.~Misiak and M.~Munz, \emph{{Beta functions and anomalous
  dimensions up to three loops}},
  \href{https://doi.org/10.1016/S0550-3213(98)00122-9}{\emph{Nucl. Phys. B}
  {\bfseries 518} (1998) 473}
  [\href{https://arxiv.org/abs/hep-ph/9711266}{{\ttfamily hep-ph/9711266}}].

\bibitem{Kuusela:2019iok}
P.~Kuusela, \emph{{``GammaMaP'' -- A Mathematica Package for Clifford Algebras,
  Gamma Matrices and Spinors}},
  \href{https://arxiv.org/abs/1905.00429}{{\ttfamily 1905.00429}}.

\bibitem{Adler:1969gk}
S.L.~Adler, \emph{{Axial vector vertex in spinor electrodynamics}},
  \href{https://doi.org/10.1103/PhysRev.177.2426}{\emph{Phys. Rev.} {\bfseries
  177} (1969) 2426}.

\bibitem{Bell:1969ts}
J.S.~Bell and R.~Jackiw, \emph{{A PCAC puzzle: $\pi^0 \to \gamma \gamma$ in the
  $\sigma$ model}}, \href{https://doi.org/10.1007/BF02823296}{\emph{Nuovo Cim.
  A} {\bfseries 60} (1969) 47}.

\bibitem{Bardeen:1969md}
W.A.~Bardeen, \emph{{Anomalous Ward identities in spinor field theories}},
  \href{https://doi.org/10.1103/PhysRev.184.1848}{\emph{Phys. Rev.} {\bfseries
  184} (1969) 1848}.

\bibitem{Bouchiat:1972iq}
C.~Bouchiat, J.~Iliopoulos and P.~Meyer, \emph{{An Anomaly Free Version of
  Weinberg's Model}},
  \href{https://doi.org/10.1016/0370-2693(72)90532-1}{\emph{Phys. Lett. B}
  {\bfseries 38} (1972) 519}.

\bibitem{Gross:1972pv}
D.J.~Gross and R.~Jackiw, \emph{{Effect of anomalies on quasirenormalizable
  theories}}, \href{https://doi.org/10.1103/PhysRevD.6.477}{\emph{Phys. Rev. D}
  {\bfseries 6} (1972) 477}.

\bibitem{DiNoi:2025uan}
S.~Di~Noi, R.~Gr{\"o}ber and P.~Olgoso, \emph{{Mapping between $\gamma_5$
  schemes in the Standard Model Effective Field Theory}},
  \href{https://doi.org/10.1007/JHEP09(2025)027}{\emph{JHEP} {\bfseries 09}
  (2025) 027} [\href{https://arxiv.org/abs/2504.00112}{{\ttfamily
  2504.00112}}].

\bibitem{Cornella:2022hkc}
C.~Cornella, F.~Feruglio and L.~Vecchi, \emph{{Gauge invariance and finite
  counterterms in chiral gauge theories}},
  \href{https://doi.org/10.1007/JHEP02(2023)244}{\emph{JHEP} {\bfseries 02}
  (2023) 244} [\href{https://arxiv.org/abs/2205.10381}{{\ttfamily
  2205.10381}}].

\bibitem{OlgosoRuiz:2024dzq}
P.~Olgoso~Ruiz and L.~Vecchi, \emph{{Spurious gauge-invariance and
  {\ensuremath{\gamma}}$_{5}$ in dimensional regularization}},
  \href{https://doi.org/10.1007/JHEP12(2024)080}{\emph{JHEP} {\bfseries 12}
  (2024) 080} [\href{https://arxiv.org/abs/2406.17013}{{\ttfamily
  2406.17013}}].

\bibitem{Feruglio:2020kfq}
F.~Feruglio, \emph{{A Note on Gauge Anomaly Cancellation in Effective Field
  Theories}}, \href{https://doi.org/10.1007/JHEP03(2021)128}{\emph{JHEP}
  {\bfseries 03} (2021) 128}
  [\href{https://arxiv.org/abs/2012.13989}{{\ttfamily 2012.13989}}].

\bibitem{Bonnefoy:2020tyv}
Q.~Bonnefoy, L.~Di~Luzio, C.~Grojean, A.~Paul and A.N.~Rossia, \emph{{Comments
  on gauge anomalies at dimension-six in the Standard Model Effective Field
  Theory}}, \href{https://doi.org/10.1007/JHEP05(2021)153}{\emph{JHEP}
  {\bfseries 05} (2021) 153}
  [\href{https://arxiv.org/abs/2012.07740}{{\ttfamily 2012.07740}}].

\bibitem{Cata:2020crs}
O.~Cata, W.~Kilian and N.~Kreher, \emph{{Gauge anomalies in the Standard-Model
  Effective Field Theory}},  \href{https://arxiv.org/abs/2011.09976}{{\ttfamily
  2011.09976}}.

\bibitem{Tracas:1982gp}
N.~Tracas and N.~Vlachos, \emph{{Two Loop Calculations in {QCD} and the $\Delta
  I = 1/2$ Rule in Nonleptonic Weak Decays}},
  \href{https://doi.org/10.1016/0370-2693(82)90530-5}{\emph{Phys. Lett. B}
  {\bfseries 115} (1982) 419}.

\bibitem{Naterop:2023dek}
L.~Naterop and P.~Stoffer, \emph{{Low-energy effective field theory below the
  electroweak scale: one-loop renormalization in the {\textquoteright}t
  Hooft-Veltman scheme}},
  \href{https://doi.org/10.1007/JHEP02(2024)068}{\emph{JHEP} {\bfseries 02}
  (2024) 068} [\href{https://arxiv.org/abs/2310.13051}{{\ttfamily
  2310.13051}}].

\bibitem{Buras:2000if}
A.J.~Buras, M.~Misiak and J.~Urban, \emph{{Two loop QCD anomalous dimensions of
  flavor changing four quark operators within and beyond the standard model}},
  \href{https://doi.org/10.1016/S0550-3213(00)00437-5}{\emph{Nucl. Phys. B}
  {\bfseries 586} (2000) 397}
  [\href{https://arxiv.org/abs/hep-ph/0005183}{{\ttfamily hep-ph/0005183}}]
  [Erratum: Nucl.Phys.B 1002, 116529 (2024)].

\bibitem{Gherardi:2020det}
V.~Gherardi, D.~Marzocca and E.~Venturini, \emph{{Matching scalar leptoquarks
  to the SMEFT at one loop}},
  \href{https://doi.org/10.1007/JHEP07(2020)225}{\emph{JHEP} {\bfseries 07}
  (2020) 225} [\href{https://arxiv.org/abs/2003.12525}{{\ttfamily 2003.12525}}]
  [Erratum: JHEP 01, 006 (2021)].

\bibitem{Denner:1991kt}
A.~Denner, \emph{{Techniques for calculation of electroweak radiative
  corrections at the one loop level and results for W physics at LEP-200}},
  \href{https://doi.org/10.1002/prop.2190410402}{\emph{Fortsch. Phys.}
  {\bfseries 41} (1993) 307} [\href{https://arxiv.org/abs/0709.1075}{{\ttfamily
  0709.1075}}].

\bibitem{Endo:2020kie}
M.~Endo, S.~Mishima and D.~Ueda, \emph{{Revisiting electroweak radiative
  corrections to $b\to s\ell\ell$ in SMEFT}},
  \href{https://doi.org/10.1007/JHEP05(2021)050}{\emph{JHEP} {\bfseries 05}
  (2021) 050} [\href{https://arxiv.org/abs/2012.06197}{{\ttfamily
  2012.06197}}].

\bibitem{Gambino:1998ec}
P.~Gambino, P.P.~Grassi and F.~Madricardo, \emph{{Fermion mixing
  renormalization and gauge invariance}},
  \href{https://doi.org/10.1016/S0370-2693(99)00321-4}{\emph{Phys. Lett. B}
  {\bfseries 454} (1999) 98}
  [\href{https://arxiv.org/abs/hep-ph/9811470}{{\ttfamily hep-ph/9811470}}].

\bibitem{Kniehl:2006bs}
B.A.~Kniehl and A.~Sirlin, \emph{{Simple Approach to Renormalize the
  Cabibbo-Kobayashi-Maskawa Matrix}},
  \href{https://doi.org/10.1103/PhysRevLett.97.221801}{\emph{Phys. Rev. Lett.}
  {\bfseries 97} (2006) 221801}
  [\href{https://arxiv.org/abs/hep-ph/0608306}{{\ttfamily hep-ph/0608306}}].

\bibitem{Aebischer:2020lsx}
J.~Aebischer and J.~Kumar, \emph{{Flavour violating effects of Yukawa running
  in SMEFT}}, \href{https://doi.org/10.1007/JHEP09(2020)187}{\emph{JHEP}
  {\bfseries 09} (2020) 187}
  [\href{https://arxiv.org/abs/2005.12283}{{\ttfamily 2005.12283}}].

\bibitem{Pospelov:2005pr}
M.~Pospelov and A.~Ritz, \emph{{Electric dipole moments as probes of new
  physics}}, \href{https://doi.org/10.1016/j.aop.2005.04.002}{\emph{Annals
  Phys.} {\bfseries 318} (2005) 119}
  [\href{https://arxiv.org/abs/hep-ph/0504231}{{\ttfamily hep-ph/0504231}}].

\bibitem{Fujikawa:1979ay}
K.~Fujikawa, \emph{{Path integral measure for gauge invariant fermion
  theories}}, \href{https://doi.org/10.1103/PhysRevLett.42.1195}{\emph{Phys.
  Rev. Lett.} {\bfseries 42} (1979) 1195}.

\bibitem{Bhattacharya:2015rsa}
T.~Bhattacharya, V.~Cirigliano, R.~Gupta, E.~Mereghetti and B.~Yoon,
  \emph{{Dimension-5 CP-odd operators: QCD mixing and renormalization}},
  \href{https://doi.org/10.1103/PhysRevD.92.114026}{\emph{Phys. Rev. D}
  {\bfseries 92} (2015) 114026}
  [\href{https://arxiv.org/abs/1502.07325}{{\ttfamily 1502.07325}}].

\bibitem{SMEFTatNLO:definitions}
C.~Degrande, G.~Durieux, F.~Maltoni, K.~Mimasu, E.~Vryonidou and C.~Zhang,
  ``{SMEFT@NLO: definition of the operator basis}.''
  \url{https://feynrules.irmp.ucl.ac.be/attachment/wiki/SMEFTatNLO/definitions.pdf}.

\bibitem{DAmbrosio:2002vsn}
G.~D'Ambrosio, G.F.~Giudice, G.~Isidori and A.~Strumia, \emph{{Minimal flavor
  violation: An Effective field theory approach}},
  \href{https://doi.org/10.1016/S0550-3213(02)00836-2}{\emph{Nucl. Phys. B}
  {\bfseries 645} (2002) 155}
  [\href{https://arxiv.org/abs/hep-ph/0207036}{{\ttfamily hep-ph/0207036}}].

\bibitem{Faroughy:2020ina}
D.A.~Faroughy, G.~Isidori, F.~Wilsch and K.~Yamamoto, \emph{{Flavour symmetries
  in the SMEFT}}, \href{https://doi.org/10.1007/JHEP08(2020)166}{\emph{JHEP}
  {\bfseries 08} (2020) 166}
  [\href{https://arxiv.org/abs/2005.05366}{{\ttfamily 2005.05366}}].

\bibitem{Aebischer:2022aze}
J.~Aebischer and M.~Pesut, \emph{{One-loop Fierz transformations}},
  \href{https://doi.org/10.1007/JHEP10(2022)090}{\emph{JHEP} {\bfseries 10}
  (2022) 090} [\href{https://arxiv.org/abs/2208.10513}{{\ttfamily
  2208.10513}}].

\bibitem{SMEFTatNLO:wiki}
C.~Degrande, G.~Durieux, F.~Maltoni, K.~Mimasu, E.~Vryonidou and C.~Zhang,
  ``{SMEFT@NLO}.'' \url{https://feynrules.irmp.ucl.ac.be/wiki/SMEFTatNLO}.

\bibitem{Zhang:2010bm}
J.J.~Zhang, C.S.~Li, J.~Gao, H.X.~Zhu, C.P.~Yuan and T.-C.~Yuan,
  \emph{{Next-to-leading order QCD corrections to the top quark decay via the
  Flavor-Changing Neutral-Current operators with mixing effects}},
  \href{https://doi.org/10.1103/PhysRevD.82.073005}{\emph{Phys. Rev. D}
  {\bfseries 82} (2010) 073005}
  [\href{https://arxiv.org/abs/1004.0898}{{\ttfamily 1004.0898}}].

\bibitem{Spira:1995rr}
M.~Spira, A.~Djouadi, D.~Graudenz and P.M.~Zerwas, \emph{{Higgs boson
  production at the LHC}},
  \href{https://doi.org/10.1016/0550-3213(95)00379-7}{\emph{Nucl. Phys. B}
  {\bfseries 453} (1995) 17}
  [\href{https://arxiv.org/abs/hep-ph/9504378}{{\ttfamily hep-ph/9504378}}].

\bibitem{Durieux:2014xla}
G.~Durieux, F.~Maltoni and C.~Zhang, \emph{{Global approach to top-quark
  flavor-changing interactions}},
  \href{https://doi.org/10.1103/PhysRevD.91.074017}{\emph{Phys. Rev. D}
  {\bfseries 91} (2015) 074017}
  [\href{https://arxiv.org/abs/1412.7166}{{\ttfamily 1412.7166}}].

\bibitem{UTfit:2007eik}
{\scshape UTfit} collaboration, \emph{{Model-independent constraints on $\Delta
  F=2$ operators and the scale of new physics}},
  \href{https://doi.org/10.1088/1126-6708/2008/03/049}{\emph{JHEP} {\bfseries
  03} (2008) 049} [\href{https://arxiv.org/abs/0707.0636}{{\ttfamily
  0707.0636}}].

\bibitem{Silvestrini:2018dos}
L.~Silvestrini and M.~Valli, \emph{{Model-independent Bounds on the Standard
  Model Effective Theory from Flavour Physics}},
  \href{https://doi.org/10.1016/j.physletb.2019.135062}{\emph{Phys. Lett. B}
  {\bfseries 799} (2019) 135062}
  [\href{https://arxiv.org/abs/1812.10913}{{\ttfamily 1812.10913}}].

\bibitem{LHCb:2017hag}
{\scshape LHCb} collaboration, \emph{{Search for the lepton-flavour violating
  decays B$_{(s)}^{0} \to e^{\pm}\mu^{\mp}$}},
  \href{https://doi.org/10.1007/JHEP03(2018)078}{\emph{JHEP} {\bfseries 03}
  (2018) 078} [\href{https://arxiv.org/abs/1710.04111}{{\ttfamily
  1710.04111}}].

\bibitem{ParticleDataGroup:2020ssz}
{\scshape Particle Data Group} collaboration, \emph{{Review of Particle
  Physics}}, \href{https://doi.org/10.1093/ptep/ptaa104}{\emph{PTEP} {\bfseries
  2020} (2020) 083C01}.

\bibitem{NNPDF:2021njg}
{\scshape NNPDF} collaboration, \emph{{The path to proton structure at 1{\%}
  accuracy}}, \href{https://doi.org/10.1140/epjc/s10052-022-10328-7}{\emph{Eur.
  Phys. J. C} {\bfseries 82} (2022) 428}
  [\href{https://arxiv.org/abs/2109.02653}{{\ttfamily 2109.02653}}].

\bibitem{Afik:2021jjh}
Y.~Afik, S.~Bar-Shalom, A.~Soni and J.~Wudka, \emph{{New flavor physics in di-
  and trilepton events from single-top production at the LHC and beyond}},
  \href{https://doi.org/10.1103/PhysRevD.103.075031}{\emph{Phys. Rev. D}
  {\bfseries 103} (2021) 075031}
  [\href{https://arxiv.org/abs/2101.05286}{{\ttfamily 2101.05286}}].

\bibitem{Altmannshofer:2025lun}
W.~Altmannshofer, Z.~Balme, C.M.~Donohue, S.~Gori and S.V.~Mukundhan,
  \emph{{Targets for flavor-violating top decays}},
  \href{https://doi.org/10.1007/JHEP08(2025)191}{\emph{JHEP} {\bfseries 08}
  (2025) 191} [\href{https://arxiv.org/abs/2504.18664}{{\ttfamily
  2504.18664}}].

\bibitem{Frixione:2008yi}
S.~Frixione, E.~Laenen, P.~Motylinski, B.R.~Webber and C.D.~White,
  \emph{{Single-top hadroproduction in association with a W boson}},
  \href{https://doi.org/10.1088/1126-6708/2008/07/029}{\emph{JHEP} {\bfseries
  07} (2008) 029} [\href{https://arxiv.org/abs/0805.3067}{{\ttfamily
  0805.3067}}].

\bibitem{Aguilar-Saavedra:2010uur}
J.A.~Aguilar-Saavedra, \emph{{Effective four-fermion operators in top physics:
  A Roadmap}},
  \href{https://doi.org/10.1016/j.nuclphysb.2011.06.003}{\emph{Nucl. Phys. B}
  {\bfseries 843} (2011) 638}
  [\href{https://arxiv.org/abs/1008.3562}{{\ttfamily 1008.3562}}] [Erratum:
  Nucl.Phys.B 851, 443--444 (2011)].

\bibitem{Degrande:2011rt}
C.~Degrande, J.-M.~Gerard, C.~Grojean, F.~Maltoni and G.~Servant, \emph{{An
  Effective approach to same sign top pair production at the LHC and the
  forward-backward asymmetry at the Tevatron}},
  \href{https://doi.org/10.1016/j.physletb.2011.08.003}{\emph{Phys. Lett. B}
  {\bfseries 703} (2011) 306}
  [\href{https://arxiv.org/abs/1104.1798}{{\ttfamily 1104.1798}}].

\bibitem{Li:2012dtz}
B.H.~Li, C.S.~Li, H.T.~Li, Y.C.~Zhan, Y.~Zhang and J.~Wang, \emph{{Signature of
  same-sign top pair production mediated by a non-universal $Z^\prime$ with QCD
  NLO accuracy at the LHC}},
  \href{https://doi.org/10.1103/PhysRevD.86.114027}{\emph{Phys. Rev. D}
  {\bfseries 86} (2012) 114027}
  [\href{https://arxiv.org/abs/1208.0418}{{\ttfamily 1208.0418}}].

\bibitem{Mertig:1990an}
R.~Mertig, M.~Bohm and A.~Denner, \emph{{Feyn Calc: Computer algebraic
  calculation of Feynman amplitudes}},
  \href{https://doi.org/10.1016/0010-4655(91)90130-D}{\emph{Comput. Phys.
  Commun.} {\bfseries 64} (1991) 345}.

\bibitem{Shtabovenko:2023idz}
V.~Shtabovenko, R.~Mertig and F.~Orellana, \emph{{FeynCalc 10: Do multiloop
  integrals dream of computer codes?}},
  \href{https://doi.org/10.1016/j.cpc.2024.109357}{\emph{Comput. Phys. Commun.}
  {\bfseries 306} (2025) 109357}
  [\href{https://arxiv.org/abs/2312.14089}{{\ttfamily 2312.14089}}].

\bibitem{Hahn:1998yk}
T.~Hahn and M.~Perez-Victoria, \emph{{Automatized one loop calculations in
  four-dimensions and D-dimensions}},
  \href{https://doi.org/10.1016/S0010-4655(98)00173-8}{\emph{Comput. Phys.
  Commun.} {\bfseries 118} (1999) 153}
  [\href{https://arxiv.org/abs/hep-ph/9807565}{{\ttfamily hep-ph/9807565}}].

\bibitem{Degrande:2024mbg}
C.~Degrande, R.~Rosenfeld and A.~Vasquez, \emph{{Collider sensitivity to SMEFT
  heavy-quark operators at one-loop in top-quark processes}},
  \href{https://doi.org/10.1007/JHEP07(2024)114}{\emph{JHEP} {\bfseries 07}
  (2024) 114} [\href{https://arxiv.org/abs/2402.06528}{{\ttfamily
  2402.06528}}].

\bibitem{Gunion:1996xu}
J.F.~Gunion and X.-G.~He, \emph{{Determining the CP nature of a neutral Higgs
  boson at the LHC}},
  \href{https://doi.org/10.1103/PhysRevLett.76.4468}{\emph{Phys. Rev. Lett.}
  {\bfseries 76} (1996) 4468}
  [\href{https://arxiv.org/abs/hep-ph/9602226}{{\ttfamily hep-ph/9602226}}].

\bibitem{Boudjema:2015nda}
F.~Boudjema, R.M.~Godbole, D.~Guadagnoli and K.A.~Mohan, \emph{{Lab-frame
  observables for probing the top-Higgs interaction}},
  \href{https://doi.org/10.1103/PhysRevD.92.015019}{\emph{Phys. Rev. D}
  {\bfseries 92} (2015) 015019}
  [\href{https://arxiv.org/abs/1501.03157}{{\ttfamily 1501.03157}}].

\bibitem{Miralles:2024huv}
V.~Miralles, Y.~Peters, E.~Vryonidou and J.K.~Winter, \emph{{Sensitivity to $
  \mathcal{CP} $-violating effective couplings in the top-Higgs sector}},
  \href{https://doi.org/10.1007/JHEP06(2025)052}{\emph{JHEP} {\bfseries 06}
  (2025) 052} [\href{https://arxiv.org/abs/2412.10309}{{\ttfamily
  2412.10309}}].

\bibitem{Hermann:2022vit}
J.~Hermann, D.~Stremmer and M.~Worek, \emph{{$ \mathcal{CP} $ structure of the
  top-quark Yukawa interaction: NLO QCD corrections and off-shell effects}},
  \href{https://doi.org/10.1007/JHEP09(2022)138}{\emph{JHEP} {\bfseries 09}
  (2022) 138} [\href{https://arxiv.org/abs/2205.09983}{{\ttfamily
  2205.09983}}].

\bibitem{Hart:2010flint}
W.B.~Hart, \emph{{Fast Library for Number Theory: An Introduction}},
  \href{https://doi.org/10.1007/978-3-642-15582-6_18}{\emph{Lect. Notes Comput.
  Sci.} {\bfseries 6327} (2010) 88}.

\bibitem{Gauld:2015lmb}
R.~Gauld, B.D.~Pecjak and D.J.~Scott, \emph{{One-loop corrections to $h\to
  b\bar b$ and $h\to \tau\bar \tau$ decays in the Standard Model Dimension-6
  EFT: four-fermion operators and the large-$m_t$ limit}},
  \href{https://doi.org/10.1007/JHEP05(2016)080}{\emph{JHEP} {\bfseries 05}
  (2016) 080} [\href{https://arxiv.org/abs/1512.02508}{{\ttfamily
  1512.02508}}].

\bibitem{Rossia:2023hen}
A.~Rossia, M.~Thomas and E.~Vryonidou, \emph{{Diboson production in the SMEFT
  from gluon fusion}},
  \href{https://doi.org/10.1007/JHEP11(2023)132}{\emph{JHEP} {\bfseries 11}
  (2023) 132} [\href{https://arxiv.org/abs/2306.09963}{{\ttfamily
  2306.09963}}].

\end{thebibliography}\endgroup
\end{document}